\documentstyle[12pt]{article}
\begin{document}
{\sf
\title{
{\normalsize
\begin{flushright}
CU-TP-1110\\
RBRC-~~413
\end{flushright}}
New Ways to Solve the Schroedinger Equation\thanks{This research
was supported in part by the U.S. Department of Energy Grant
DE-FG02-92ER-40699 and by the RIKEN-BNL Research Center,
Brookhaven National Laboratory}}

\author{
R. Friedberg\\
{\small \it Physics Department, Columbia University, New York, NY 10027}\\
and\\
T. D. Lee\\
{\small \it Physics Department, Columbia University, New York, NY 10027}\\
{\small \it China Center of Advanced Science and Technology (CCAST)}\\
{\small \it (World Laboratory), P.O. Box 8730, Beijing 100080, People's Republic of China}\\
{\small \it RIKEN BNL Research Center (RBRC), Brookhaven National Laboratory, Upton, NY 11973}}
\maketitle

\begin{abstract}

We discuss a new approach to solve the low lying states of the
Schroedinger equation. For a fairly large class of problems, this
new approach leads to convergent iterative solutions, in contrast
to perturbative series expansions. These convergent solutions
include the long standing difficult problem of a quartic potential
with either symmetric or asymmetric minima.

\end{abstract}
\vspace{2cm}

~~~~PACS{:~~11.10.Ef,~~03.65.Ge}

\newpage

\section*{\bf 1. Introduction}
\setcounter{section}{1}
\setcounter{equation}{0}

Quantum physics is largely governed by the Schroedinger equation.
Yet, exact solutions of the equation are relatively few. Besides
lattice and other numerical calculations, we rely mostly on
perturbative expansions. Such expansion quite often leads to a
divergent series with zero radius of convergence, as in quantum
electrodynamics, quantum chromodynamics and problems involving
tunnelling and instantons. In a series of previous papers[1-4] we
have presented a new approach to solve the low lying states of the
Schroedinger equation. In the special case of one dimensional
problems, this new approach leads to explicit convergent iterative
solutions, in contrast to perturbative series expansions. These
convergent solutions include the long standing difficult problem
[5-14] of a quartic potential with symmetric minima.

In this paper, we discuss some additional results bearing on the
new method. In the one-dimensional case, we show that by changing
the boundary condition to be applied at each iteration, we can
obtain a convergent alternating sequence for the groundstate
energy and wave function instead of the monotonic sequence found
before[4]. This result will be spelled out later in this section
and proved in Section 3. We also find that the asymmetric quartic
double-well potential can be treated by an extension of the
procedure used previously for the symmetric case. This extension
is treated in Section 4.

In addition, we have begun the exploration of higher dimensional
problems along the same line. Although the same kind of iterative
procedure can be set up, the linear inhomogeneous equation to be
solved at each step cannot now be reduced to simple quadratures,
as was done for one dimension. However, it is of interest that
this equation is identical in form to an electrostatic analog
problem with a given position dependent dielectric constant media;
at each $n$th iteration, there is an external electrostatic charge
distribution determined by the $(n-1)$th iterated solution, as we
shall discuss in this section.

Consider the Schroedinger equation
\begin{eqnarray}\label{e1.1}
H\psi = E \psi
\end{eqnarray}
where $H$ is the Hamiltonian operator, $\psi$ the wave function
and $E$ its energy. For different physics problems, $H$ assumes
different forms. For example, for a system of $n$ non-relativistic
particles in three dimensions, $H$ may be written as
\begin{eqnarray}\label{e1.2}
H=\sum_{i,j}C_{ij}p_ip_j +V(x)
\end{eqnarray}
where $x$ stands for $x_1,~x_2,~\cdots,~x_{3n}$ the coordinate
components of these $n$ particles, $V(x)$ is the potential
function, $C_{ij}$ are constants and $p_1,~p_2,~\cdots,~p_{3n}$
are the momentum operators satisfying the commutation relation
\begin{eqnarray}\label{e1.3}
[p_i,~x_j]=-i\delta_{ij}.
\end{eqnarray}
(Throughout the paper, we set Planck's constant $\hbar=1$.) For a
relativistic field theory, the Hamiltonian usually takes on a
different form. Let $\Phi({\bf r})$ be a scalar boson field at a
three-dimensional position vector ${\bf r}$, and $\Pi({\bf r})$ be
the corresponding conjugate momentum operator. In this case we may
write
\begin{eqnarray}\label{e1.4}
H=\int d^3 r [\Pi^2({\bf r})+V(\Phi({\bf r}))]
\end{eqnarray}
with $\Pi({\bf r})$ and $\Phi({\bf r}')$ satisfying the
commutation relation
\begin{eqnarray}\label{e1.5}
[\Pi({\bf r}),~ \Phi({\bf r}')]=-i\delta^3({\bf r}-{\bf r}').
\end{eqnarray}

In both cases, the dependence of $H$ on the momentum operators
$p_i$ and $\Pi({\bf r})$ are quadratic. Consequently, they can be
brought into an identical standard form. In the above case of a
system of non-relativistic particles, through a linear
transformation
\begin{eqnarray}\label{e1.6}
\{x_i\} \rightarrow \{q_i\},
\end{eqnarray}
the Hamiltonian (\ref{e1.2}) can be written in the standard form
\begin{eqnarray}\label{e1.7}
H=-\frac{1}{2}\nabla^2 + V(q_1,~q_2,~\cdots,~q_N)
\end{eqnarray}
with
\begin{eqnarray}\label{e1.8}
\nabla^2 =\sum_{i=1}^N \frac{\partial^2}{\partial q_i^2}.
\end{eqnarray}
Likewise for the relativistic boson field Hamiltonian
(\ref{e1.4}), we can use the Fourier-components of $\Phi({\bf r})$
and $\Pi({\bf r})$ as the set $\{x_i\}$ and $\{p_i\}$. Through a
similar transformation (\ref{e1.6}), the field Hamiltonian
(\ref{e1.4}) can also be brought into the standard form
(\ref{e1.7}) and (\ref{e1.8}), but with the number of variables
$N=\infty$. All our subsequent discussions will start from the
Schroedinger equation in this standard form (\ref{e1.7}) -
(\ref{e1.8}). Furthermore, in this paper, we shall limit our
discussions only to the groundstate.

In order to solve
\begin{eqnarray}\label{e1.9}
(-\frac{1}{2}\nabla^2 + V({\bf q}))\psi({\bf q})=E\psi({\bf q})
\end{eqnarray}
where ${\bf q}$ stands for the set $\{q_i\}$, we proceed as
follows:\\
1. Construct a good trial function $\phi({\bf q})$. A rather
efficient way to find such trial functions is given in the next
section.\\
2. By differentiating $\phi$, we define
\begin{eqnarray}\label{e1.10}
U({\bf q})-E_0 \equiv \phi({\bf q})^{-1}(\frac{1}{2}\nabla^2
\phi({\bf q})),
\end{eqnarray}
in which the constant $E_0$ may be determined by, e.g., setting
the minimum value of $U({\bf q})$ to be zero. Thus, $\phi({\bf
q})$ satisfies a different Schroedinger equation
\begin{eqnarray}\label{e1.11}
(-\frac{1}{2}\nabla^2 + U({\bf q}))\phi({\bf q})=E_0 \phi({\bf
q}).
\end{eqnarray}
Define $w({\bf q})$ and ${\cal E}$ by
\begin{eqnarray}\label{e1.12}
U({\bf q})=V({\bf q})+w({\bf q})
\end{eqnarray}
and
\begin{eqnarray}\label{e1.13}
E_0=E+{\cal E}.
\end{eqnarray}
The original Schroedinger equation (\ref{e1.9}) can then be
written as
\begin{eqnarray}\label{e1.14}
(-\frac{1}{2}\nabla^2 + U({\bf q})-E_0)\psi({\bf q})=(w({\bf
q})-{\cal E}) \psi({\bf q}).
\end{eqnarray}
Multiplying this equation on the left by $\phi({\bf q})$ and
(\ref{e1.11}) by $\psi({\bf q})$, we find their difference to be
\begin{eqnarray}\label{e1.15}
-\frac{1}{2}\nabla\cdot(\phi\nabla\psi-\psi\nabla\phi)=(w-{\cal
E})\psi\phi.
\end{eqnarray}
The integration of its lefthand side over all space is zero, which
yields
\begin{eqnarray}\label{e1.16}
{\cal E}=\frac{\int w \psi \phi~d^Nq}{\int \psi \phi~d^Nq}.
\end{eqnarray}
3. The above equation (\ref{e1.14}) will be solved iteratively by
considering the sequences
\begin{eqnarray}\label{e1.17}
\psi_1,~\psi_2,~\cdots,~\psi_n,~\cdots~~~~{\sf and}~~~~{\cal
E}_1,~{\cal E}_2,~\cdots,~{\cal E}_n,~\cdots
\end{eqnarray}
that satisfy
\begin{eqnarray}\label{e1.18}
(-\frac{1}{2}\nabla^2 + U({\bf q})-E_0)\psi_n({\bf q})=(w({\bf
q})-{\cal E}_n) \psi_{n-1}({\bf q})
\end{eqnarray}
with
\begin{eqnarray}\label{e1.19}
\psi_0({\bf q})=\phi({\bf q}).
\end{eqnarray}
As in (\ref{e1.15}) and (\ref{e1.16}), we multiply (\ref{e1.11})
by $\psi_n$ and (\ref{e1.18}) by $\phi$; their difference gives
\begin{eqnarray}\label{e1.20}
-\frac{1}{2}\nabla\cdot(\phi\nabla\psi_n-\psi_n\nabla\phi)=(w-{\cal
E}_n)\psi_{n-1}\phi
\end{eqnarray}
and therefore
\begin{eqnarray}\label{e1.21}
{\cal E}_n=\frac {\int w \psi_{n-1}\phi~d^Nq}{\int \psi_{n-1}
\phi~ d^Nq}.
\end{eqnarray}
As we shall show, for many interesting problems
\begin{eqnarray}\label{e1.22}
\lim_{n\rightarrow \infty} {\cal E}_n={\cal E}~~~~~{\sf
and}~~~~\lim_{n\rightarrow \infty} \psi_n=\psi,
\end{eqnarray}
{\it in contrast} to the perturbative series expansion using
$w({\bf q})$ as the perturbation. The key difference lies in the
above expression (\ref{e1.21}) of ${\cal E}_n$, which is a ratio,
with both its numerator and denominator depending on the
$(n-1)^{{\sf th}}$ iterative solution $\psi_{n-1}$.\\
4. There exists a simple electrostatic analog problem for the
iterative equation (\ref{e1.18}). Assuming that $\psi_{n-1}({\bf
q})$ has already been solved, we can determine ${\cal E}_n$
through (\ref{e1.21}). The righthand side of (\ref{e1.20}),
defined by
\begin{eqnarray}\label{e1.23}
\sigma_n({\bf q}) \equiv (w({\bf q})-{\cal E}_n)\psi_{n-1}({\bf
q})\phi({\bf q}),
\end{eqnarray}
is then a known function. Introduce
\begin{eqnarray}\label{e1.24}
f_n({\bf q}) \equiv \psi_n({\bf q})/\phi({\bf q}).
\end{eqnarray}
In terms of $f_n({\bf q})$, the $n^{{\sf th}}$ order iterative
equation (\ref{e1.20}) becomes
\begin{eqnarray}\label{e1.25}
-\frac{1}{2}\nabla\cdot(\phi^2\nabla f_n)=\sigma_n.
\end{eqnarray}

Consider a dielectric medium with a dielectric constant dependent
on ${\bf q}$, given by
\begin{eqnarray}\label{e1.26}
\kappa({\bf q})\equiv \phi^2({\bf q}).
\end{eqnarray}
Interpret $\sigma_n({\bf q})$ as  the external electrostatic
charge distribution, $\frac{1}{2} f_n$ the electrostatic
potential, $-\frac{1}{2}\nabla f_n$ the electrostatic field and
\begin{eqnarray}\label{e1.27}
D_n \equiv -\frac{1}{2}\kappa \nabla f_n
\end{eqnarray}
the corresponding displacement vector field. Thus (\ref{e1.25})
becomes
\begin{eqnarray}\label{e1.28}
\nabla\cdot D_n =\sigma_n,
\end{eqnarray}
the Maxwell equation for this electrostatic analog problem.

At infinity, $\phi(\infty)=0$. In accordance with (\ref{e1.26}) -
(\ref{e1.27}), we also have $D_n(\infty)=0$. Hence the integration
of (\ref{e1.28}) leads to the total external electrostatic charge
to be also zero; i.e.,
\begin{eqnarray}\label{e1.29}
\int\sigma_n({\bf q})~ d^Nq=0
\end{eqnarray}
which is the same result given by (\ref{e1.21}) for the
determination of ${\cal E}_n$. Because the dielectric constant
$\kappa({\bf q})$ in this analog problem is zero at ${\bf
q}=\infty$, the dielectric media becomes a perfect dia-electric at
$\infty$. Thus, the equation of zero total charge, given by
(\ref{e1.29}), may serve as a much simplified model of charge
confinement, analogous to color confinement in quantum
chromodynamics.

We note that (\ref{e1.25}) can also be derived from a minimal
principle by defining
\begin{eqnarray}\label{e1.30a}
I(f_n({\bf q})) \equiv \int\{\frac{1}{4}\kappa(\nabla f_n)^2 +
\sigma_n f_n\}d^N q.
\end{eqnarray}
Because of (\ref{e1.29}), the functional $I(f_n({\bf q}))$ is
invariant under
\begin{eqnarray}\label{e1.31a}
f_n({\bf q}) \rightarrow f_n({\bf q}) + {\sf constant}.
\end{eqnarray}
Since the quadratic part of $I(f_n({\bf q}))$ is the integral of
the positive definite $\frac{1}{4}\kappa(\nabla f_n)^2$, the
curvature of $I(f_n({\bf q}))$ in the functional space $f_n({\bf
q})$ is always positive. Hence, $I(f_n({\bf q}))$ has a minimum,
and that minimum determines a unique electrostatic field
$-\frac{1}{2}\nabla f_n$, as we shall see. To establish the
uniqueness, let us assume two different $\nabla f_n$, both satisfy
(\ref{e1.25}), with the same $\kappa=\phi^2$ and the same
$\sigma_n$; their difference would then satisfy (\ref{e1.25}) with
a zero external charge distribution. For $\sigma_n=0$, the minimum
of $I(f_n({\bf q}))$ is clearly zero with the corresponding
$\nabla f_n=0$. To derive $f_n({\bf q})$ from $\nabla f_n$, there
remains an additive constant at each iteration. As we shall show,
this arbitrariness allows us the freedom to derive different types
of convergent series.

To illustrate this freedom, let us consider a one-dimensional
problem in which we may replace the variables $\{q_i\}$ by a
single $x$. Furthermore, for this discussion, let us assume the
potential $V(x)$ to be an even function, with
\begin{eqnarray}\label{e1.30}
V(x)=V(-x)
\end{eqnarray}
(a condition that will be relaxed in our later analysis). The
evenness of $V(x)$ requires $\psi(x)=\psi(-x)$ and therefore also
$\phi(x)=\phi(-x)$. Thus, we need only to consider the half-space
\begin{eqnarray}\label{e1.31}
x\geq 0.
\end{eqnarray}
Equations (\ref{e1.24}), (\ref{e1.27}) and (\ref{e1.28}) can be
written now as
\begin{eqnarray}\label{e1.32}
f_n(x) = \psi_n(x)/\phi(x),
\end{eqnarray}
\begin{eqnarray}\label{e1.33}
D_n = -\frac{1}{2}\kappa(x)f'_n(x)
\end{eqnarray}
and
\begin{eqnarray}\label{e1.34}
D'_n(x) =\sigma_n(x)
\end{eqnarray}
where
\begin{eqnarray}\label{e1.35}
\kappa(x)=\phi^2(x)
\end{eqnarray}
and
\begin{eqnarray}\label{e1.36}
\sigma_n(x)=(w(x)-{\cal E}_n)\phi(x)\psi_{n-1}(x)
\end{eqnarray}
same as before. Throughout the paper, ' denote $\frac{d}{dx}$.

From (\ref{e1.34}) and $D_n(\infty)=0$, we have
\begin{eqnarray}\label{e1.37}
D_n(x) = -\int_x^\infty \sigma_n(z)dz
\end{eqnarray}
and, since $\sigma_n(x)$ is even in $x$, we have from
(\ref{e1.29}),
\begin{eqnarray}\label{e1.38}
\int_0^\infty \sigma_n(z)dz=0.
\end{eqnarray}
It follows then from (\ref{e1.32}) and
(\ref{e1.36})-(\ref{e1.38}),
\begin{eqnarray}\label{e1.39}
&D_n(x) = -\int_x^\infty (w(z)-{\cal
E}_n)\phi^2(z)f_{n-1}(z)~dz~~~~~~~~~~~~~~~~~~~~~~~~~~~~\nonumber\\
{\sf and}~~~~~~~~~~~~~~~~~~~~~~&\\
&D_n(x) = \int_0^x (w(z)-{\cal
E}_n)\phi^2(z)f_{n-1}(z)~dz,~~~~~~~~~~~~~~~~~~~~~~~~~~~~~~\nonumber
\end{eqnarray}
which lead to, through (\ref{e1.33}),
\begin{eqnarray}\label{e1.40}
f_n(x)=f_n(\infty) - 2 \int_{x}^{\infty}\phi^{-2}(y)
\int_{y}^{\infty}\phi^{2}(z) (w(z)- {\cal E}_n) f_{n-1}(z) dz
\end{eqnarray}
and
\begin{eqnarray}\label{e1.41}
f_n(x)=f_n(0) - 2 \int_{0}^{x}\phi^{-2}(y) \int_{0}^{y}\phi^{2}(z)
(w(z)- {\cal E}_n) f_{n-1}(z) dz.
\end{eqnarray}
Consider first the case that $w(x)$
in (\ref{e1.36}) is positive and satisfies
\begin{eqnarray}\label{e1.42}
w'(x)<0~~~~~{\sf for}~~~x>0.
\end{eqnarray}
The hierarchy theorem that will be proved in Section 3 states that
if $w(x)$ satisfies (\ref{e1.42}) then the iterative solution of
(\ref{e1.40}) with the boundary condition
\begin{eqnarray}\label{e1.43}
f_n(\infty)=1~~~~~~~~~~~~~~~{\sf for~all}~~~n
\end{eqnarray}
gives a convergent monotonic sequence ${\cal E}_1,~{\cal
E}_2,~{\cal E}_3,~\cdots$, where for all $n$,
\begin{eqnarray}\label{e1.44}
{\cal E}_n > {\cal E}_{n-1},
\end{eqnarray}
and
\begin{eqnarray}\label{e1.45}
{\cal E}_n \rightarrow {\cal E}~~~~~~~~~~~~~~~~~~~{\sf
as}~~~n\rightarrow\infty;
\end{eqnarray}
likewise, the sequence $f_0(x)=1,~f_1(x),~f_2(x),~\cdots$ is also
monotonic and convergent at any $x\geq 0$ with
\begin{eqnarray}\label{e1.46}
f_n(x) > f_{n-1}(x)
\end{eqnarray}
and
\begin{eqnarray}\label{e1.47}
f_n(x) \rightarrow f(x)~~~~~~~~~~~~~~~~~~~{\sf
as}~~~n\rightarrow\infty.
\end{eqnarray}
Furthermore, the convergence of (\ref{e1.45}) and (\ref{e1.47})
can hold for {\it arbitrarily} large but finite $w(x)$. A result that
is surprising, but pleasant.

On the other hand, if instead of (\ref{e1.43}), we impose a
different boundary condition, one given by
\begin{eqnarray}\label{e1.48}
f_n(0) =1~~~~~~~~~~~~~~~~~~~{\sf for~all}~~~n,
\end{eqnarray}
then instead of (\ref{e1.45}), we have for all odd $n=2m+1$ an
ascending sequence
\begin{eqnarray}\label{e1.49}
{\cal E}_1 < {\cal E}_3 < {\cal E}_5< \cdots;\
\end{eqnarray}
however, for the even $n=2m$ series, we have a descending sequence
\begin{eqnarray}\label{e1.50}
{\cal E}_2 > {\cal E}_4 > {\cal E}_6 > \cdots;
\end{eqnarray}
furthermore, between any even $n=2m$ and any odd $n=2l+1$, we have
\begin{eqnarray}\label{e1.51}
{\cal E}_{2m} > {\cal E}_{2l+1}.
\end{eqnarray}
Since according to (\ref{e1.13}), ${\cal E}_n$ is the $n^{{\sf
th}}$ order iteration towards
\begin{eqnarray}\label{e1.52}
{\cal E} =E_0-E,
\end{eqnarray}
each odd member ${\cal E}_{2l+1}$ in (\ref{e1.49}) gives an upper
bound of $E$, whereas each even member ${\cal E}_{2m}$ in
(\ref{e1.50}) leads to a lower bound of $E$. Both sequences
approach the correct ${\cal E}$ as $n \rightarrow \infty$, one
from above and the other from below. For the boundary condition
(\ref{e1.48}), our proof of convergence requires a condition on
the magnitude of $w(x)$. Still this is quite a remarkable result.

In Section~2, we discuss the details of how to construct a good
trial function $\phi({\bf q})$ for the N-dimensional problem.
Section~3 gives the proof of the hierarchy theorem for the
one-dimensional problem in which $V(x)=V(-x)$ is an even function
of $x$ and the potential-difference function $w(x)$ is assumed to
satisfy (\ref{e1.42}); i.e., $w'(x)<0$ for $x>0$. The extension to
the asymmetric case $V(x)\neq V(-x)$ is discussed in Section~4.
The hierarchy theorem is also applicable to Mathieu's equation,
which has infinite number of maxima and minima. In the Appendix,
we give a soluble example in one dimension.

In dimensions greater than 1, at each iteration Eq.(\ref{e1.21})
gives a fine tuning of the energy, just like the one-dimensional
problem. Hence, there are good reasons to expect our approach to
yield convergent solutions in any higher dimension. In Section 5,
we formulate an explicit conjecture to this effect. We describe an
attempt to prove this conjecture by generalizing the steps used to
prove the hierarchy theorem in one dimension. The attempt fails at
present because the proof of one of the lemmas does not appear to
generalize in higher dimension.

The present paper represents the synthesis and generalization of
results, some of which have appeared in our earlier
publications[1-4]. The function $D_n$ introduced in this paper is
identical to the function $h_n$ used in Ref.[4].

\newpage

\section*{\bf 2. Construction of Trial Functions}
\setcounter{section}{2} \setcounter{equation}{0}

\noindent {\bf 2.1 A New Formulation of Perturbative Expansion}\\

In many problems of interest, perturbative expansion lead to
asymptotic series, which is not the aim of this paper.
Nevertheless, the first few terms of such an expansion could
provide important insight to what a good trial function might be.
For our purpose, a particularly convenient way is to follow the
method developed in Refs.[1] and [2]. As we shall see, in  this
new method to each order of the perturbation, the wave function is
always expressible in terms of a single line-integral in the
N-dimensional  coordinate space, which can be readily used for the
construction of the trial wave function.

We begin with the Hamiltonian $H$ in its standard form
(\ref{e1.7}). Assume $V({\bf q})$ to be positive definite, and
choose its minimum to be at ${\bf q}=0$, with
\begin{eqnarray}\label{e2.1}
V({\bf q}) \geq V(0)=0.
\end{eqnarray}
Introduce a scale factor $g^2$ by writing
\begin{eqnarray}\label{e2.2}
V({\bf q}) = g^2 v({\bf q})
\end{eqnarray}
and correspondingly
\begin{eqnarray}\label{e2.3}
\psi({\bf q}) = e^{-g S({\bf q})}.
\end{eqnarray}
Thus, the Schroedinger equation (\ref{e1.9}) becomes
\begin{eqnarray}\label{e2.4}
(-\frac{1}{2}\nabla^2 + g^2 v({\bf q}))e^{-g S({\bf q})}=E e^{-g
S({\bf q})}
\end{eqnarray}
where, as before, ${\bf q}$ denotes $q_1,~q_2,~\cdots,~q_N$ and
$\nabla$ the corresponding gradient operator. Hence $S({\bf q})$
satisfies
\begin{eqnarray}\label{e2.5}
-\frac{1}{2}g^2(\nabla S)^2 + \frac{1}{2}g\nabla^2 S + g^2 v = E.
\end{eqnarray}
Considering the case of large $g$, we expand
\begin{eqnarray}\label{e2.6}
S({\bf q}) = S_0({\bf q}) + g^{-1}S_1({\bf q}) + g^{-2}S_2({\bf
q}) + \cdots
\end{eqnarray}
and
\begin{eqnarray}\label{e2.7}
E = gE_0 + E_1 + g^{-1}E_2  + \cdots.
\end{eqnarray}
Substituting (\ref{e2.6}) - (\ref{e2.7}) into (\ref{e2.5}) and
equating the coefficients of $g^{-n}$ on both sides, we find
\begin{eqnarray}\label{e2.8}
(\nabla S_0)^2 &=& 2v, \nonumber\\
\nabla S_0 \cdot \nabla S_1 &=& \frac{1}{2}~\nabla^2 S_0
-E_0,\nonumber\\
\nabla S_0 \cdot \nabla S_2 &=& \frac{1}{2} ~[\nabla^2
S_1 - (\nabla S_1)^2] -E_1,\\
\nabla S_0 \cdot \nabla S_3 &=& \frac{1}{2} ~[\nabla^2 S_2 - 2~
\nabla S_1 \cdot \nabla S_2] -E_2, \nonumber
\end{eqnarray}
etc. In this way, the second order partial differential equation
(\ref{e2.5}) is reduced to a series of first order partial
differential equations (\ref{e2.8}). The first of this set of
equations can be written as
\begin{eqnarray}\label{e2.9}
\frac{1}{2}[\nabla S_0({\bf q})]^2 - v({\bf q})=0+.
\end{eqnarray}

As noted in Ref.[1], this is precisely the Hamilton-Jacobi
equation of a single particle with unit mass moving in a potential
"$-v({\bf q})$" in the N-dimensional ${\bf q}$-space. Since ${\bf
q}=0$ is the {\it maximum} of the classical potential energy
function $-v({\bf q})$, for any point ${\bf q} \neq 0$ there is
always a classical trajectory with a total energy $0+$, which
begins from ${\bf q}=0$ and ends at the other point ${\bf q} \neq
0$, with $S_0({\bf q})$ given by the corresponding classical
action integral. Furthermore, $S_0({\bf q})$ increases along the
direction of the trajectory, which can be extended beyond the
selected point ${\bf q} \neq 0$, towards $\infty$. At infinity, it
is easy to see that $S_0({\bf q}) = \infty$, and therefore the
corresponding wave amplitude $e^{-gS_0({\bf q})}$ is zero. To
solve the second equation in (\ref{e2.8}), we note that, in
accordance with (\ref{e2.1}) - (\ref{e2.2}) at ${\bf q}=0$,
$\nabla S_0 \propto v^{\frac{1}{2}}(0)=0$. By requiring $S_1({\bf
q})$ to be analytic at ${\bf q}=0$, we determine
\begin{eqnarray}\label{e2.10}
E_0 = \frac{1}{2}(\nabla^2 S_0)_{{\sf at}~{\bf q}= 0}.
\end{eqnarray}
It is convenient to consider the surface
\begin{eqnarray}\label{e2.11}
S_0({\bf q}) = {\sf constant};
\end{eqnarray}
its normal is along the corresponding classical trajectory passing
through ${\bf q}$. Characterize  each classical trajectory by the
$S_0$-value along the trajectory and a set of $N-1$ angular
variables
\begin{eqnarray}\label{e2.12}
\alpha = (\alpha_1({\bf q}),\alpha_2({\bf q}), \cdots,
\alpha_{N-1}({\bf q})),
\end{eqnarray}
so that each $\alpha$ determines one classical trajectory with
\begin{eqnarray}\label{e2.13}
\nabla \alpha_j \cdot \nabla S_0 = 0,
\end{eqnarray}
where
\begin{eqnarray}\label{e2.14}
j = 1, 2, \cdots, N-1.
\end{eqnarray}
(As an example, we note that  as ${\bf q} \rightarrow 0$, $v({\bf
q}) \rightarrow \frac{1}{2} \sum\limits_i \omega_i^2 q_i^2$ and
therefore $ S_0 \rightarrow \frac{1}{2} \sum\limits_i \omega_i
q_i^2$. Consider  the ellipsoidal surface $ S_0 =$ constant. For
$S_0$ sufficiently small, each classical trajectory is normal to
this ellipsoidal surface. A convenient choice of $\alpha$ could be
simply any $N-1$ orthogonal parametric coordinates on the
surface.) Each $\alpha$ designates one classical trajectory, and
vice versa. Every $(S_0, \alpha)$ is mapped into a unique set
$(q_1,~q_2,~\cdots,~q_N)$ with $ S_0 \geq 0$ by construction. In
what follows, we regard the points in the ${\bf q}$-space as
specified by the coordinates $( S_0, \alpha)$. Depending on the
problem, the mapping $(q_1,~q_2,~\cdots,~q_N) \rightarrow
(S_0,\alpha)$ may or may not be one-to-one. We note that, for
${\bf q}$ near $0$, different trajectories emanating  from ${\bf
q}=0$ have to go along different directions, and therefore must
associate with different $\alpha$. Later on, as $S_0$ increases
each different trajectory retains its initially different
$\alpha$-designation; consequently, using $(S_0,\alpha)$ as the
primary coordinates, different trajectories never cross each
other. The trouble-some complications of trajectory-crossing in
${\bf q}$-space is automatically resolved by using $(S_0,\alpha)$
as coordinates. Keeping $\alpha$ fixed, the set of first order
partial differential equation can be further reduced to a set of
first order ordinary differential equation, which are readily
solvable, as we shall see.

Write
\begin{eqnarray}\label{e2.15}
S_1({\bf q}) = S_1(S_0, \alpha),
\end{eqnarray}
the second line of (\ref{e2.8}) becomes
\begin{eqnarray}\label{e2.16}
(\nabla S_0)^2 (\frac{\partial S_1}{\partial S_0})_{\alpha} =
\frac{1}{2}\nabla^2 S_0 -E_0,
\end{eqnarray}
and leads to, besides(\ref{e2.10}), also
\begin{eqnarray}\label{e2.17}
S_1({\bf q}) = S_1(S_0, \alpha) = \int\limits_0^{S_0} \frac{d
S_0}{(\nabla S_0)^2} [\frac{1}{2} \nabla^2 S_0 - E_0],
\end{eqnarray}
where the integration is taken along the classical trajectory of
constant $\alpha$. Likewise, the third, fourth and other lines of
(\ref{e2.8}) lead to
\begin{eqnarray}\label{e2.18}
E_1 &=& \frac{1}{2} [ \nabla^2 S_1 - ( \nabla S_1)^2 ]_
{{\rm at}~{\bf q}= 0},\\
 S_2({\bf q}) = S_2( S_0, \alpha) &=&
\int\limits_0^{S_0} \frac{d S_0}{(\nabla S_0)^2} \{\frac{1}{2}
[\nabla^2
S_1 - ( \nabla S_1)^2] - E_1\},\\
E_2 &=& \frac{1}{2} [\nabla^2 S_2 - 2 ( \nabla S_1)\cdot
( \nabla S_2) ]_{{\rm at}~{\bf q}= 0},\\
S_3({\bf q}) =  S_3( S_0, \alpha) &=& \int\limits_0^{ S_0} \frac{d
S_0}{( \nabla S_0)^2} \{\frac{1}{2} [ \nabla^2 S_2 - 2 ( \nabla
S_1) \cdot ( \nabla S_2)] - E_2\},
\end{eqnarray}
etc. These solutions give the convenient normalization convention
at ${\bf q} = 0$,
\begin{eqnarray}
S(0) = 0 \nonumber
\end{eqnarray}
and
\begin{eqnarray}\label{e2.22}
e^{-S(0)} = 1.
\end{eqnarray}

\newpage

\noindent
\underline{Remarks}\\

\noindent (i) As an example, consider an N-dimensional harmonic
oscillator with
\begin{eqnarray}\label{e2.23}
V({\bf q}) = \frac{g^2}{2}(q_1^2+q_2^2+\cdots +q_N^2).
\end{eqnarray}
From (\ref{e2.2}), one sees that the Hamilton-Jacobi equation
(\ref{e2.9}) is for a particle moving in a potential given by
\begin{eqnarray}\label{e2.24}
-v({\bf q}) = -\frac{1}{2}(q_1^2+q_2^2+\cdots +q_N^2).
\end{eqnarray}
Thus, for any point ${\bf q} \neq 0$ the classical trajectory of
interest is simply a straight line connecting the origin and the
specific point, with the action
\begin{eqnarray}\label{e2.25}
S_0({\bf q}) = \frac{1}{2}(q_1^2+q_2^2+\cdots +q_N^2).
\end{eqnarray}
The corresponding energy is, in accordance with (\ref{e2.10}),
\begin{eqnarray}\label{e2.26}
E_0 = \frac{N}{2}.
\end{eqnarray}
By using (\ref{e2.8}), one can readily show that
$E_1=E_2=\cdots=0$ and $S_1=S_2=\cdots=0$. The result is the well
known exact answer with the groundstate wave function for the
Schroedinger equation (\ref{e2.4}) given by
\begin{eqnarray}\label{e2.27}
e^{-gS({\bf q})} = exp~[-\frac{g}{2}(q_1^2+q_2^2+\cdots +q_N^2)]
\end{eqnarray}
and the corresponding energy
\begin{eqnarray}\label{e2.28}
E = \frac{N}{2}g.
\end{eqnarray}

\noindent (ii) From this example, it is clear that the above
expression (\ref{e2.6}) - (\ref{e2.8}) is {\it not} the well-known
WKB method. The new formalism uses $-v({\bf q})$ as the potential
for the Hamilton-Jacobi equation, and its "classical" trajectory
carries a $0+$ energy; consequently, unlike the WKB method, there
is no turning point along the classical trajectory, and the
formalism is applicable to arbitrary dimensions.

\newpage

\noindent {\bf 2.2 Trial Function for the Quantum Double-well
Potential}\\

To illustrate how to construct a trial function, consider the
quartic potential in one dimension with degenerate minima:
\begin{eqnarray}\label{e2.29}
V(x) = \frac{1}{2}g^2 (x^2-a^2)^2.
\end{eqnarray}
An alternative form of the same problem can be obtained by setting
$q=\sqrt{2ga}(a-x)$ so that the Hamiltonian becomes
\begin{eqnarray}\label{e2.30}
H=-\frac{1}{2}~\frac{d^2}{dx^2}+ V(x) \equiv 2gah,
\end{eqnarray}
where
\begin{eqnarray}\label{e2.31}
h=-\frac{1}{2}~\frac{d^2}{dq^2} + \frac{1}{2} q^2
\Big(1-\frac{q}{\sqrt{8ga^3}}\Big)^2.
\end{eqnarray}
This shows that the dimensionless (small) expansion parameter is
related to $1/\sqrt{8ga^3}$; as it turns out, the relevant
parameter is its square. In the following, we shall take $a=1$ so
that the expansion parameter is $1/g$; in the literature[5-14] one
often finds the assumption $2ga=1$ (placing the second minimum of
the potential at $q=1/g$) so that $1/\sqrt{8ga^3}$ reduces to $g$
and the anharmonic potential appears as $(1/2)q^2(1-gq)^2$. Then
$g$ appears with positive powers instead of negative, but the
coefficients of the power series are the same as with our form of
the potential, apart from the overall factor $2ga$.

For the above potential (\ref{e2.29}), the Schroedinger equation
(\ref{e2.4}) is (with $a=1$)
\begin{eqnarray}\label{e2.32}
(-\frac{1}{2}\frac{d^2}{dx^2} + \frac{1}{2}g^2
(x^2-1)^2)\psi(x)=E\psi(x)
\end{eqnarray}
where, as before, $\psi(x)=e^{-gS(x)}$ is the groundstate wave
function and $E$ its energy. Using the expansions (\ref{e2.6}) -
(\ref{e2.7}) and following the steps (\ref{e2.8}), (\ref{e2.10})
and (\ref{e2.15}) - (2.21), we find the well-known perturbative
series
\begin{eqnarray}\label{e2.33}
S_0(x) = \frac{1}{3}(x-1)^2(x+2),~~~~~S_1(x) = \ln
\frac{x+1}{2},~~~~~ S_2(x) =
\frac{3}{16}-\frac{x+2}{4(x+1)^2},~~~\cdots
\end{eqnarray}
and
\begin{eqnarray}\label{e2.34}
E_0=1,~~~E_1=-\frac{1}{4},~~~E_2=-\frac{9}{64},~~~\cdots.
\end{eqnarray}
Both expansions $S=S_0+g^{-1}S_1+g^{-2}S_2+\cdots$ and
$E=gE_0+E_1+g^{-1}E_2+\cdots$ are divergent, furthermore, at
$x=-1$ and for $n\geq 1$, each $S_n(x)$ is infinite. The
reflection $x\rightarrow -x$ gives a corresponding asymptotic
expansion $S_n(x)\rightarrow S_n(-x)$, in which each $S_n(-x)$ is
regular at $x=-1$, but singular at $x=+1$.

We note that for $g$ large, the first few terms of the
perturbative series (with (\ref{e2.33}) for $x$ positive and the
corresponding expansion $S_n(x)\rightarrow S_n(-x)$ for $x$
negative) give a fairly good description of the true wave function
$\psi(x)$ whenever $\psi(x)$ is large (i.e. for $x$ near $\pm 1$).
However, for $x$ near zero, when $\psi(x)$ is exponentially small,
the perturbative series becomes totally unreliable. This suggests
the use of first few terms of the perturbative series for regions
whenever $\psi(x)$ is expected to be large. In regions where
$\psi(x)$ is exponentially small, simple interpolations by hand
may already be adequate for a trial function, as we shall see.
Since the quartic potential (\ref{e2.29}) is even in $x$, so is
the groundstate wave function; likewise, we require the trial
function $\phi(x)$ also to satisfy $\phi(x)=\phi(-x)$. At $x=0$,
we require
\begin{eqnarray}\label{e2.35}
(\frac{d\phi}{dx})_{x=0} = \phi'(0)=0.
\end{eqnarray}
To construct $\phi(x)$, we start with the first two functions
$S_0(x)$ and $S_1(x)$ in (\ref{e2.33}). Introduce, for $x\geq 0$,
\begin{eqnarray}\label{e2.36}
\phi_+(x) \equiv e^{-gS_0(x)-S_1(x)}=(\frac{2}{1+x})e^{-gS_0(x)}
\end{eqnarray}
and
\begin{eqnarray}\label{e2.37}
\phi_-(x) \equiv
e^{-gS_0(-x)-S_1(x)}=(\frac{2}{1+x})e^{-\frac{4}{3}g+gS_0(x)}.
\end{eqnarray}
In order to satisfy (\ref{e2.35}), we define
\begin{eqnarray}\label{e2.38}
\phi(x)=\phi(-x) \equiv \left\{
\begin{array}{ll}
\phi_+(x) + \frac{g-1}{g+1}\phi_-(x), &~~~~ {\sf for} ~~~0\leq x<1\\
(1+ \frac{g-1}{g+1}e^{-\frac{4}{3}g})\phi_+(x), &~~~~ {\sf for}
~~~x>1
\end{array}
\right.
\end{eqnarray}
Thus, by construct $\phi'(0)=0$, $\phi(x)$ is continuous
everywhere, for $x$ from $-\infty$ to $\infty$, and so is its
derivative.

By differentiating $\phi_+(x)$ and $\phi(x)$, we see that they
satisfy
\begin{eqnarray}\label{e2.39}
(T+V+u)\phi_+=g\phi_+
\end{eqnarray}
and
\begin{eqnarray}\label{e2.40}
(T+V+w)\phi=g\phi,
\end{eqnarray}
where
\begin{eqnarray}\label{e2.41}
u(x)=\frac{1}{(1+x)^2}
\end{eqnarray}
and
\begin{eqnarray}\label{e2.42}
w(x)=w(-x)
\end{eqnarray}
with, for $x \geq 0$
\begin{eqnarray}\label{e2.43}
w(x)=u(x) + \hat{g} (x)
\end{eqnarray}
where
\begin{eqnarray}\label{e2.44}
\hat{g} (x) = \left\{\begin{array}{ll} 2g\frac{(g-1) e^{2g
S_0(x)-\frac{4}{3} g}}{(g+1)+(g-1)e^{2g S_0(x)-\frac{4}{3} g}},
&~~~~~~{\sf for}~~0 \leq x<1\\
0&~~~~~~{\sf for}~~x>1.
\end{array}
\right.
\end{eqnarray}
Note that for $g>1$, $\hat{g}(x)$ is positive, and has a
discontinuity at $x=1$. Furthermore, for $x$ positive both $u(x)$
and $\hat{g}(x)$ are decreasing functions of $x$. Therefore,
$w(x)$ also satisfies for $x>0$,
\begin{eqnarray}\label{e2.45}
w'(x)<0,
\end{eqnarray}
a property that is very useful in our proof of convergence which
will be discussed in the next section.

\newpage

\section*{\bf 3. Hierarchy Theorem and Its Generalization}
\setcounter{section}{3} \setcounter{equation}{0}

In this section, we restrict our discussions to a one-dimensional
problem, in which the potential $V(x)$ is an even function of $x$,
as in the example given in the previous Section 2.2. The
Schroedinger equation (\ref{e1.9}) becomes
\begin{eqnarray}\label{e3.1}
-\frac{1}{2}\psi''(x) + V(x)\psi(x)=E\psi(x)
\end{eqnarray}
with $\psi(x)$ as its groundstate wave function, $E$ the
groundstate energy and ' denoting $\frac{d}{dx}$, as before. For
the one-dimensional problem, the trial function $\phi(x)$
satisfies
\begin{eqnarray}\label{e3.2}
-\frac{1}{2}\phi''(x) + U(x)\phi(x)=E_0\phi(x),
\end{eqnarray}
as in (\ref{e1.11}); therefore  (\ref{e3.1}) can be written as
\begin{eqnarray}\label{e3.3}
-\frac{1}{2}\psi''(x) + (U(x)-E_0)\psi(x)=(w(x)-{\cal E})\psi(x),
\end{eqnarray}
in which
\begin{eqnarray}\label{e3.4}
U(x)=V(x)+w(x)
\end{eqnarray}
and
\begin{eqnarray}\label{e3.5}
E_0=E+{\cal E},
\end{eqnarray}
as before. Throughout this section, we assume
\begin{eqnarray}\label{e3.6}
V(x) &=& V(-x),~~~~~~~~~U(x)=U(-x),\nonumber\\
\psi(x) &=& \psi(-x)~~~ {\sf and}~~~\phi(x)=\phi(-x);
\end{eqnarray}
hence, we need only to consider
\begin{eqnarray}\label{e3.7}
x \geq 0.
\end{eqnarray}
Furthermore, as in the  example of the symmetric quartic
double-well potential given in Section~2.2, we assume $w(x)$ to
satisfy
\begin{eqnarray}\label{e3.8}
w'(x)<0 ~~~{\sf for}~~x > 0
\end{eqnarray}
and
\begin{eqnarray}\label{e3.9}
w(\infty)= 0.
\end{eqnarray}
Therefore, $w(x)$ is positive for $x$ positive. Otherwise, the
shape of $w(x)$ can be arbitrary. The Schroedinger equation
(\ref{e3.1}) will be solved through the iterative steps
(\ref{e1.32}) - (\ref{e1.41}), using the sequences
\begin{eqnarray}\label{e3.10}
{\cal E}_1,~ {\cal E}_2,~\cdots,~{\cal E}_n,~\cdots
\end{eqnarray}
for the energy difference ${\cal E}=E_0-E$, and the sequence
\begin{eqnarray}\label{e3.11}
f_1(x),~ f_2(x),~\cdots,~f_n(x),~\cdots
\end{eqnarray}
for the ratio $f(x)=\psi(x)/\phi(x)$ with, for $n=0$,
\begin{eqnarray}\label{e3.12}
f_0(x)=1.
\end{eqnarray}
In this section, we differentiate two sets of sequences, labelled
$A$ and $B$, satisfying different boundary conditions:
\begin{eqnarray}
&f_n(\infty)=1~~~~~{\sf for~~all}~~~n,~~~~~~~~~~{\sf
in~~Case}~~~(A)~~~~~~~~~~~~~~~~~~~~~~~~~~~~\nonumber\\
{\sf or}~~~~~~~~~~~~~~~~~~~~~&\nonumber
\end{eqnarray}
\begin{eqnarray}
f_n(0)=1~~~~~{\sf for~~all}~~~n.~~~~~~~~~~{\sf
in~~Case}~~~(B)\nonumber
\end{eqnarray}
Thus, in accordance with (\ref{e1.40})-(\ref{e1.41}), we have in
Case $(A)$
\begin{eqnarray}\label{e3.13A}
~~~~~~~~~~~f_n(x)=1 - 2 \int_{x}^{\infty}\phi^{-2}(y)dy
\int_{y}^{\infty}\phi^{2}(z) (w(z)- {\cal E}_n) f_{n-1}(z)
dz,~~~~~~~~~~~~~~~~~~~(3.13A)\nonumber
\end{eqnarray}
whereas in Case $(B)$
\begin{eqnarray}\label{e3.13B}
~~~~~~~~~~f_n(x)=1 - 2 \int_{0}^{x}\phi^{-2}(y)dy
\int_{0}^{y}\phi^{2}(z) (w(z)- {\cal E}_n) f_{n-1}(z)
dz.~~~~~~~~~~~~~~~~~~~ ~~~(3.13B)\nonumber
\end{eqnarray}
\setcounter{equation}{13} In both cases, ${\cal E}_n$ is
determined by the corresponding $f_{n-1}(x)$ through (\ref{e1.36})
and (\ref{e1.38}); i.e.,
\begin{eqnarray}\label{e3.14}
{\cal E}_n=[w~f_{n-1}]/[f_{n-1}]
\end{eqnarray}
in which $[F]$ of any function $F(x)$ is defined to be
\begin{eqnarray}\label{e3.15}
[F]=\int_0^{\infty}\phi^2(x) F(x) dx.
\end{eqnarray}
Eqs. (\ref{e1.33}) and (\ref{e1.35}) give
\begin{eqnarray}\label{e3.16}
f'_n(x) = -2 \phi^{-2}D_n(x).
\end{eqnarray}
Likewise, (\ref{e1.36}) - (\ref{e1.37}) lead to
\begin{eqnarray}\label{e3.17}
D_n(x)=- \int_{x}^{\infty}\phi^{2}(z) (w(z)- {\cal E}_n)
f_{n-1}(z)dz
\end{eqnarray}
which, on account of (\ref{e1.38}), is identical to
\begin{eqnarray}\label{e3.18}
D_n(x)= \int_0^{x}\phi^{2}(z) (w(z)- {\cal E}_n) f_{n-1}(z)dz.
\end{eqnarray}
These two expressions of $D_n(x)$ are valid for both cases $(A)$
and $(B)$. Let $x_n$ be defined by
\begin{eqnarray}\label{e3.19}
w(x)-{\cal E}_n=0 ~~~~~~~{\sf at}~~~x=x_n.
\end{eqnarray}
Since $w'(x)<0$, (\ref{e3.19}) has one and only one solution, with
$w(x)-{\cal E}_n$ negative for $x>x_n$ and positive for $x<x_n$.
Thus, if
\begin{eqnarray}\label{e3.20}
f_{n-1}(x)>0
\end{eqnarray}
for all $x>0$, we have from (\ref{e3.17}) - (\ref{e3.18})
\begin{eqnarray}\label{e3.21}
D_n(x)>0
\end{eqnarray}
and therefore, on account of (\ref{e3.16}),
\begin{eqnarray}\label{e3.22}
f'_n(x)<0.
\end{eqnarray}
In terms of the electrostatic analog introduced in Section~1,
through (\ref{e1.26}) - (\ref{e1.29}), one can form a simple
physical picture of these expressions. Represent $D_n(x)$ by the
standard flux of lines of force. Because the dielectric constant
$\kappa(x)=\phi^2(x)$ is zero at $x=\infty$, so is the
displacement field. Hence, $D_n(\infty)=0$; therefore each line of
force must terminate at a finite point. Since the electric charge
density is $\sigma_n(x)=\phi^2(x)(w(x)-{\cal E}_n)f_{n-1}(x)$, the
total electric charge to the right of $x$ is
\begin{eqnarray*}
Q_n(x)=\int_x^{\infty} \sigma_n(z)dz.
\end{eqnarray*}
It must also be the negative of the flux $D_n(x)$ passing through
the same point $x$: i.e.,
\begin{eqnarray*}
D_n(x)=-Q_n(x)=-\int_x^{\infty} \sigma_n(z)dz,
\end{eqnarray*}
which gives (\ref{e3.17}). In the whole range from $x=0$ to
$\infty$, the total electric charge $\int_0^{\infty}
\sigma_n(z)dz$ is zero; therefore, we have
\begin{eqnarray*}
Q_n(0)=0~~~~~{\sf and}~~~~~D_n(0)=0.
\end{eqnarray*}
Furthermore, at any point $x>0$, the total charge from the origin
to the point $x$ is
\begin{eqnarray*}
\int_0^x \sigma_n(z)dz,
\end{eqnarray*}
which must also be the negative of the above $Q_n(x)$, and
therefore the same as $D_n(x)$; that leads to (\ref{e3.18}). From
(\ref{e3.19}) and $w'(x)<0$, one sees that the charge distribution
$\sigma_n(x)$ is negative for $x>x_n$, $0$ at $x=x_n$ and positive
for $0<x<x_n$. Correspondingly. moving from $x=\infty$ towards the
left, the displacement field increases from $D_n(\infty)=0$ to
$D_n(x)>0$, reaching its maximum at $x=x_n$, then as $x$ further
decreases, so does $D_n(x)$, and finally reaches $D_n(0)=0$ at
$x=0$.

In Case $(A)$, because of $f_n(\infty)=1$, (\ref{e3.22}) leads to
\begin{eqnarray}\label{e3.23A}
~~~~~~~~~~~~~~~~~~~~~~~~~~~~~~~~~~~~~~~~f_n(0)>f_n(x)>f_n(\infty)=1.~~~~~~~~~~~~~~~~~~~~
~~~~~~~~~~(3.23A)\nonumber
\end{eqnarray}
Since for $n=0$, $f_0(x)=1$, (\ref{e3.20}) - (3.23A) are valid for
$n=1$; by induction these expressions also hold for all $n$; in
Case $(A)$, their validity imposes no restriction on the magnitude
of $w(x)$. In Case $(B)$ we assume $w(x)$ to be not too large, so
that (3.13B) is consistent with
\begin{eqnarray}
f_n(x) > 0~~~~~~~~{\sf for}~~~x > 0\nonumber
\end{eqnarray}
and therefore
\begin{eqnarray}\label{e3.23B}
~~~~~~~~~~~~~~~~~~~~~~~~~~~~~~~~f_n(0)=1>f_n(x)>f_n(\infty)>0.~~~~~~~~~~~~~~~~~~~~~
~~~~~~~~~~~(3.23B)\nonumber
\end{eqnarray}
\setcounter{equation}{23} As we shall see, these two boundary
conditions $(A)$ and $(B)$ produce sequences that have very
different behavior. Yet, they also share a number of common
properties.


\underline{Hierarchy Theorem} $(A)$ With the boundary condition
$f_n(\infty)=1$, we have for all $n$
\begin{eqnarray}\label{e2.15n}
{\cal E}_{n+1} > {\cal E}_n
\end{eqnarray}
and
\begin{eqnarray}\label{e2.16n}
\frac{d}{dx}\left( \frac{f_{n+1}(x)}{f_n(x)}\right)<0 ~~~~~{\sf
at~~any}~~~x>0.
\end{eqnarray}
Thus, the sequences $\{{\cal E}_n\}$ and $\{f_n(x)\}$ are all
monotonic, with
\begin{eqnarray}\label{e2.17n}
{\cal E}_1<{\cal E}_2<{\cal E}_3<\cdots
\end{eqnarray}
and
\begin{eqnarray}\label{e2.18n}
1<f_1(x)<f_2(x)<f_3(x)<\cdots
\end{eqnarray}
at all finite and positive $x$.

$(B)$ With the boundary condition $f_n(0)=1$, we have for all odd
$n=2m+1$ an ascending sequence
\begin{eqnarray}\label{e2.19n}
{\cal E}_1<{\cal E}_3<{\cal E}_5<\cdots,
\end{eqnarray}
but for all even $n=2m$, a descending sequence
\begin{eqnarray}\label{e2.20n}
{\cal E}_2>{\cal E}_4>{\cal E}_6>\cdots;
\end{eqnarray}
furthermore, between any even $n=2m$ and any odd $n=2l+1$
\begin{eqnarray}\label{e2.21n}
{\cal E}_{2m}>{\cal E}_{2l+1}.
\end{eqnarray}
Likewise, at any $x>0$, for any even $n=2m$
\begin{eqnarray}\label{e2.22n}
\frac{d}{dx}\left( \frac{f_{2m+1}(x)}{f_{2m}(x)}\right)<0,
\end{eqnarray}
whereas for any odd $n=2l+1$
\begin{eqnarray}\label{e2.23n}
\frac{d}{dx}\left( \frac{f_{2l+2}(x)}{f_{2l+1}(x)}\right)>0.
\end{eqnarray}

\noindent \underline{Remarks.}

\noindent 1. The validity of Eqs. (\ref{e2.15n}) and
(\ref{e2.16n}) for the boundary condition $f_n(\infty)=1$ was
established in Ref.[4]. The validity of Eqs.
(\ref{e2.19n})-(\ref{e2.23n}) for the boundary condition
$f_n(0)=1$ is the new result of this paper, which we shall
establish.

\noindent 2. As we shall also show, the lowest eigenvalue $E$ of
the Hamiltonian $T+V$ is the limit of the sequence $\{E_n\}$ with
\begin{eqnarray}\label{e2.24n}
E_n=E_0-{\cal E}_n.
\end{eqnarray}
Thus, the boundary condition $f_n(\infty)=1$ yields a sequence, in
accordance with (\ref{e2.17n}),
\begin{eqnarray}\label{e2.25n}
E_1>E_2>E_3>\cdots>E,
\end{eqnarray}
with each member $E_n$ an {\it upper} bound of $E$, similar to the
usual variational iterative sequence.

\noindent 3. On the other hand, with the boundary condition
$f_n(0) =1$, while the sequence of its odd members $n=2l+1$ yields
a similar one, like (\ref{e2.25n}), with
\begin{eqnarray}\label{e2.26n}
E_1>E_3>E_5> \cdots >E,
\end{eqnarray}
its even members $n=2m$ satisfy
\begin{eqnarray}\label{e2.27n}
E_2<E_4<E_6< \cdots <E.
\end{eqnarray}
It is unusual   to have an iterative sequence of {\it lower}
bounds of the eigenvalue $E$. Together,  these sequences may be
quite efficient to pinpoint the limiting $E$.

The proof of  the above generalized hierarchy theorem depends on
several lemmas  that are applicable to {\it both} boundary
conditions: $(A)$ $f_n(\infty)=1$ and $(B)$ $f_n(0)=1$; these
lemmas will be established first, and then followed by the proof
of the theorem.

\noindent \underline{Lemma 1} ~~For any pair $f_m(x)$ and $f_l(x)$

\begin{eqnarray}\label{e2.28n}
{\sf i)~~if~~at~all}~x>0,~~~\frac{d}{dx}\left(
\frac{f_m(x)}{f_l(x)}\right) <0~~~~~{\sf then}~~~{\cal
E}_{m+1}>{\cal E}_{l+1},
\end{eqnarray}
and
\begin{eqnarray}\label{e2.29n}
{\sf ii)~~if~~at~all}~x>0,~~~\frac{d}{dx}\left(
\frac{f_m(x)}{f_l(x)}\right)
>0~~~~~{\sf then}~~~{\cal E}_{m+1}<{\cal E}_{l+1}.
\end{eqnarray}
\underline{Proof}

According to (\ref{e3.14})
\begin{eqnarray}\label{e2.30n}
{\cal E}_{m+1}[f_m]=[w~f_m].
\end{eqnarray}
Also by definition (\ref{e3.15}),
\begin{eqnarray}\label{e2.31n}
{\cal E}_{l+1}[f_m]=[{\cal E}_{l+1}~f_m].
\end{eqnarray}
Their difference gives
\begin{eqnarray}\label{e2.32n}
({\cal E}_{m+1}-{\cal E}_{l+1})[f_m]=[(w-{\cal E}_{l+1})~f_m].
\end{eqnarray}
From (\ref{e3.14}),
\begin{eqnarray}\label{e2.33n}
0=[(w-{\cal E}_{l+1})~f_l].
\end{eqnarray}
Let $x_{l+1}$ be defined by (\ref{e3.19}). Multiplying
(\ref{e2.32n}) by $f_l(x_{l+1})$ and (\ref{e2.33n}) by
$f_m(x_{l+1})$ and taking their difference, we have
\begin{eqnarray}\label{e2.34n}
f_l(x_{l+1})({\cal E}_{m+1}-{\cal E}_{l+1})[f_m]=[(w-{\cal
E}_{l+1})~(f_m(x)f_l(x_{l+1})-f_l(x)f_m(x_{l+1}))],
\end{eqnarray}
in which the unsubscripted $x$ acts as a dummy variable; thus
$[f_m(x)]$ means $[f_m]$ and $[f_m(x_{l+1})]$ means
$f_m(x_{l+1})\cdot [\hspace*{.04cm}1\hspace*{.04cm}]$, etc.

(i) If $(f_m(x)/f_l(x))'<0$, then for $x<x_{l+1}$
\begin{eqnarray}\label{e2.35n}
\frac{f_m(x)}{f_l(x)} > \frac{f_m(x_{l+1})}{f_l(x_{l+1})}.
\end{eqnarray}
In addition, since $w'(x)<0$ and $w(x_{l+1})={\cal E}_{l+1}$, we
also have for $x<x_{l+1}$
\begin{eqnarray}\label{e2.36n}
w(x)>{\cal E}_{l+1}.
\end{eqnarray}
Thus,  the function inside  the square bracket on the right hand
side of (\ref{e2.34n}) is positive for $x<x_{l+1}$. Also, the
inequalities (\ref{e2.35n}) and (\ref{e2.36n}) both reverse their
signs for $x>x_{l+1}$. Consequently, the right hand side of
(\ref{e2.34n}) is positive definite, and so is its left side.
Therefore, on account of (3.23A)-(3.23B), (\ref{e2.28n}) holds.

(ii) If $(f_m(x)/f_l(x))'>0$, we see   that for $x<x_{l+1}$,
(\ref{e2.35n}) reverses its sign but not (\ref{e2.36n}). A similar
reversal of sign happens for $x>x_{l+1}$. Thus, the right hand
side of (\ref{e2.34n}) is now negative definite and therefore
${\cal E}_{m+1}<{\cal E}_{l+1}$. Lemma 1 is proved.

The following lemma was already proved in Ref.[4]. For the
convenience of the readers, we also include it in this paper. Let
\begin{eqnarray}\label{e2.37n}
\eta = \eta(\xi)
\end{eqnarray}
be a single valued differentiable function of $\xi$ in the range
between $a$ and $b$ with
\begin{eqnarray}\label{e2.38n}
0 \leq a \leq \xi \leq b
\end{eqnarray}
and with
\begin{eqnarray}\label{e2.39n}
\eta(a) \geq 0.
\end{eqnarray}

\noindent \underline{Lemma 2}.

(i) The ratio $\eta/\xi$ is a decreasing function of $\xi$ for
$a<\xi<b$ if
\begin{eqnarray}\label{e2.40n}
\frac{d\eta}{d\xi} \leq \frac{\eta}{\xi}~~~~~{\sf at}~~~\xi=a
\end{eqnarray}
and
\begin{eqnarray}\label{e2.41n}
\frac{d^2\eta}{d\xi^2} < 0~~~~~~{\sf for}~~~a<\xi<b.
\end{eqnarray}

(ii) The ratio $\eta/\xi$ is an increasing function of $\xi$ for
$a<\xi<b$ if
\begin{eqnarray}\label{e2.42n}
\frac{d\eta}{d\xi} \geq \frac{\eta}{\xi}~~~~~{\sf at}~~~\xi=a
\end{eqnarray}
and
\begin{eqnarray}\label{e2.43n}
\frac{d^2\eta}{d\xi^2} > 0~~~~~~{\sf for}~~~a<\xi<b.
\end{eqnarray}
\underline{Proof} ~~Define
\begin{eqnarray}\label{e2.44n}
L \equiv \xi \frac{d\eta}{d\xi} - \eta
\end{eqnarray}
to be the Legendre transform $L(\xi)$. We have
\begin{eqnarray}\label{e2.45n}
\frac{dL}{d\xi} = \xi\frac{d^2\eta}{d\xi^2}
\end{eqnarray}
and
\begin{eqnarray}\label{e2.46}
\frac{d}{d\xi}\left(\frac{\eta}{\xi}\right) = \frac{L}{\xi^2}.
\end{eqnarray}
Since (\ref{e2.40n}) says that $L(a) \leq 0$ and (\ref{e2.41n})
says that $\frac{dL}{d\xi}<0$  for $a<\xi<b$, these two conditions
imply $L(\xi)<0$ for $a<\xi<b$, which proves (i) in view of
(\ref{e2.46}). The proof of (ii) is the same, but with
inequalities reversed.\\

\newpage

\noindent \underline{Lemma 3} ~~For any pair $f_m(x)$ and $f_l(x)$

(i) if over all $x>0$,
\begin{eqnarray}\label{e2.47}
\frac{d}{dx}\left(\frac{f_m(x)}{f_l(x)}\right)<0~~~~ {\sf
then~~at~all}~x>0,~~~~
\frac{d}{dx}\left(\frac{D_{m+1}(x)}{D_{l+1}(x)}\right)<0,
\end{eqnarray}
and (ii) if over all $x>0$,
\begin{eqnarray}\label{e2.48}
\frac{d}{dx}\left(\frac{f_m(x)}{f_l(x)}\right)>0~~~~{\sf
then~~at~all}~x>0,~~~~
\frac{d}{dx}\left(\frac{D_{m+1}(x)}{D_{l+1}(x)}\right)>0.
\end{eqnarray}
\underline{Proof} ~~From (\ref{e3.17})-(\ref{e3.18}), we have
\begin{eqnarray}\label{e2.49}
D'_{m+1}(x)=(w(x)-{\cal E}_{m+1})~\phi^2(x)~f_m(x)
\end{eqnarray}
and
\begin{eqnarray}\label{e2.50}
D'_{l+1}(x)=(w(x)-{\cal E}_{l+1})~\phi^2(x)~f_l(x).
\end{eqnarray}
Define
\begin{eqnarray}\label{e2.51}
\xi=D_{l+1}(x)~~~~~~~{\sf and}~~~~~~~~~\eta=D_{m+1}(x).
\end{eqnarray}
In any local region of $x$ where $D'_{l+1}(x) \neq 0$, we can
regard $\eta = \eta(\xi)$ through $\eta(x) = \eta(\xi(x))$. Hence,
we have
\begin{eqnarray}\label{e2.52}
\frac{d\eta}{d\xi}=\frac{D'_{m+1}(x)}{D'_{l+1}(x)}=r(x)
\frac{f_{m}(x)}{f_{l}(x)}
\end{eqnarray}
where
\begin{eqnarray}\label{e2.53}
r(x)=\frac{w(x)-{\cal E}_{m+1}}{w(x)-{\cal E}_{l+1}},
\end{eqnarray}
and
\begin{eqnarray}\label{e2.54}
\frac{d}{d\xi}\left(\frac{d\eta}{d\xi}\right) &=&
\frac{1}{D'_{l+1}} \left(\frac{D'_{m+1}}{D'_{l+1}}\right)'=
\frac{1}{D'_{l+1}}\left( r \frac{f_{m}}{f_{l}}\right)'\nonumber\\
&=&\frac{1}{D'_{l+1}}\left(r' \frac{f_{m}}{f_{l}}+r\left(
\frac{f_{m}}{f_{l}}\right)'\right)
\end{eqnarray}
where
\begin{eqnarray}\label{e2.55}
r'(x)= \frac{{\cal E}_{m+1}-{\cal E}_{l+1}}{(w(x)-{\cal
E}_{l+1})^2}w'(x).
\end{eqnarray}
(i) If $(f_m/f_l)'<0$, from Lemma 1, we have
\begin{eqnarray}\label{e2.56}
{\cal E}_{m+1}>{\cal E}_{l+1}.
\end{eqnarray}
From $w'(x)<0$ and the definition of $x_{m+1}$ and $x_{l+1}$,
given by (\ref{e3.19}), we have
\begin{eqnarray}\label{e2.57}
x_{m+1}<x_{l+1},
\end{eqnarray}
\begin{eqnarray}\label{e2.58}
w(x_{m+1})={\cal E}_{m+1}~~~~~{\sf and}~~~~~w(x_{l+1})={\cal
E}_{l+1}.
\end{eqnarray}

We note that from (\ref{e3.17}) - (\ref{e3.18}) $D_{m+1}(x)$ and
$D_{l+1}(x)$ are both positive continuous   functions of $x$,
varying from at $x=0$,
\begin{eqnarray}\label{e2.59}
D_{m+1}(0)=D_{l+1}(0)=0
\end{eqnarray}
to at $x=\infty$
\begin{eqnarray}\label{e2.60}
D_{m+1}(\infty)=D_{l+1}(\infty)=0
\end{eqnarray}
with their maxima at $x_{m+1}$ for $D_{m+1}(x)$ and $x_{l+1}$ for
$D_{l+1}(x)$, since in accordance with (\ref{e2.49})-(\ref{e2.50})
and (\ref{e2.58}),
\begin{eqnarray}\label{e2.61}
D'_{m+1}(x_{m+1})=0~~~~{\sf and}~~~~D'_{l+1}(x_{l+1})=0.
\end{eqnarray}
From (\ref{e2.55})-(\ref{e2.56}), we see that $r'(x)$ is always
$<0$. Furthermore, from (\ref{e2.53}), we also find that the
function $r(x)$ has a discontinuity at $x=x_{l+1}$. At $x=0$,
$r(0)$ satisfies
\begin{eqnarray}\label{e2.62}
0<r(0)=\frac{w(0)-{\cal E}_{m+1}}{w(0)-{\cal E}_{l+1}}<1.
\end{eqnarray}
As $x$ increases from $0$, $r(x)$ decreases from $r(0)$, through
\begin{eqnarray}\label{e2.63}
r(x_{m+1})=0,
\end{eqnarray}
to $-\infty$ at $x=x_{l+1}-$; $r(x)$ then switches to $+\infty$ at
$x=x_{l+1}+$,  and continues to decrease as $x$ increases from
$x_{l+1}+$. At $x=\infty$, $r(x)$ becomes
\begin{eqnarray}\label{e2.64}
r(\infty)=\frac{{\cal E}_{m+1}}{{\cal E}_{l+1}}>1.
\end{eqnarray}

It is convenient to divide the positive $x$-axis into three
regions:
\begin{eqnarray}\label{e2.65}
{\rm (I)}~~~&~ 0 < x < x_{m+1},\nonumber\\
{\rm (II)}~~&~x_{m+1} < x < x_{l+1}\\
{\rm (III)}~&~ x_{l+1} < x.\nonumber
\end{eqnarray}

\newpage

\begin{tabular}{|c|c|c|c|c|c|c|}
\hline
region& $D'_{m+1}(x)$ & $D'_{l+1}(x)$ & $w(x)-{\cal E}_{m+1}$ & $w(x)-{\cal E}_{l+1}$ & ~~~~$r(x)$~~~~ & ~~~~$r'(x)$~~~~\\
\hline
{\rm I} & $>0$    & $>0$          &  $>0$             &  $>0$                 &  $>0$  & $<0$\\
{\rm II} & $<0$    & $>0$          &  $<0$             &  $>0$                 &  $<0$  & $<0$\\
{\rm III} & $<0$    & $<0$          &  $<0$             &  $<0$                 &  $>0$  & $<0$\\
\hline
\end{tabular}

\begin{center}
Table 1. The signs of $D'_{m+1}(x)$, $D'_{l+1}(x)$, $w(x)-{\cal
E}_{m+1}$, $w(x)-{\cal E}_{l+1}$, \\
$r(x)$ and $r'(x)$ in the three regions defined by (\ref{e2.65}),
when ${\cal E}_{m+1} >  {\cal E}_{l+1}$.
\end{center}

\vspace{1cm}

Table 1 summarizes the signs of $D'_{m+1}$, $D'_{l+1}$, $r$ and
$r'$ in these regions. Assuming $(f_m/f_l)'<0$ we shall show
separately the validity of (\ref{e2.47}), $(D_{m+1}/D_{l+1})'<0$,
in each of these three regions.

Since
\begin{eqnarray}\label{e2.66}
{\cal E}_{l+1}<w(x)<{\cal E}_{m+1}~~~~~{\sf in}~~~{\rm II},
\end{eqnarray}
$D_{m+1}(x)$ is decreasing and $D_{l+1}(x)$ is increasing; it is
clear that (\ref{e2.47}) holds in {\rm II}.

In each of regions ({\rm I}) and ({\rm III}), we have $r(x)>0$
from (\ref{e2.53}) and $r'(x)<0$ from (\ref{e2.55}). Since
$(f_m/f_l)'$ is always negative by the assumption in
(\ref{e2.47}), both terms inside the big parenthesis of
(\ref{e2.54}) are negative; hence the same (\ref{e2.54}) states
that $d^2\eta/d\xi^2$ has the opposite sign from $D'_{l+1}$. From
the sign of $D'_{l+1}$ listed in Table 1, we see that
\begin{eqnarray}\label{e2.67}
\frac{d^2\eta}{d\xi^2}<0~~~~~~~{\sf in}~~~({\rm I})
\end{eqnarray}
and
\begin{eqnarray}\label{e2.68}
\frac{d^2\eta}{d\xi^2}>0~~~~~~~{\sf in}~~~({\rm III}).
\end{eqnarray}
Within each region, $\eta=D_{m+1}(x)$ and $\xi=D_{l+1}(x)$ are
both monotonic in $x$; therefore, $\eta$ is a single-valued
function of $\xi$ and we can apply Lemma 2. In ({\rm I}), at
$x=0$, both $D_{m+1}(0)$ and $D_{l+1}(0)$ are $0$ according to
(\ref{e3.18}), but their ratio is given by
\begin{eqnarray}\label{e2.69}
\frac{D_{m+1}(0)}{D_{l+1}(0)}=\frac{D'_{m+1}(0)}{D'_{l+1}(0)}.
\end{eqnarray}
Therefore,
\begin{eqnarray}\label{e2.70}
\left(\frac{d\eta}{d\xi}\right)_{x=0}=\left(\frac{\eta}{\xi}\right)_{x=0}.
\end{eqnarray}
Furthermore, from (\ref{e2.67}), $\frac{d^2\eta}{d\xi^2}<0$ in
({\rm I}), it follows from Lemma 2, case(i), the ratio $\eta/\xi$
is a decreasing function of $\xi$. Since $\xi'=D'_{l+1}$ is $>0$
in ({\rm I}), according to (\ref{e2.50}), we have
\begin{eqnarray}\label{e2.71}
\frac{d}{dx}\left(\frac{D_{m+1}}{D_{l+1}}\right)<0~~~~~{\sf
in}~~~({\rm I})..
\end{eqnarray}

In ({\rm III}), at $x=\infty$,   both $D_{m+1}(\infty)$ and
$D_{l+1}(\infty)$ are $0$ according to (\ref{e2.60}). Their ratio
is
\begin{eqnarray}
\frac{D_{m+1}(\infty)}{D_{l+1}(\infty)}=\frac{D'_{m+1}(\infty)}{D'_{l+1}(\infty)},\nonumber
\end{eqnarray}
which gives at $x=\infty$,
\begin{eqnarray}\label{e2.72}
\left(\frac{d\eta}{d\xi}\right)_{x=\infty}=\left(\frac{\eta}{\xi}\right)_{x=\infty}.
\end{eqnarray}
As $x$ decreases from $x=\infty$ to $x=x_{l+1}$, from
(\ref{e2.68}) we have $\frac{d^2\eta}{d\xi^2}>0$ in ({\rm III}).
It follows from Lemma 2, case (ii), $\eta/\xi$ is an increasing
function of $\xi$. Since $\xi'=h'_{l+1}$ is $<0$, because
$x>x_{l+1}$, we have
\begin{eqnarray}\label{e2.73}
\frac{d}{dx}\left(\frac{D_{m+1}}{D_{l+1}}\right)<0~~~~~{\sf
in}~~~({\rm III})..
\end{eqnarray}
Thus, we prove case(i) of Lemma 3. Case(ii) of Lemma 3 follows
from case (i) through the interchange of the subscripts $m$ and
$l$. Lemma 3 is  then established.

\noindent \underline{Lemma 4} ~~Take any pair $f_m(x)$ and
$f_l(x)$

\noindent $(A)$ For the boundary condition $f_n(\infty)=1$, if at
all $x>0$,
\begin{eqnarray}\label{e3.83A}
~~~~~~~~~~~~~\frac{d}{dx}\left(\frac{f_m(x)}{f_l(x)}\right)<0~~~~{\sf
then~~at~all}~x>0,~~~~
\frac{d}{dx}\left(\frac{f_{m+1}(x)}{f_{l+1}(x)}\right)<0;~~~~~~~~~~~~~~(3.83A)\nonumber
\end{eqnarray}
therefore, if at all $x>0$,
\begin{eqnarray}\label{e3.84A}
~~~~~~~~~~~~~\frac{d}{dx}\left(\frac{f_m(x)}{f_l(x)}\right)>0~~~~{\sf
then~~at~all}~x>0,~~~~
\frac{d}{dx}\left(\frac{f_{m+1}(x)}{f_{l+1}(x)}\right)>0.~~~~~~~~~~~~~~(3.84A)\nonumber
\end{eqnarray}
$(B)$ For the boundary condition $f_n(0)=1$, if at all $x>0$,
\begin{eqnarray}\label{e3.83B}
~~~~~~~~~~~~~\frac{d}{dx}\left(\frac{f_m(x)}{f_l(x)}\right)<0~~~~{\sf
then~~at~all}~x>0,~~~~
\frac{d}{dx}\left(\frac{f_{m+1}(x)}{f_{l+1}(x)}\right)>0;~~~~~~~~~~~~~~(3.83B)\nonumber
\end{eqnarray}
therefore, if at all $x>0$,
\begin{eqnarray}\label{e3.84B}
~~~~~~~~~~~~~\frac{d}{dx}\left(\frac{f_m(x)}{f_l(x)}\right)>0~~~~{\sf
then~~at~all}~x>0,~~~~
\frac{d}{dx}\left(\frac{f_{m+1}(x)}{f_{l+1}(x)}\right)<0.~~~~~~~~~~~~~~(3.84B)\nonumber
\end{eqnarray}
\setcounter{equation}{84} \underline{Proof}~~Define
\begin{eqnarray}\label{e2.76}
\hat{\xi} = f_{l+1}(x)~~~~~~~{\sf and}~~~~~~\hat{\eta}
=f_{m+1}(x).
\end{eqnarray}
From (\ref{e1.33}) we see that
\begin{eqnarray}\label{e2.77}
\frac{d\hat{\eta}}{d\hat{\xi}} = \frac{f'_{m+1}(x)}{f'_{l+1}(x)} =
\frac{D_{m+1}(x)}{D_{l+1}(x)}
\end{eqnarray}
and
\begin{eqnarray}\label{e2.78}
\frac{d}{d\hat{\xi}}\left( \frac{d\hat{\eta}}{d\hat{\xi}}\right)
=\frac{1}{f'_{l+1}}\frac{d}{dx}\left(
\frac{D_{m+1}(x)}{D_{l+1}(x)}\right).
\end{eqnarray}
$(A)$ In this case. $f_n(\infty)=1$  for all $n$. Thus, at
$x=\infty$, $\hat{\xi}=f_{l+1}(\infty)=1$,
$\hat{\eta}=f_{m+1}(\infty)=1$, and their ratio
\begin{eqnarray}\label{e2.79}
\left(\frac{\hat{\eta}}{\hat{\xi}}\right)_{x=\infty} = 1.
\end{eqnarray}
At the same point $x=\infty$, in accordance with (\ref{e3.17}),
$D_{l+1}(\infty)=D_{m+1}(\infty)=0$, but their ratio is, on
account of $w(\infty)=0$ and (\ref{e2.28n}) of Lemma 1,
\begin{eqnarray}\label{e2.80}
\frac{D_{m+1}(\infty)}{D_{l+1}(\infty)}=\frac{D'_{m+1}(\infty)}{D'_{l+1}(\infty)}=
\frac{w(\infty)-{\cal E}_{m+1}}{w(\infty)-{\cal
E}_{l+1}}=\frac{{\cal E}_{m+1}}{{\cal E}_{l+1}}>1,
\end{eqnarray}
in which the last inequality follows from the same assumption, if
$(f_m/f_l)'<0$, shared by (\ref{e2.28n}) of Lemma 1 and the
present (3.83A) that we wish to prove. Thus, from (\ref{e2.77}),
at $x=\infty$
\begin{eqnarray}\label{e2.81}
\left(\frac{d\hat{\eta}}{d\hat{\xi}}\right)_{x=\infty}=
\frac{{\cal E}_{m+1}}{{\cal E}_{l+1}} >
\left(\frac{\hat{\eta}}{\hat{\xi}}\right)_{x=\infty}=1.
\end{eqnarray}
As $x$ decreases from $\infty$ to $0$, $\hat{\xi}$ increases from
$f_{l+1}(\infty)=1$ to $f_{l+1}(0)>1$, in accordance with
(\ref{e3.22}) and (3.23A). On account of (\ref{e2.47}) of Lemma 3,
we have $(D_{m+1}/D_{l+1})'<0$, which when combined with
(\ref{e2.78}) and $f'_n(x)<0$ leads to
\begin{eqnarray}\label{e2.82}
\frac{d}{d\hat{\xi}}\left(\frac{d\hat{\eta}}{d\hat{\xi}}\right)>0.
\end{eqnarray}
Thus, by using (\ref{e2.42n})-(\ref{e2.43n}) of Lemma 2, we have
$\hat{\eta}/\hat{\xi}$ to be an increasing function of
$\hat{\xi}$; i.e.,
\begin{eqnarray}\label{e2.83}
\frac{d}{d\hat{\xi}}\left(\frac{\hat{\eta}}{\hat{\xi}}\right)>0.
\end{eqnarray}
Because
\begin{eqnarray}\label{e2.84}
\frac{d}{dx}\left(\frac{\hat{\eta}}{\hat{\xi}}\right)=
\hat{\xi}'\frac{d}{d\hat{\xi}}\left(\frac{\hat{\eta}}{\hat{\xi}}\right)=
f'_{l+1}\frac{d}{d\hat{\xi}}\left(\frac{\hat{\eta}}{\hat{\xi}}\right)
\end{eqnarray}
and $f'_{l+1}<0$, we find
\begin{eqnarray}\label{e2.85}
\frac{d}{dx}\left( \frac{f_{m+1}}{f_{l+1}}\right)=
\frac{d}{dx}\left(\frac{\hat{\eta}}{\hat{\xi}}\right)<0,
\end{eqnarray}
which establishes (3.83A). Through the interchange of the
subscripts $m$ and $l$, we also obtain (3.84A).

Next, we turn to Case $(B)$  with the boundary condition
$f_n(0)=1$ for all $n$. Therefore at $x=0$,
\begin{eqnarray}\label{e2.86}
\frac{f_{m+1}(0)}{f_{l+1}(0)}=1.
\end{eqnarray}
Furthermore from (\ref{e3.16}) and (3.18B), we also have
$f'_{m+1}(0)=f'_{l+1}(0)=0$ and $D_{m+1}(0)=D_{l+1}(0)=0$, with
their ratio given by
\begin{eqnarray}\label{e2.87}
\left(\frac{df_{m+1}}{df_{l+1}}\right)_{x=0}&=&\frac{f'_{m+1}(0)}{f'_{l+1}(0)}
=\frac{D_{m+1}(0)}{D_{l+1}(0)}
=\frac{D'_{m+1}(0)}{D'_{l+1}(0)}\nonumber\\
&=& \frac{w(0)-{\cal E}_{m+1}}{w(0)-{\cal E}_{l+1}}.
\end{eqnarray}
From (\ref{e2.28n}) of Lemma 1, we see that if $(f_m/f_l)'<0$,
then ${\cal E}_{m+1}>{\cal E}_{l+1}$ and therefore
\begin{eqnarray}\label{e2.88}
\left(\frac{df_{m+1}}{df_{l+1}}\right)_{x=0}&<&1.\\
\left( \frac{\hat{\eta}}{\hat{\xi}}\right)_{x=0}&=&1.
\end{eqnarray}
Thus,
\begin{eqnarray}\label{e2.90}
\left( \frac{d\hat{\eta}}{d\hat{\xi}}\right)_{x=0}<\left(
\frac{\hat{\eta}}{\hat{\xi}}\right)_{x=0}.
\end{eqnarray}
Analogously to (\ref{e2.44n}), define
\begin{eqnarray}\label{e2.91}
L(x) \equiv \hat{\xi} \frac{d\hat{\eta}}{d\hat{\xi}} - \hat{\eta}
=f_{l+1}(x) \frac{f'_{m+1}(x)}{f'_{l+1}(x)}-f_{m+1}(x);
\end{eqnarray}
therefore
\begin{eqnarray}\label{e2.92}
\frac{dL(x)}{dx}&=&\hat{\xi}' \frac{dL}{d\hat{\xi}}
=\hat{\xi}'\hat{\xi}
\frac{d}{d\hat{\xi}}\frac{d\hat{\eta}}{d\hat{\xi}}\nonumber\\
&=&\hat{\xi}
\frac{d}{dx}\left(\frac{d\hat{\eta}}{d\hat{\xi}}\right)=
f_{l+1}\frac{d}{dx}\left(\frac{f'_{m+1}}{f'_{l+1}} \right)
=f_{l+1}\frac{d}{dx}\left(\frac{D_{m+1}}{D_{l+1}} \right).
\end{eqnarray}
From (\ref{e2.47}) of Lemma 3, we know    that if $(f_m/f_l)'<0$
then $(D_{m+1}/D_{l+1})'<0$, which leads to
\begin{eqnarray}\label{e2.93}
\frac{dL(x)}{dx}<0.
\end{eqnarray}
From (\ref{e2.91}), we have
\begin{eqnarray}\label{e2.94}
L(x) = \hat{\xi}\left( \frac{d\hat{\eta}}{d\hat{\xi}} - \frac{
\hat{\eta}}{\hat{\xi}} \right) =f_{l+1}(x) \left(
\frac{d\hat{\eta}}{d\hat{\xi}} - \frac{ \hat{\eta}}{\hat{\xi}}
\right),
\end{eqnarray}
and therefore at $x=0$, because of (\ref{e2.90}),
\begin{eqnarray}\label{e2.95}
L(0)<0.
\end{eqnarray}
Combining (\ref{e2.93}) and (\ref{e2.95}), we derive
\begin{eqnarray}\label{e2.96}
L(x)<0~~~~~~~{\sf for}~~~x\geq 0.
\end{eqnarray}

Multiplying (\ref{e2.91}) by $f'_{l+1}(x)$, we have
\begin{eqnarray}\label{e2.97}
f'_{l+1}(x) L(x) &=&f_{l+1}(x)
f'_{m+1}(x)-f_{m+1}(x)f'_{l+1}(x)\nonumber\\
&=& f^2_{l+1}(x)\left(\frac{f_{m+1}(x)}{f_{l+1}(x)}\right)'.
\end{eqnarray}
Because $f'_{l+1}(x)$ and $L(x)$ are  both negative, it follows
then
\begin{eqnarray}
\left(\frac{f_{m+1}(x)}{f_{l+1}(x)}\right)'>0,\nonumber
\end{eqnarray}
which gives (3.83B) for Case $(B)$, with  the boundary condition
$f_n(0)=1$. Interchanging the subscripts $m$ and $l$, (3.83B)
becomes (3.84B), and Lemma 4 is established.

We now turn to the proof of the theorem stated in
(\ref{e2.15n})-(\ref{e2.23n}).\\

\noindent \underline{Proof of the Hierarchy Theorem}

When $n=0$, we have
\begin{eqnarray}\label{e2.98}
f_0(x)=1.
\end{eqnarray}
From (\ref{e3.20})-(\ref{e3.22}), we find for $n=1$
\begin{eqnarray}\label{e2.99}
f'_1(x)<0,
\end{eqnarray}
and therefore
\begin{eqnarray}\label{e2.100}
(f_1/f_0)'<0.
\end{eqnarray}

In Case $(A)$, by using (3.83A) and by setting $m=1$ and $l=0$, we
derive $(f_2/f_1)'<0$; through induction, it follows then
$(f_{n+1}/f_n)'<0$ for all $n$. From Lemma 1, we also find ${\cal
E}_{n+1}>{\cal E}_n$ for all n. Thus,
(\ref{e2.15n})-(\ref{e2.18n}) are established.

In Case $(B)$, by using (\ref{e2.100}) and (3.83B), and setting
$m=1$ and $l=0$, we find $(f_2/f_1)'>0$, which in turn leads to
$(f_3/f_2)'<0$, $\cdots$, and (\ref{e2.22n})-(\ref{e2.23n}).
Inequalities (\ref{e2.19n})-(\ref{e2.21n}) now follow from
(\ref{e2.28n})-(\ref{e2.29n}) of Lemma 1. The Hierarchy Theorem is
proved.

Assuming that $w(0)$ is finite, we have for any $n$
$$
0<{\cal E}_n<w(0).
\eqno(3.110)
$$
Therefore, each of the monotonic sequences
$$
{\cal E}_1<{\cal E}_2<{\cal E}_3<\cdots,~~~~{\sf in}~~~(A)
$$
$$
{\cal E}_1<{\cal E}_3<{\cal E}_5<\cdots,~~~~{\sf in}~~~(B)
$$
and
$$
{\cal E}_2>{\cal E}_4>{\cal E}_6>\cdots,~~~~{\sf in}~~~(B)
$$
converges to a finite limit ${\cal E}$. By following the
discussions in Section~5 of Ref.[4], one can show that each of the
corresponding monotonic sequences of $f_n(x)$ also converges to a
finite limit $f(x)$. The interchange of the limit $n\rightarrow
\infty$ and the integrations in (3.13A) completes the proof that
in Case (A) the limits ${\cal E}$ and $f(x)$ satisfy
$$
f(x)=1 - 2 \int_{x}^{\infty}\phi^{-2}(y)dy
\int_{y}^{\infty}\phi^{2}(z) (w(z)- {\cal E}) f(z) dz.
\eqno(3.111A)
$$
As noted before, the convergence in Case (A) can hold for any
large but finite $w(x)$, provided that $w'(x)$ is negative for
$x>0$. In Case (B), a large $w(x)$ may yield a negative $f_n(x)$,
in violation of (3.23B) . Therefore, the convergence does depend
on the smallness of $w(x)$. One has to follow discussions similar
to those given in Ref.[3] to ensure that the limits ${\cal E}$ and
$f(x)$ satisfy
$$
f(x)=1 - 2 \int_{0}^{x}\phi^{-2}(y)dy \int_0^{y}\phi^{2}(z) (w(z)-
{\cal E}) f(z) dz.
\eqno(3.111B)
$$

\newpage

\section*{\bf 4. Asymmetric Quartic Double-well Problem}
\setcounter{section}{4} \setcounter{equation}{0}

The hierarchy theorem established in the previous section has two
restrictions: (i) the limitation of half space $x\geq 0$ and (ii)
the requirement of a monotonically decreasing perturbative
potential $w(x)$. In this section, we shall remove these two
restrictions.

Consider the specific example of an asymmetric quadratic
double-well potential
\begin{eqnarray}\label{e4.1}
V(x) = \frac{1}{2}g^2 (x^2-1)^2 + g \lambda x
\end{eqnarray}
with the constant $\lambda >0$. The groundstate wave function
$\psi(x)$ and energy $E$ satisfy the Schroedinger equation
\begin{eqnarray}\label{e4.2}
(T+V(x))\psi(x) = E \psi(x),
\end{eqnarray}
where $T=-\frac{1}{2} \frac{d^2}{dx^2}$, as before. In the
following, we shall present our method in two steps: We first
construct a trial function $\phi(x)$ of the form
\begin{eqnarray}\label{e4.3}
\phi(x) = \left\{\begin{array}{ccc} \phi^+(x)~~~~~~~{\sf for}~~x
\geq 0\\ \phi^-(x)~~~~~~{\sf for}~~x\leq 0.
\end{array}
\right.
\end{eqnarray}
At $x=0$, $\phi(x) $ and $\phi'(x)$ are both continuous, given by
\begin{eqnarray}\label{e4.4}
\phi(0)=\phi^+(0)=\phi^-(0)
\end{eqnarray}
and
\begin{eqnarray}\label{e4.5}
\phi'(0)=\phi^{+\prime}(0)=\phi^{-\prime}(0)=0,
\end{eqnarray}
with prime denoting $\frac{d}{dx}$, as before. As we shall see,
for $x>0$, the trial function $\phi(x)=\phi^+(x)$ satisfies
\begin{eqnarray}\label{e4.6a}
~~~~~~~~~~~~~~~~~~~~~~~~~~~~~~~(T+V(x)+v_+(x))\phi^+(x) = E_0^+
\phi^+(x)~~~~~~~~~~~~~~~~~~~~~~~~~~~~~(4.6a)\nonumber
\end{eqnarray}
with
\begin{eqnarray}\label{e4.7a}
~~~~~~~~~~~~~~~~~~~~~~~~~~~~~~~~~~~~~~~~~~~~~v'_+(x)<0,~~~~~~~~~~~~~
~~~~~~~~~~~~~~~~~~~~~~~~~~~~~~~~~~~~~(4.7a)\nonumber
\end{eqnarray}
whereas for $x<0$, $\phi(x)=\phi^-(x)$ satisfies
\begin{eqnarray}\label{e4.6b}
~~~~~~~~~~~~~~~~~~~~~~~~~~~~~~~(T+V(x)+v_-(x))\phi^-(x) = E_0^-
\phi^-(x)~~~~~~~~~~~~~~~~~~~~~~~~~~~~~(4.6b)\nonumber
\end{eqnarray}
with
\begin{eqnarray}\label{e4.7b}
~~~~~~~~~~~~~~~~~~~~~~~~~~~~~~~~~~~~~~~~~~~~~v'_-(x)>0.~~~~~~~~~~~~~
~~~~~~~~~~~~~~~~~~~~~~~~~~~~~~~~~~~~~(4.7b)\nonumber
\end{eqnarray}
\setcounter{equation}{7} Furthermore, at $x=\pm\infty$
\begin{eqnarray}\label{e4.8}
v_+(\infty)=v_-(-\infty)=0.
\end{eqnarray}
Starting separately from $\phi^+(x)$ and $\phi^-(x)$ and applying
the hierarchy theorem, as we shall show, we can construct from
$\phi(x)$ another trial function
\begin{eqnarray}\label{e4.9}
\chi(x) = \left\{\begin{array}{ccc} \chi^+(x)~~~~~~~{\sf
for}~~x>0\\ \chi^-(x)~~~~~~{\sf for}~~x < 0
\end{array}
\right.
\end{eqnarray}
with $\chi(x)$ and $\chi'(x)$ both continuous at $x=0$, given by
\begin{eqnarray}\label{e4.10}
\chi(0)=\chi^+(0)=\chi^-(0)
\end{eqnarray}
and
\begin{eqnarray}\label{e4.11}
\chi'(0)=\chi^{+\prime}(0)=\chi^{-\prime}(0)=0.
\end{eqnarray}
In addition, they satisfy the following Schroedinger equations
\begin{eqnarray}\label{e4.12}
(T+V(x))\chi^+(x) = E^+ \chi^+(x)~~~~~~~~~{\sf for}~~~x>0
\end{eqnarray}
and
\begin{eqnarray}\label{e4.13}
(T+V(x))\chi^-(x) = E^- \chi^-(x)~~~~~~~~~{\sf for}~~~x<0.
\end{eqnarray}
From $V(x)$ given by (\ref{e4.1}) with $\lambda$ positive, we see
that at any $x>0$, $V(x)>V(-x)$; therefore, $E^+>E^-$.

Our second step is to regard $\chi(x)$ as a new trial function,
which satisfies
\begin{eqnarray}\label{e4.14}
(T+V(x)+w(x))\chi(x) = E_0 \chi(x)
\end{eqnarray}
with $w(x)$ being a step function,
\begin{eqnarray}\label{e4.15}
w(x)=\frac{1}{2}(E^+-E^-)\cdot
 \left\{\begin{array}{ccc}
-1~~~~~~~{\sf for}~~x > 0\\ 1~~~~~~~~{\sf for}~~x< 0
\end{array}
\right.
\end{eqnarray}
and
\begin{eqnarray}\label{e4.16}
E_0=\frac{1}{2}(E^++E^-).
\end{eqnarray}
We see that $w(x)$ is now monotonic, with
\begin{eqnarray}\label{e4.17}
w'(x) \leq 0
\end{eqnarray}
for the entire range of $x$ from $-\infty$ to $+\infty$. The
hierarchy theorem can be applied again, and that will lead from
$\chi(x)$ to $\psi(x)$, as we shall see.

\newpage

\noindent {\bf 4.1 Construction of the First Trial Function}

We consider first the positive $x$ region. Following Sec.~2.1, we
begin with the usual perturbative power series expansion for
\begin{eqnarray}\label{e4.18}
\psi(x) = e^{-gS(x)}
\end{eqnarray}
with
\begin{eqnarray}\label{e4.19}
g S(x) = g S_0(+) + S_1(+) + g^{-1} S_2(+) + \cdots
\end{eqnarray}
and
\begin{eqnarray}\label{e4.20}
E = gE_0(+) + E_1(+) + g^{-1}E_2(+) +  \cdots,
\end{eqnarray}
in which $S_n(+)$ and $E_n(+)$ are $g$-independent. Substituting
(\ref{e4.18})-(\ref{e4.20}) into the Schroedinger equation
(\ref{e4.2}) and equating both sides, we find
\begin{eqnarray}\label{e4.21}
S'_0(+) &=& x^2-1,\\
S'_0(+) S'_1(+) &=& \frac{1}{2} S''_0(+) +\lambda x -E_0(+),
\end{eqnarray}
etc. Thus, (\ref{e4.21}) leads to
\begin{eqnarray}\label{e4.23}
S_0(+)=\frac{1}{3}(x-1)^2(x+2).
\end{eqnarray}
Since the left side of (4.22) vanishes at $x=1$, so is   the right
side; hence, we determine
\begin{eqnarray}\label{e3.24}
E_0(+)=1+\lambda,
\end{eqnarray}
which leads to
\begin{eqnarray}\label{e3.25}
S_1(+)=(1+\lambda) \ln (1+x).
\end{eqnarray}
Of course, the power series expansion (\ref{e4.19}) and
(\ref{e4.20}) are both divergent. However, if we retain the first
two terms in (\ref{e4.19}), the function
\begin{eqnarray}\label{e3.26}
\phi(+)=e^{-gS_0(+)-S_1(+)}=\left(\frac{1}{1+x}\right)^{1+\lambda}e^{-gS_0(+)}
\end{eqnarray}
serves as a reasonable approximation of $\psi(x)$ for $x>0$,
except when $x$ is near zero. By differentiating $\phi(+)$, we
find $\phi(+)$ satisfies
\begin{eqnarray}\label{e3.27}
(T+V(x)+u_+(x))\phi(+) = g E_0(+) \phi(+)
\end{eqnarray}
where
\begin{eqnarray}\label{e3.28}
u_+(x)=\frac{1}{2}\left(S'_1(+)^2-S''_1(+)\right)=\frac{(1+\lambda)(2+\lambda)}{2(1+x)^2}.
\end{eqnarray}

In order to construct the trial function $\phi(x)$ that satisfies
the boundary condition (\ref{e4.5}), we introduce for $x \geq 0$,
\begin{eqnarray}\label{e3.29}
\hat{\phi}(+) \equiv \left(\frac{1}{1+x}\right)^{1+\lambda} e^{-
\frac{4}{3}g +g S_0(+)},
\end{eqnarray}
and
\begin{eqnarray}\label{e3.30}
\phi^+(x) \equiv \frac{1}{2g}\cdot \left\{\begin{array}{ccc}
(g+1+\lambda) \phi(+) +(g-1-\lambda)\hat{\phi}(+),~~~~~~~{\sf for}~~0<x<1\\
\left((g+1+\lambda) +(g-1-\lambda)
e^{-\frac{4}{3}g}\right)\phi(+), ~~~~~~~{\sf for}~~x>1
\end{array}
\right.
\end{eqnarray}
so that $\phi^+(x)$ and its derivative $\phi^{+\prime}(x)$ are both
continuous at $x=1$, and in addition, at $x=0$ we have
$\phi^{+\prime}(0)=0$. For $x\leq 0$, we observe that $V(x)$ is invariant
under
\begin{eqnarray}\label{e3.31}
\lambda \rightarrow -\lambda~~~~~~~~{\sf and}~~~~~~~x\rightarrow
-x.
\end{eqnarray}
The same transformation converts $\phi^+(x)$ for $x$ positive to
$\phi^-(x)$ for $x$ negative. Define
\begin{eqnarray}\label{e3.32}
\phi^-(x) \equiv \frac{1}{2g}\cdot \left\{\begin{array}{ccc}
(g+1-\lambda) \phi(-) +(g-1+\lambda)\hat{\phi}(-),~~~~~~~{\sf for}~~-1<x\leq 0\\
\left((g+1-\lambda) +(g-1+\lambda)
e^{-\frac{4}{3}g}\right)\phi(-), ~~~~~~~{\sf for}~~x<-1
\end{array}
\right.
\end{eqnarray}
where
\begin{eqnarray}\label{e3.33}
\phi(-) &=& \left(\frac{1}{1-x}\right)^{1-\lambda} e^{-g S_0(-)},\\
\hat{\phi}(-) &=& \left(\frac{1}{1-x}\right)^{1-\lambda} e^{-
\frac{4}{3}g +g S_0(-)}
\end{eqnarray}
and
\begin{eqnarray}\label{e3.35}
S_0(-)=\frac{1}{3}(x+1)^2(-x+2).
\end{eqnarray}
Both $\phi^-(x)$ and its derivative $\phi^{-\prime}(x)$ are continuous at
$x=-1$; furthermore, $\phi^+(x)$ and $\phi^-(x)$ also satisfy the
continuity condition (\ref{e4.4}) and (\ref{e4.5}), as well as the
Schroedinger equation (4.6a) and (4.6b), with the perturbative
potentials $v_+(x)$ and $v_-(x)$ given by
\begin{eqnarray}\label{e4.36a}
~~~~~~~~~~~~~~~~~~~~~~~~~~~~v_+(x)=u_+(x) +
\hat{g}_+(x)~~~~~~~~~~~~~~~~~~~~~~~~~~~~~~~~~~~~~~~~~~~~~~~~~(4.36a)\nonumber
\end{eqnarray}
and
\begin{eqnarray}\label{e4.36b}
~~~~~~~~~~~~~~~~~~~~~~~~~~~v_-(x)=u_-(x) +
\hat{g}_-(x),~~~~~~~~~~~~~~~~~~~~~~~~~~~~~~~~~~~~~~~~~~~~~~~~~(4.36b)\nonumber
\end{eqnarray}
\setcounter{equation}{36} in which $u_+(x)$ is  given by
(\ref{e3.28}),
\begin{eqnarray}\label{e3.37}
u_-(x)=\frac{(1-\lambda)(2-\lambda)}{2(1-x)^2},
\end{eqnarray}
\begin{eqnarray}\label{e4.38a}
~~~~~~\hat{g}_+(x) = \left\{\begin{array}{ccc}
\frac{2g(g-1-\lambda)(1+\lambda-\lambda
x)~e^{-\frac{4}{3}g+2gS_0(+)}}
{(g+1+\lambda)+(g-1-\lambda)~e^{-\frac{4}{3}g+2gS_0(+)}} ,~~~~~~~{\sf for}~~0<x<1\\
0~~,~~~~~~~~~~~~~~~~~~~~~~~~~~~~~~~~~~~~~~~~~~{\sf for}~~~~x>1
\end{array}
\right.~~~~~~~~~~~~~~~~~~~~~~(4.38a)\nonumber
\end{eqnarray}
and
\begin{eqnarray}\label{e4.38b}
~~~~\hat{g}_-(x) = \left\{\begin{array}{ccc}
\frac{2g(g-1+\lambda)(1-\lambda-\lambda
x)~e^{-\frac{4}{3}g+2gS_0(-)}}
{(g+1-\lambda)+(g-1+\lambda)~e^{-\frac{4}{3}g+2gS_0(-)}}
,~~~~~~~{\sf for}~~-1<x<0\\
0~.~~~~~~~~~~~~~~~~~~~~~~~~~~~~~~~~~~~~~~~~~~{\sf for}~~~~x<-1
\end{array}
\right.~~~~~~~~~~~~~~~~~~~~~(4.38b)\nonumber
\end{eqnarray}
\setcounter{equation}{38} In order that $u_+(x)$, $\hat{g}_+(x)$
be positive for $x>0$ and $u_-(x)$, $\hat{g}_-(x)$ positive for
$x<0$, we impose
\begin{eqnarray}\label{e3.39}
\lambda <1~~~~~~~~~~~~~~{\sf and}~~~~~~~~~~~~~~1+\lambda<g,
\end{eqnarray}
\setcounter{equation}{39} in addition to the earlier condition
$\lambda >0$. From (\ref{e3.28}) and (\ref{e3.37}), we have
\begin{eqnarray}\label{e4.40a}
~~~~~~~~~~~~~~~~~~~~~~~u_+'(x)=-\frac{(1+\lambda)(2+\lambda)}{(1+x)^3}<0~~~~~~~~~~~~~~~{\sf
for}~~~x>0~~~~~~~~~~~~~~~~~(4.40a)\nonumber
\end{eqnarray}
and
\begin{eqnarray}\label{e4.40b}
~~~~~~~~~~~~~~~~~~~~~~~u_-'(x)=\frac{(1+\lambda)(2+\lambda)}{(1-x)^3}>0
~~~~~~~~~~~~~~~~~{\sf for}~~~x<0.~~~~~~~~~~~~~~~~~
(4.40b)\nonumber
\end{eqnarray}
Likewise, from (4.38a) and (4.38b), we find
\begin{eqnarray}\label{e4.41a}
\hat{g}'_+(x) = \left\{\begin{array}{lcc}
\left(-\frac{\lambda}{1+\lambda-\lambda
x}-\frac{2g(1-x^2)(g+1+\lambda)}
{(g+1+\lambda)+(g-1-\lambda)~e^{-\frac{4}{3}g+2gS_0(+)}}\right)
\hat{g}_+(x) ,~~~{\sf for}~~0<x<1\\
~~~~~~~~~~0~,~~~~~~~~~~~~~~~~~~~~~~~~~~~~~~~~~~~~~{\sf for}~~~~x>1
\end{array}
\right.~~~~~~~~(4.41a)\nonumber
\end{eqnarray}
and
\begin{eqnarray}\label{e4.41b}
\hat{g}'_-(x) = \left\{\begin{array}{lcc}
\left(-\frac{\lambda}{1-\lambda-\lambda
x}+\frac{2g(1-x^2)(g+1-\lambda)}
{(g+1-\lambda)+(g-1+\lambda)~e^{-\frac{4}{3}g+2gS_0(-)}}\right)
\hat{g}_-(x) ,~~~{\sf for}~~-1<x<0\\
~~~~~~~~~~0~,~~~~~~~~~~~~~~~~~~~~~~~~~~~~~~~~~~ ~{\sf
for}~~~~x<-1.
\end{array}
\right.~~~~(4.41b)\nonumber
\end{eqnarray}
Furthermore, as $x \rightarrow \pm 1$,
\begin{eqnarray}\label{e4.42a}
~~~~~~~~~~~~~~~\lim_{x\rightarrow 1}\hat{g}'_+(x) =
-\frac{2g(g-1-\lambda)~e^{-\frac{4}{3}g}}{(g+1+\lambda)+(g-1-\lambda)~e^{-\frac{4}{3}g}}\delta
(x-1) ~~~~~~~~~~~~~~~~~~~~~~(4.42a)\nonumber
\end{eqnarray}
and
\begin{eqnarray}\label{e4.42b}
~~~~~~~~~~~~~~\lim_{x\rightarrow -1}\hat{g}'_-(x) =
\frac{2g(g-1+\lambda)~e^{-\frac{4}{3}g}}{(g+1-\lambda)+(g-1+\lambda)~e^{-\frac{4}{3}g}}\delta
(x+1).~~~~~~~~~~~~~~~~~~~~~~~(4.42b)\nonumber
\end{eqnarray}
\setcounter{equation}{42} Thus, for $x\geq 0$, we have
\begin{eqnarray}\label{e3.43}
g'_+(x) \leq 0
\end{eqnarray}
and, together with (4.36a) and (4.40a),
\begin{eqnarray}\label{e4.44a}
~~~~~~~~~~~~~~~~~~~~~~~~~~~~~~v'_+(x)\leq
0~~~~~~~~~~~~~~~~~~~~~~~~~~~~~~~~~~~~~~~~~~~~~~~~~~~~~~~~~~~~~~~(4.44a)\nonumber
\end{eqnarray}
for $x$ positive. On the other hand for $x \leq 0$,
$\hat{g}'_-(x)$ is not always positive; e.g., at $x=0$,
\begin{eqnarray}
\hat{g}'_-(0)=\left(-\frac{\lambda}{1-\lambda}+
(g+1-\lambda)\right)(1-\lambda)(g-1+\lambda)\nonumber
\end{eqnarray}
which is positive  for $g>\frac{3\lambda-1-\lambda^2}{1-\lambda}$,
but at $x=-1+$,
\begin{eqnarray}
\hat{g}'_-(-1+)=-\lambda \hat{g}_-(-1+)=-\frac{2\lambda
g(g-1+\lambda)~e^{-\frac{4}{3}g}}{(g+1-\lambda)+(g-1+\lambda)~e^{-\frac{4}{3}g}}<0.\nonumber
\end{eqnarray}
However, at $x=-1$, $u'_-(-1)=\frac{1}{8}(1+\lambda)(2+\lambda)$.
It is easy to see that the sum $u_-+\hat{g}_-=v_-$ can satisfy for
$x \leq 0$,
\begin{eqnarray}\label{e4.44b}
~~~~~~~~~~~~~~~~~~~~~~~~~~~v'_-(x)>0~~~~~~~~~~~~~{\sf
if}~~g>>1.~~~~~~~~~~~~~~~~~~~~~~~~~~~~~~~~~~~~~~~~(4.44b)\nonumber
\end{eqnarray}

To summarize: $\phi^+(x)$ and $\phi^-(x)$ satisfy the Schroedinger
equation (4.6a) and (4.6b), with $v_{\pm}(x)$ given by (4.36a) and
(4.36b),
\begin{eqnarray}\label{e4.45a}
~~~~~~~~~~~~~~~~~~~~~~~~~~~~~~E_0^+=g E_0(+) \equiv g(1+\lambda)
~~~~~~~~~~~~~~~~~~~~~~~~~~~~~~~~~~~~~~~~~~~~(4.45a)\nonumber
\end{eqnarray}
and
\begin{eqnarray}\label{e4.45b}
~~~~~~~~~~~~~~~~~~~~~~~~~~~~~~E_0^-=g E_0(-) \equiv g(1-\lambda)
~~~~~~~~~~~~~~~~~~~~~~~~~~~~~~~~~~~~~~~~~~~~~(4.45b)\nonumber
\end{eqnarray}
\setcounter{equation}{45} and the boundary conditions (\ref{e4.4})
and (\ref{e4.5}). In addition, $v_{\pm}(x)$ satisfies
\begin{eqnarray}\label{e3.46}
v_+(\infty)=0,~~~~~~~~v_-(-\infty)=0
\end{eqnarray}
and the monotonicity conditions (4.7a) and (4.7b).

\newpage

\noindent {\bf 4.2 Construction of the Second Trial Function}

To construct the second trial function $\chi(x)$ introduced in
(\ref{e4.9}), we define $f^\pm(x)$ by
\begin{eqnarray}\label{e4.47a}
~~~~~~~~~~~~~~~~~~~~~~~\chi^+(x)=\phi^+(x)f^+(x)~~~~~~~{\sf
for}~~x \geq 0
~~~~~~~~~~~~~~~~~~~~~~~~~~~~~~~~~~~~~(4.47a)\nonumber
\end{eqnarray}
and
\begin{eqnarray}\label{e4.47b}
~~~~~~~~~~~~~~~~~~~~~~~\chi^-(x)=\phi^-(x)f^-(x)~~~~~~~{\sf
for}~~x \leq 0
~~~~~~~~~~~~~~~~~~~~~~~~~~~~~~~~~~~~~(4.47b)\nonumber
\end{eqnarray}
\setcounter{equation}{47} To retain flexibility it is convenient
to impose only the boundary condition (\ref{e4.11}) first, but not
(\ref{e4.10}); i.e., at $x=0$
\begin{eqnarray}\label{e3.48}
\chi^{+\prime}(0)=\chi^{-\prime}(0)=0,
\end{eqnarray}
but leaving the choice of the overall normalization of $\chi^+(0)$
and $\chi^-(0)$ to be decided later. We rewrite the Schroedinger
equations (\ref{e4.12}) and (\ref{e4.13}) in their equivalent
forms
\begin{eqnarray}\label{e4.49a}
~~~~~~~~~~~~~~(T+V(x)+v_+(x)-E_0^+)\chi^+(x)=(v_+(x)-{\cal
E}^+)\chi^+(x) ~~~~~~~~~~~~~~~~~~~~~~~(4.49a)\nonumber
\end{eqnarray}
and
\begin{eqnarray}\label{e4.49b}
~~~~~~~~~~~~~~(T+V(x)+v_-(x)-E_0^-)\chi^-(x)=(v_-(x)-{\cal
E}^-)\chi^-(x) ~~~~~~~~~~~~~~~~~~~~~~~(4.49b)\nonumber
\end{eqnarray}
where
\begin{eqnarray}\label{e4.50a}
~~~~~~~~~~~~~~~~~~~~~~~~~~~~~~~~~{\cal E}^+=E_0^+ - E^+
~~~~~~~~~~~~~~~~~~~~~~~~~~~~~~~~~~~~~~~~~~~~~~~~~~~~~~(4.50a)\nonumber
\end{eqnarray}
and
\begin{eqnarray}\label{e4.50b}
~~~~~~~~~~~~~~~~~~~~~~~~~~~~~~~~~{\cal E}^-=E_0^- - E^-~.
~~~~~~~~~~~~~~~~~~~~~~~~~~~~~~~~~~~~~~~~~~~~~~~~~~~~~(4.50b)\nonumber
\end{eqnarray}
\setcounter{equation}{50} Because at $x=0$,
$\phi^{+\prime}(0)=\phi^{-\prime}(0)=0$, in accordance with (\ref{e4.5}), we
have, on account of (\ref{e3.48}),
\begin{eqnarray}\label{e3.51}
f^{+\prime}(0)=f^{-\prime}(0)=0.
\end{eqnarray}
So far, the overall normalization of $f^+(x)$ and $f^-(x)$ are
still free. We may choose
\begin{eqnarray}\label{e3.52}
f^+(\infty)=1~~~~~{\sf and}~~~~~f^- (-\infty)=1.
\end{eqnarray}
From (4.6a), (4.47a), (4.49a) and (4.50a), we see that $f^+(x)$
satisfies the integral equation (for $x\geq 0$)
\begin{eqnarray}\label{e4.53a}
~~~~~~~~~~~f^+(x) =1- 2\int\limits_x^\infty (\phi^+(y))^{-2} dy
\int\limits_y^\infty (\phi^+(z))^2 (v_+(z)-{\cal E}^+)f^+(z)
dz.~~~~~~~~~~~~~~~~~(4.53a)\nonumber
\end{eqnarray}
Furthermore, from (4.6a) and (4.49a), we also have
\begin{eqnarray}\label{e4.54a}
~~~~~~~~~~~~~~~~~~~~~~~~~~~~~~~ \int\limits_0^\infty (\phi^+(x))^2
(v_+(x)-{\cal E}^+)f^+(x)
dx.~~~~~~~~~~~~~~~~~~~~~~~~~~~~~~~~~~(4.54a)\nonumber
\end{eqnarray}
Likewise, $f^-(x)$ satisfies (for $x \leq 0$)
\begin{eqnarray}\label{e4.53b}
~~~~~~~~~~~f^-(x) =1- 2\int\limits_{-\infty}^x (\phi^-(y))^{-2} dy
\int\limits_{-\infty}^y (\phi^-(z))^2 (v_-(z)-{\cal E}^-)f^-(z)
dz~~~~~~~~~~~~~~~~~(4.53b)\nonumber
\end{eqnarray}
and
\begin{eqnarray}\label{e4.54b}
~~~~~~~~~~~~~~~~~~~~~~~~~~~~~~~ \int\limits_{-\infty}^0
(\phi^-(x))^2 (v_-(x)-{\cal E}^-)f^-(x)
dx.~~~~~~~~~~~~~~~~~~~~~~~~~~~~~~~~~~(4.54b)\nonumber
\end{eqnarray}
The function $f^+(x)$ and $f^-(x)$ will be solved through the
iterative process described in Section~1. We introduce the
sequences $\{f_n^{\pm}(x)\}$ and $\{{\cal E}_n^{\pm}\}$ for
$n=1,~2,~3,~\cdots$, with
\begin{eqnarray}\label{e4.55a}
~~~~~~~~~~~f_n^+(x) =1- 2\int\limits_x^\infty (\phi^+(y))^{-2} dy
\int\limits_y^\infty (\phi^+(z))^2 (v_+(z)-{\cal
E}_n^+)f_{n-1}^+(z) dz.~~~~~~~~~~~~~~~~(4.55a)\nonumber
\end{eqnarray}
for $x \geq 0$, and
\begin{eqnarray}\label{e4.55b}
~~~~~~~~~~~f_n^-(x) =1- 2\int\limits_{-\infty}^x (\phi^-(y))^{-2}
dy \int\limits_{-\infty}^y (\phi^-(z))^2 (v_-(z)-{\cal
E}_n^-)f_{n-1}^-(z) dz~~~~~~~~~~~~~~(4.55b)\nonumber
\end{eqnarray}
for $x \leq 0$, where ${\cal E}_n^{\pm}$ satisfies
\begin{eqnarray}\label{e4.56a}
~~~~~~~~~~~~~~~~~~~~~~~~~~~~~~~ \int\limits_0^\infty (\phi^+(x))^2
(v_+(x)-{\cal E}_n^+)f_{n-1}^+(x)
dx~~~~~~~~~~~~~~~~~~~~~~~~~~~~~~~~~(4.56a)\nonumber
\end{eqnarray}
and
\begin{eqnarray}\label{e4.56b}
~~~~~~~~~~~~~~~~~~~~~~~~~~~~~~~ \int\limits_{-\infty}^0
(\phi^-(x))^2 (v_-(x)-{\cal E}_n^-)f_{n-1}^-(x)
dx.~~~~~~~~~~~~~~~~~~~~~~~~~~~~~~~(4.56b)\nonumber
\end{eqnarray}
Thus, (4.55a) and (4.55b) can also be written in their equivalent
expressions
\begin{eqnarray}\label{e4.57a}
~~~~~~~~~~~f_n^+(x) =f_n^+(0)- 2\int\limits_0^x(\phi^+(y))^{-2} dy
\int\limits_0^y(\phi^+(z))^2 (v_+(z)-{\cal E}_n^+)f_{n-1}^+(z)
dz.~~~~~~~~~(4.57a)\nonumber
\end{eqnarray}
for $x \geq 0$, and
\begin{eqnarray}\label{e4.57b}
~~~~~~~~~~~f_n^-(x) =f_n^-(0)- 2\int\limits_x^0 (\phi^-(y))^{-2}
dy \int\limits_y^0 (\phi^-(z))^2 (v_-(z)-{\cal E}_n^-)f_{n-1}^-(z)
dz~~~~~~~~~~(4.57b)\nonumber
\end{eqnarray}
for $x \leq 0$.

\newpage

\setcounter{equation}{57} For $n=0$, we set
\begin{eqnarray}\label{e3.58}
f_0^+(x)=f_0^-(x)=1;
\end{eqnarray}
through induction and by using (4.55a)-(4.56b), we derive all
subsequent $f_n^{\pm}(x)$ and ${\cal E}_n^{\pm}$. Because
$v_{\pm}(x)$ satisfies (4.44a), (4.44b) and (\ref{e3.46}), the
Hierarchy theorem proved in Section~3 applies. The boundary
conditions $f^+(\infty)=f^-(-\infty)=1$, given by (\ref{e3.52}),
lead to $f_n^+(\infty)=f_n^-(-\infty)=1$, in agreement with
(4.55a) and (4.55b). According to (\ref{e2.15n})-(\ref{e2.18n}) of
Case~$(A)$ of the theorem, we have
\begin{eqnarray}\label{4.59a}
~~~~~~~~~~~~~~~~~~~~~~~~~~~~~~{\cal E}_1^+<{\cal E}_2^+<{\cal
E}_3^+<\cdots,~~~~~~~~~~~~~~~~~~~~~~~~~~~~~~~~~~~~~~~~~~~~~~(4.59a)\nonumber\\
~~~~~~~~~~~~~~~~~~~~~~~~~~~~~~{\cal E}_1^-<{\cal E}_2^-<{\cal
E}_3^-<\cdots,~~~~~~~~~~~~~~~~~~~~~~~~~~~~~~~~~~~~~~~~~~~~~~(4.59b)\nonumber\\
~~~~~~~~~~~~~~~~~~~~~~~~~~~~~~1<f_1^+(x)<f_2^+(x)<f_3^+(x)<\cdots
~~~~~~~~~~~~~~~~~~~~~~~~~~~~~~~~~(4.60a)\nonumber
\end{eqnarray}
at all finite and positive $x$, and
\begin{eqnarray}\label{4.60b}
~~~~~~~~~~~~~~~~~~~~~~~~~~~~~~1<f_1^-(x)<f_2^-(x)<f_3^-(x)<\cdots
~~~~~~~~~~~~~~~~~~~~~~~~~~~~~~~~~(4.60b)\nonumber
\end{eqnarray}
at all finite and negative $x$. Since
\begin{eqnarray}\label{4.61a}
~~~~~~~~~~~~~~~~~~~~~~~~~~~~~~~~~~~~~~~~~{\cal E}_n^+<v_+(0)
~~~~~~~~~~~~~~~~~~~~~~~~~~~~~~~~~~~~~~~~~~~~~~~~~~~~(4.61a)\nonumber
\end{eqnarray}
and
\begin{eqnarray}\label{4.61b}
~~~~~~~~~~~~~~~~~~~~~~~~~~~~~~~~~~~~~~~~~{\cal E}_n^-<v_-(0)
~~~~~~~~~~~~~~~~~~~~~~~~~~~~~~~~~~~~~~~~~~~~~~~~~~~~~(4.61b)\nonumber
\end{eqnarray}
with both $v_{\pm}(0)$ finite, \setcounter{equation}{61}
\begin{eqnarray}\label{e3.62}
\lim_{n\rightarrow \infty}{\cal E}_n^+={\cal E}^+~~~~~~~{\sf
and}~~~~~~~ \lim_{n\rightarrow \infty}{\cal E}_n^-={\cal E}^-
\end{eqnarray}
both exist. Furthermore, by using the integral equations
(4.55a)-(4.55b)  for $f_n^{\pm}(x)$ and by following the arguments
similar to those given in Section~5 of Ref.~13, we can show  that
\begin{eqnarray}\label{e3.63}
\lim_{n\rightarrow \infty}f_n^+(x)=f^+(x)~~~~~~~{\sf and}~~~~~~~
\lim_{n\rightarrow \infty}f_n^-(x)=f^-(x)
\end{eqnarray}
also exist. This leads us from the first trial function $\phi(x)$
given by (\ref{e4.3}) to $\phi f^+(x)$ and $\phi f^-(x)$ which are
solutions of
\begin{eqnarray}\label{4.64a}
~~~~~~~~~~~~~~~~~~~(T+V(x)-E^+)\phi f^+(x)=0~~~~~~~{\sf
for}~~~x>0~~~~~~~~~~~~~~~~~~~~~~~~~~~~~~(4.64a)\nonumber
\end{eqnarray}
and
\begin{eqnarray}\label{4.64b}
~~~~~~~~~~~~~~~~~~~(T+V(x)-E^-)\phi f^-(x)=0~~~~~~~{\sf
for}~~~x<0~~~~~~~~~~~~~~~~~~~~~~~~~~~~~~(4.64b)\nonumber
\end{eqnarray}
with \setcounter{equation}{64}
\begin{eqnarray}\label{e3.65}
E^+=E_0^+-{\cal E}^+~~~~~~~~~{\sf and}~~~~~~~~E^-=E_0^- -{\cal
E}^-~,
\end{eqnarray}
and the boundary conditions at $x=0$,
\begin{eqnarray}\label{e3.66}
\phi'(0)=f^{+\prime}(0)=f^{-\prime}(0)=0.
\end{eqnarray}
An additional normalization factor multiplying, say, $f^-(x)$
would enable us to construct the second trial function $\chi(x)$
that satisfies (\ref{e4.9})-(\ref{e4.13}).

\newpage

\noindent {\bf 4.3 Symmetric vs Asymmetric Potential}

As we shall discuss, the general description leading from the
trial function $\chi(x)$ to the final wave function $\psi(x)$ that
satisfies the Schroedinger equation (\ref{e4.2}) may be set in a
more general framework. Decompose any potential $V(x)$ into two
parts
\begin{eqnarray}\label{e3.67}
V(x) \equiv \left\{\begin{array}{ccc}
V_a(x),~~~~~~~~~~~~~~~~~~{\sf for}~~x\geq 0\\
V_b(x),~~~~~~~~~~~~~~~~~~{\sf for}~~x\leq 0.
\end{array}
\right.
\end{eqnarray}
Next, extend the functions $V_a(x)$ and $V_b(x)$ by defining
\begin{eqnarray}\label{e3.68}
~~~~~~~~~~~~~~~~~~~~~~~~&V_a(x) \equiv
V_a(-x)~~~~~~~~~~~~~~~~~~~~~~~~{\sf for}~~~x<0~~~~~~~~~~~~~~~~\nonumber\\
{\sf
and}~~~~~~~~~~~~~~~~~~~~~~~~~&\\
~~~~~~~~~~~~~~~~~~~~~~~~&V_b(x) \equiv
V_b(-x)~~~~~~~~~~~~~~~~~~~~~~~~{\sf
for}~~~x>0.~~~~~~~~~~~~~~~~\nonumber
\end{eqnarray}
Thus, both $V_a(x)$ and $V_b(x)$ are symmetric potential covering
the entire $x$-axis. Let $\chi_a(x)$ and $\chi_b(x)$ be the
groundstate wave functions of the Hamiltonians $T+V_a$ and
$T+V_b$:
\begin{eqnarray}\label{e4.69a}
~~~~~~~~~~~~~~~~~~~~~~~~~~~~~~~~~(T+V_a(x))\chi_a(x)=E_a\chi_a(x)
~~~~~~~~~~~~~~~~~~~~~~~~~~~~~~~~~~~~(4.69a)\nonumber
\end{eqnarray}
and
\begin{eqnarray}\label{e4.69b}
~~~~~~~~~~~~~~~~~~~~~~~~~~~~~~~~~(T+V_b(x))\chi_b(x)=E_b\chi_b(x).
~~~~~~~~~~~~~~~~~~~~~~~~~~~~~~~~~~~~(4.69b)\nonumber
\end{eqnarray}
The symmetry (\ref{e3.68}) implies that \setcounter{equation}{69}
\begin{eqnarray}\label{e3.70}
\chi_a(x)=\chi_a(-x),~~~~~~~~~~~~~~\chi_b(x)=\chi_b(-x)
\end{eqnarray}
and at $x=0$
\begin{eqnarray}\label{e3.71}
\chi'_a(0)=\chi'_b(0)=0.
\end{eqnarray}
Choose the relative normalization factors of $\chi_a$ and
$\chi_b$, so that at $x=0$
\begin{eqnarray}\label{e3.72}
\chi_a(0)=\chi_b(0).
\end{eqnarray}
The same trial function (\ref{e4.9}) for the specific quartic
potential (\ref{e4.1}) is a special example of
\begin{eqnarray}\label{e3.73}
\chi(x) \equiv \left\{\begin{array}{ccc}
\chi_a(x),~~~~~~~~~~~~~~~~~~{\sf for}~~x\geq 0\\
\chi_b(x),~~~~~~~~~~~~~~~~~~{\sf for}~~x\leq 0
\end{array}
\right.
\end{eqnarray}
with
\begin{eqnarray}\label{e3.74}
~~~~~~~~~~~~~~~~~~~~~~~~&\chi_a(x) =
\chi^+(x)~~~~~~~~~~~~~~~~~~~~~~~~{\sf for}~~~x\geq 0~~~~~~~~~~~~~~~~\nonumber\\
{\sf and}~~~~~~~~~~~~~~~~~~~~~~~~~&\\
~~~~~~~~~~~~~~~~~~~~~~~~&\chi_b(x) =
\chi^-(x)~~~~~~~~~~~~~~~~~~~~~~~~{\sf for}~~~x\leq
0.~~~~~~~~~~~~~~~~\nonumber
\end{eqnarray}

In general, from (4.69a)-(4.69b) we see that $\chi(x)$ satisfies
\begin{eqnarray}\label{e3.75}
(T+V(x)+\hat{w}(x))\chi(x)=\hat{E}_0\chi(x).
\end{eqnarray}
Depending on the relative magnitude of $E_a$ and $E_b$, we define,
in the case of $E_a>E_b$
\begin{eqnarray}\label{e4.76a}
~~~~~~~~~~~~~~~~~\hat{w}(x)= \left\{\begin{array}{ccc}
0~~~~~~~~~~~~~~~~~~~~~~~~~~~~~~~~{\sf for}~~x> 0\\
E_a-E_b~~~~~~~~~~~~~~~~~~~~~~~~{\sf for}~~x<0
\end{array}
\right.~~~~~~~~~~~~~~~~~~~~~~~~~~~~(4.76a)\nonumber
\end{eqnarray}
and
\begin{eqnarray}\label{e4.77a}
~~~~~~~~~~~~~~~~~~~~~~~~~~~~~~~~~~~~~~~~~~~\hat{E}_0=E_a;
~~~~~~~~~~~~~~~~~~~~~~~~~~~~~~~~~~~~~~~~~~~~~~~~~~~~~(4.77a)\nonumber
\end{eqnarray}
otherwise, if $E_b>E_a$, we set
\begin{eqnarray}\label{e4.76b}
~~~~~~~~~~~~~~~~~\hat{w}(x)= \left\{\begin{array}{ccc}
E_b-E_a~~~~~~~~~~~~~~~~~~~~~~~~{\sf for}~~x>0\\
0~~~~~~~~~~~~~~~~~~~~~~~~~~~~~~~~{\sf for}~~x<0\\
\end{array}
\right.~~~~~~~~~~~~~~~~~~~~~~~~~~~~~(4.76b)\nonumber
\end{eqnarray}
and
\begin{eqnarray}\label{e4.77b}
~~~~~~~~~~~~~~~~~~~~~~~~~~~~~~~~~~~~~~~~~~~\hat{E}_0=E_b.
~~~~~~~~~~~~~~~~~~~~~~~~~~~~~~~~~~~~~~~~~~~~~~~~~~~~~(4.77b)\nonumber
\end{eqnarray}
Thus, we have either
\begin{eqnarray}\label{e4.78a}
~~~~~~~~~~~~~~~~~~~~~~~\hat{w}(\infty)=0~~~~~~{\sf
and}~~~~~~\hat{w}'(x)<0
~~~~~~~~~~~~~~~~~~~~~~~~~~~~~~~~~~~~~~~~~~(4.78a)\nonumber
\end{eqnarray}
at all finite $x$, or
\begin{eqnarray}\label{e4.78b}
~~~~~~~~~~~~~~~~~~~~~~~\hat{w}(-\infty)=0~~~~~~{\sf
and}~~~~~~\hat{w}'(x)>0
~~~~~~~~~~~~~~~~~~~~~~~~~~~~~~~~~~~~~~~~(4.78b)\nonumber
\end{eqnarray}
at all finite $x$. A comparison between (\ref{e4.9})-(\ref{e4.17})
and (\ref{e3.73})-(4.77a) shows that $w(x)$ of (\ref{e4.14}) and
the above $\hat{w}(x)$ differs only by a constant.

As in (\ref{e4.2}), $\psi(x)$ is the groundstate wave function
that satisfies \setcounter{equation}{78}
\begin{eqnarray}\label{e3.79}
(T+V(x))\psi(x)=E\psi(x),
\end{eqnarray}
which can also be written in the same form as (\ref{e1.14})
\begin{eqnarray}\label{e3.80}
(T+V(x)+\hat{w}(x)-\hat{E}_0)\psi(x)=(\hat{w}(x)-\hat{{\cal
E}})\psi(x),
\end{eqnarray}
with
\begin{eqnarray}\label{e3.81}
E=\hat{E}_0-\hat{{\cal E}}.
\end{eqnarray}
Here, unlike (\ref{e1.30}), $V(x)$ can now also be asymmetric.
Taking the difference between $\psi(x)$ times (\ref{e3.75}) and
$\chi(x)$ times (\ref{e3.80}), we derive
\begin{eqnarray}\label{e3.82}
\int_{-\infty}^{\infty}\chi(x)\psi(x)(\hat{w}(x)-\hat{{\cal
E}})dx=0.
\end{eqnarray}
Introduce
\begin{eqnarray}\label{e3.83}
\psi(x)=\chi(x)f(x),
\end{eqnarray}
in which $f(x)$ satisfies
\begin{eqnarray}\label{e3.84}
f(x)=f(\infty)-2\int_{x}^{\infty}\chi^{-2}(y)dy
\int_{y}^{\infty}\chi^2(z)(\hat{w}(z)-\hat{{\cal E}})f(z)dz.
\end{eqnarray}
On account of (\ref{e3.82})-(\ref{e3.83}), the same equation can
also be written as
\begin{eqnarray}\label{e3.85}
f(x)=f(-\infty)-2\int_{-\infty}^x\chi^{-2}(y)dy
\int_{-\infty}^y\chi^2(z)(\hat{w}(z)-\hat{{\cal E}})f(z)dz.
\end{eqnarray}

Eq.~(\ref{e3.80}) will again be solved iteratively by introducing
\begin{eqnarray}\label{e3.86}
\psi_n(x)=\chi(x)f_n(x)
\end{eqnarray}
with $\psi_n$ and its associated energy $\hat{{\cal E}}_n$
determined by
\begin{eqnarray}\label{e3.87}
(T+V(x)+\hat{w}(x)-\hat{E}_0)\psi_n(x)=(\hat{w}(x)-\hat{{\cal
E}}_n)\psi_{n-1}(x)
\end{eqnarray}
and
\begin{eqnarray}\label{e3.88}
\int_{-\infty}^{\infty}\chi(x)\psi_{n-1}(x)(\hat{w}(x)-\hat{{\cal
E}}_n)dx=0.
\end{eqnarray}
In terms of $f_n(x)$, we have
\begin{eqnarray}\label{e3.89}
f_n(x)=f_n(\infty)-2\int_{x}^{\infty}\chi^{-2}(y)dy
\int_{y}^{\infty}\chi^2(z)(\hat{w}(z)-\hat{{\cal
E}}_n)f_{n-1}(z)dz.
\end{eqnarray}
On account of (\ref{e3.88}), we also have
\begin{eqnarray}\label{e3.90}
\int_{-\infty}^{\infty}\chi^2(x)(\hat{w}(x)-\hat{{\cal
E}}_n)f_n(x)dx=0
\end{eqnarray}
and
\begin{eqnarray}\label{e3.91}
f_n(x)=f_n(-\infty)-2\int_{-\infty}^x\chi^{-2}(y)dy
\int_{-\infty}^y\chi^2(z)(\hat{w}(z)-\hat{{\cal
E}}_n)f_{n-1}(z)dz.
\end{eqnarray}

For definiteness, let us assume that
\begin{eqnarray}\label{e3.92}
E_a>E_b
\end{eqnarray}
in (4.69a)-(4.69b); therefore $\hat{w}'(x)<0$ and
$\hat{w}(\infty)=0$, in accordance with (4.76a). Start with, for
$n=0$,
\begin{eqnarray}\label{e3.93}
f_0(x)=1,
\end{eqnarray}
we can derive $\{E_n\}$ and $\{f_n(x)\}$, with
\begin{eqnarray}\label{e3.94}
E_n \equiv \hat{E}_0-\hat{{\cal E}}_n,
\end{eqnarray}
by using the boundary conditions, either
\begin{eqnarray}\label{A}
f_n(\infty)=1~~~~~~~~{\sf
for~all}~~n,~~~~~~~~~~~~~~~~~~~~~~~~~~(A)\nonumber
\end{eqnarray}
or
\begin{eqnarray}\label{B}
f_n(-\infty)=1~~~~~~~~{\sf
for~all}~~n.~~~~~~~~~~~~~~~~~~~~~~~~~~(B)\nonumber
\end{eqnarray}
It is straightforward to generalize the Hierarchy theorem to the
present case. As in Section~3, in Case $(A)$, the validity of the
Hierarchy theorem imposes no condition on the magnitude of
$\hat{w}(x)$. But in Case $(B)$ we assume $\hat{w}(x)$ to be not
too large so that (\ref{e3.91}) and the boundary condition
$f_n(-\infty)=1$ is consistent with
\begin{eqnarray}\label{e3.95}
f_n(x)>0
\end{eqnarray}
for all finite $x$. From the Hierarchy theorem, we find in Case
$(A)$
\begin{eqnarray}\label{e3.96}
E_1>E_2>E_3>\cdots
\end{eqnarray}
and
\begin{eqnarray}\label{3.97}
1\leq f_1(x)\leq f_2(x)\leq f_3(x)\leq \cdots,
\end{eqnarray}
while in Case (B)
\begin{eqnarray}\label{e3.98}
E_1>E_3>E_5>\cdots~~~~,\\
E_2<E_4<E_6<\cdots~~~~,\\
1\leq f_1(x)\leq f_3(x)\leq f_5(x)\leq \cdots
\end{eqnarray}
and
\begin{eqnarray}\label{e3.101}
1\geq f_2(x)\geq f_4(x)\geq f_6(x)\geq \cdots~.
\end{eqnarray}

A soluble model of an asymmetric square-well potential is given in
Appendix A to illustrate these properties.

\newpage

\section*{\bf 5. The N-Dimensional Problem}
\setcounter{section}{5} \setcounter{equation}{0}

The $N$-dimensional case will be discussed in this Section. We
begin with the electrostatic analog introduced in Section 1.
Suppose that the $(n-1)$th iterative solution $f_{n-1}({\bf q})$
is already known. The $n$th order charge density $\sigma_n({\bf
q})$ is
$$
\sigma_n({\bf q}) = (w({\bf q})-{\cal E}_n) \phi^2({\bf
q})f_{n-1}({\bf q}), \eqno(5.1)
$$
in accordance with (1.23)-(1.24). Likewise, from (1.26) and (1.29)
the dielectric-constant $\kappa$ of the medium is related to the
trial function $\phi({\bf q})$ by
$$
\kappa({\bf q})=\phi^2({\bf q}), \eqno(5.2)
$$
and the $n$th order energy shift ${\cal E}_n$ is determined by
$$
\int \sigma_n({\bf q})d^N q=0. \eqno(5.3)
$$
In the following we assume the range of $w({\bf q})$ to be finite,
with
$$
w(\infty)=0 \eqno(5.4)
$$
and
$$
0\leq w({\bf q})\leq W_{max}. \eqno(5.5)
$$
Introduce
$$
K(W)\equiv \int \kappa ({\bf q}) \delta (w({\bf q})-W)d^N q,
\eqno(5.6)
$$
where $\delta (w({\bf q})-W)$ is Dirac's $\delta$-function, $W$ is
a constant parameter and the integrations in (5.3) and (5.6) are
over all ${\bf q}$-space. Similarly, for any function $F({\bf
q})$, we define
$$
F^{av}(W) \equiv K(W)^{-1} \int F({\bf q})\kappa({\bf q})
\delta(w({\bf q})-W)d^N q. \eqno(5.7)
$$
In  the $N$-dimensional case, the generalization of $[F]$,
introduced by (3.15), is
$$
[F] = \int F({\bf q})\phi^2({\bf q}) d^N q. \eqno(5.8)
$$
In terms of $F^{av}(W)$, (5.8) can also be written as
$$
[F] = \int^{W_{max}}_0 F^{av}(W) K(W) dW. \eqno(5.9)
$$
Thus from (5.1) and (5.3) we have
$$
{\cal E}_n = [w({\bf q})f_{n-1}({\bf q})]/[f_{n-1}({\bf q})],
\eqno(5.10)
$$
the $n$-dimensional extension of (3.14).

Following (1.27)-(1.28), the $n$th order electric field is
$-\frac{1}{2} \nabla f_n$ and the displacement field is
$$
D_n=-\frac{1}{2} \kappa \nabla f_n. \eqno(5.11)
$$
The corresponding Maxwell equation is
$$
\nabla \cdot D_n = \sigma_n. \eqno(5.12)
$$
Eqs.(5.11) and (5.12) determine $f_n$ except for an additive
constant, which can be chosen by requiring
$$
{\sf minimum~of}~~f_n({\bf q})=1. \eqno(5.13)
$$
Therefore,
$$
f_n({\bf q}) \geq 1. \eqno(5.14)
$$
As in the one dimensional case discussed in Section 3, (5.10)
gives the same condition of fine energy tuning at each order of
iteration. It is this condition that leads to convergent iterative
solutions derived in Section 3.

We now conjecture that
$$
\lim\limits_{n\rightarrow \infty} {\cal E}_n = {\cal E} =E_0-E
\eqno(5.15)
$$
and
$$
\lim\limits_{n\rightarrow \infty} f_n({\bf q}) = f({\bf q})
=\psi({\bf q})/\phi({\bf q}). \eqno(5.16)
$$
also hold in higher dimensions. Although we are not able to
establish this conjecture, in the following we present the proofs
of the $N$-dimensional generalizations of some of the lemmas
proved in Section 3.

\noindent Lemma 1. For any pair $f_{m}({\bf q})$ and $f_{l}({\bf
q})$ if at all $W$ within the range (5.5),
$$
\frac{d}{dW}\left(\frac{f^{av}_m(W)}{f^{av}_l(W)}\right) >0,~~~~~~
{\sf then}~~{\cal E}_{m+1}>{\cal E}_{l+1}, \eqno(5.17)
$$
and
$$
\frac{d}{dW}\left(\frac{f^{av}_m(W)}{f^{av}_l(W)}\right) <0,~~~~~~
{\sf then}~~{\cal E}_{m+1}<{\cal E}_{l+1}. \eqno(5.18)
$$
\noindent \underline{Proof}. For any function ${\cal F}(W)$,
define
$$
<{\cal F}(W)> \equiv \int\limits^{W_{max}}_0 K(W) {\cal F}(W) dW,
\eqno(5.19)
$$
Thus for any function $F({\bf q})$, we have
$$
[F({\bf q})]=<F^{av}(W)>; \eqno(5.20)
$$
therefore,
$$
{\cal E}_n[f_{n-1}({\bf q})]={\cal E}_n<f^{av}_{n-1}(W)>
\eqno(5.21)
$$
and
$$
[w({\bf q})f_{n-1}({\bf q})]=<W f^{av}_{n-1}(W)>. \eqno(5.22)
$$
By setting the subscript $n$ in (5.10) to be $m+1$, we obtain
$$
{\cal E}_{m+1}<f^{av}_m(W)>=<W f^{av}_m(W)>. \eqno(5.23)
$$
Also by definition (5.19),
$$
{\cal E}_{l+1}<f^{av}_m(W)>=<{\cal E}_{l+1} f^{av}_m(W)>.
\eqno(5.24)
$$
The difference of (5.23) and (5.24) gives
$$
({\cal E}_{m+1}-{\cal E}_{l+1})<f^{av}_m(W)>=<(W-{\cal E}_{l+1})
f^{av}_m(W)>. \eqno(5.25)
$$
From (5.10) and setting the subscript $n$ to be $l+1$, we have
$$
0=<(W-{\cal E}_{l+1}) f^{av}_l(W)>. \eqno(5.26)
$$
Regard $f^{av}_l({\cal E}_{l+1})$ and $f^{av}_m({\cal E}_{l+1})$
as two constant parameters. Multiply (5.25) by $f^{av}_l({\cal
E}_{l+1})$, (5.26) by $f^{av}_m({\cal E}_{l+1})$ and take their
difference. The result is
$$
({\cal E}_{m+1}-{\cal E}_{l+1})f^{av}_l({\cal
E}_{l+1})<f^{av}_m(W)> =<(W-{\cal E}_{l+1})(f^{av}_l({\cal
E}_{l+1}) f^{av}_m(W) -f^{av}_l(W)f^{av}_m({\cal E}_{l+1})>,
\eqno(5.27)
$$
analogous to (3.43).

\noindent (i) If $\frac{d}{dW}(f^{av}_m(W)/f^{av}_l(W)) >0$, then
for $W<{\cal E}_{l+1}$
$$
\frac{f^{av}_m(W)}{f^{av}_l(W)} < \frac{f^{av}_m({\cal
E}_{l+1})}{f^{av}_l({\cal E}_{l+1})}~. \eqno(5.28)
$$
Thus, the function inside the bracket $<~>$ in (5.21) is positive,
being the product of two negative factors, $(W-{\cal E}_{l+1})$
and $(f^{av}_l({\cal E}_{l+1})f^{av}_m(W)- f^{av}_l(W)
f^{av}_m({\cal E}_{l+1}))$. Also, when $W>{\cal E}_{l+1}$, these
two factors both reverse their signs. Consequently (5.17) holds.

\noindent (ii) If $\frac{d}{dW}(f^{av}_m(W)/f^{av}_l(W)) <0$, we
see that for $W<{\cal E}_{l+1}$, (5.28) reverses its sign, and
therefore the function inside the bracket $<~>$ in (5.27) is now
negative. The same negative sign can be readily established for
$W>{\cal E}_{l+1}$. Consequently, (5.18) holds and Lemma 1 is
established.

\noindent Lemma 2. Identical to Lemma 2 of Section 3.

In order to establish the $N$-dimensional generalization of
Lemma~3 of Section 3, we define
$$
Q_n(W) \equiv \int\limits^{W_{max}}_W dW_1 \int \sigma_n({\bf q})
\delta(w({\bf q})-W_1) d^N q. \eqno(5.29)
$$
Because of (5.3), $Q_n(W)$ is also given by
$$
Q_n(W) = -\int\limits^{W}_0 dW_1 \int \sigma_n({\bf q})
\delta(w({\bf q})-W_1) d^N q. \eqno(5.30)
$$
We may picture that the entire $q$-space is divided into two
regions
$$
I.~~~~~~~~w({\bf q})>W \eqno(5.31)
$$
and
$$
II.~~~~~~~w({\bf q})<W, \eqno(5.32)
$$
with $Q_n(W)$ the total charge in $I$, which is also the negative
of the total charge in $II$. By using (5.1) and (5.7), we see that
$$
-\frac{dQ_n(W)}{dW}=(W-{\cal E}_n)f^{av}_{n-1}(W)K(W). \eqno(5.33)
$$

\noindent Lemma 3. For any pair $f_m({\bf q})$ and $f_l({\bf q})$
if at all $W$ within the range (5.5)
$$
(i)~~~~~~\frac{d}{dW}\left(\frac{f^{av}_m(W)}{f^{av}_l(W)}\right)
>0~~~~~{\sf then}
~~~~~\frac{d}{dW}\left(\frac{Q_{m+1}(W)}{Q_{l+1}(W)}\right) >0
\eqno(5.34)
$$
$$
(ii)~~~~~\frac{d}{dW}\left(\frac{f^{av}_m(W)}{f^{av}_l(W)}\right)
<0~~~~~{\sf then}
~~~~~\frac{d}{dW}\left(\frac{Q_{m+1}(W)}{Q_{l+1}(W)}\right) <0.
\eqno(5.35)
$$
\noindent \underline{Proof} Note that (5.34) and (5.35) are very
similar to (3.56) and (3.57). As in (3.60), define
$$
\xi = Q_{l+1}(W)~~~~~~{\sf and}~~~~~~\eta =Q_{m+1}(W). \eqno(5.36)
$$
From (5.33), we have
$$
-\frac{dQ_{m+1}(W)}{dW}=(W-{\cal E}_{m+1})f^{av}_{m}(W)K(W)
\eqno(5.37)
$$
and
$$
-\frac{dQ_{l+1}(W)}{dW}=(W-{\cal E}_{l+1})f^{av}_{l}(W)K(W).
\eqno(5.38)
$$
Therefore,
$$
\frac{d\eta}{d\xi}=\frac{dQ_{m+1}(W)}{dW}\Big/\frac{dQ_{l+1}(W)}{dW}
=r(W)\frac{f^{av}_{m}(W)}{f^{av}_{l}(W)} \eqno(5.39)
$$
where
$$
r(W)=\frac{W-{\cal E}_{m+1}}{W-{\cal E}_{l+1}}~. \eqno(5.40)
$$
Furthermore,
$$
\frac{d}{d\xi}\left(\frac{d\eta}{d\xi}\right)
=\left(\frac{dQ_{l+1}}{dW}\right)^{-1}\frac{d}{dW}
\left(\frac{dQ_{m+1}}{dW}\Big/\frac{dQ_{l+1}}{dW}\right)
=\left(\frac{dQ_{l+1}}{dW}\right)^{-1}\frac{d}{dW}
\left(r\frac{f^{av}_m}{f^{av}_l}\right)
$$
$$
~~~~~~~~~~~~~~~~=\left(\frac{dQ_{l+1}}{dW}\right)^{-1}
\left(\frac{dr}{dW}\frac{f^{av}_m}{f^{av}_l}+r\frac{d}{dW}
\left(\frac{f^{av}_m}{f^{av}_l}\right)\right), \eqno(5.41)
$$
where
$$
\frac{dr}{dW}=\frac{{\cal E}_{m+1}-{\cal E}_{l+1}}{(W-{\cal
E}_{l+1})^2}, \eqno(5.42)
$$
analogous to (3.61)-(3.64).

According to (5.30), at $W=0$
$$
Q_{m+1}(0)=Q_{l+1}(0)=0 \eqno(5.43)
$$
and according to (5.29), at $W=W_{max}$
$$
Q_{m+1}(W_{max})=Q_{l+1}(W_{max})=0. \eqno(5.44)
$$
From (5.37), we see that the derivative $\frac{d}{dW}Q_{m+1}(W)$
is positive when $W<{\cal E}_{m+1}$, zero at $W={\cal E}_{m+1}$
and negative when $W>{\cal E}_{m+1}$. Likewise, from (5.38),
$\frac{d}{dW}Q_{l+1}(W)$ is positive when $W<{\cal E}_{l+1}$, zero
at $W={\cal E}_{l+1}$ and negative when $W>{\cal E}_{l+1}$. Their
ratio determines $\frac{d\eta}{d\xi}$

\noindent (i) If $\frac{d}{dW}(f^{av}_m/f^{av}_l)>0$, from lemma
1, we have
$$
{\cal E}_{m+1}>{\cal E}_{l+1} \eqno(5.45)
$$
and therefore, on account of (5.42)
$$
\frac{dr}{dW} >0. \eqno(5.46)
$$
At $W=0$,
$$
r(0)=\frac{{\cal E}_{m+1}}{{\cal E}_{l+1}}>1. \eqno(5.47)
$$
As $W$ increases, so does $r(W)$. At $W={\cal E}_{l+1}$, $r(W)$
has a discontinuity, with
$$
r({\cal E}_{l+1}-)=\infty \eqno(5.48)
$$
and
$$
r({\cal E}_{l+1}+)=-\infty. \eqno(5.49)
$$
As $W$ increases from ${\cal E}_{l+1}$, $r(W)$ continues to
increase, with
$$
r({\cal E}_{m+1})=0 \eqno(5.50)
$$
and
$$
r(W_{max})=\frac{W_{max}-{\cal E}_{m+1}}{W_{max}-{\cal
E}_{l+1}}<1. \eqno(5.51)
$$

It is convenient to divide the range $0<W<W_{max}$ into three
regions:
$$
{\sf A.}~~~~~~~0<W<{\cal E}_{l+1} \eqno(5.52)
$$
$$
~~~~~{\sf B.}~~~~~~~{\cal E}_{l+1}<W<{\cal E}_{m+1} \eqno(5.53)
$$

$$
~~~~~~~~{\sf C.}~~~~~~~{\cal E}_{m+1}<W<W_{max}. \eqno(5.54)
$$
Assuming $\frac{d}{dW}(f^{av}_m/f^{av}_l)>0$, we shall show
separately $\frac{d}{dW}(Q_{m+1}/Q_{l+1})>0$ in these three
regions.

In region B, $Q_{l+1}$ is decreasing, but $Q_{m+1}$ is increasing.
Clearly,
$$
\frac{d}{dW}(Q_{m+1}/Q_{l+1})>0. \eqno(5.55)
$$
In region A, $\frac{dQ_{l+1}}{dW}>0$, $r(W)$ is positive according
to (5.47)-(5.48) and $\frac{dr}{dW}$ is always $>0$ from (5.46).
Therefore from (5.41),
$$
\frac{d^2\eta}{d\xi^2}>0~~~~~{\sf in }~~~A. \eqno(5.56)
$$
In region C, $\frac{dQ_{l+1}}{dW}<0$, but $r(W)$ and
$\frac{dr}{dW}$ are both positive. Hence,
$$
\frac{d^2\eta}{d\xi^2}<0~~~~~{\sf in }~~~C. \eqno(5.57)
$$

Within each region, $\eta=Q_{m+1}(W)$ and $\xi=Q_{l+1}(W)$ are
both monotonic in $W$; therefore $\eta$ is a single-valued
function of $\xi$ and we can apply Lemma 2 of Section 3.

In region A, at $W=0$ both $Q_{m+1}(0)$ and $Q_{l+1}(0)$ are $0$
according to (5.43), but their ratio is given by
$$
\frac{Q_{m+1}(0)}{Q_{l+1}(0)}=\left(\frac{dQ_{m+1}(W)}{dW}\Big/
\frac{dQ_{l+1}(W)}{dW}\right)_{W=0}. \eqno(5.58)
$$
Therefore
$$
\left(\frac{d\eta}{d\xi}\right)_{W=0}=\left(\frac{\eta}{\xi}\right)_{W=0}.
\eqno(5.59)
$$
Furthermore, from (5.56), $\frac{d^2\eta}{d\xi^2}>0$. It follows
from Lemma 2 of Section 3, the ratio $\eta/\xi$ is an increasing
function of $\xi$. Since
$$
\frac{d\xi}{dW}=\frac{dQ_{l+1}(W)}{dW} >0~~~~{\sf in}~~~A,
\eqno(5.60)
$$
as also have
$$
\frac{d}{dW}\left(\frac{Q_{m+1}}{Q_{ml+1}}\right)=
\frac{d\xi}{dW}\frac{d}{d\xi}\left(\frac{\eta}{\xi}\right)>0~~~{\sf
in}~~~A. \eqno(5.61)
$$

In region C, at $W=W_{max}$, both $Q_{m+1}(W_{max})$ and
$Q_{l+1}(W_{max})$ are $0$ according to (5.44). Their ratio is
$$
\frac{Q_{m+1}(W_{max})}{Q_{l+1}(W_{max})}=
\left(\frac{dQ_{m+1}(W)}{dW}\Big/
\frac{dQ_{l+1}(W)}{dW}\right)_{W=W_{max}}, \eqno(5.62)
$$
Which gives at $W=W_{max}$
$$
\left(\frac{d\eta}{d\xi}\right)_{W=W_{max}}=
\left(\frac{\eta}{\xi}\right)_{W=W_{max}}. \eqno(5.63)
$$
As $W$ decreases from $W_{max}$ to ${\cal E}_{m+1}$ in region C,
since ${\cal E}_{m+1}>{\cal E}_{l+1}$, we have
$$
\frac{d\xi}{dW}=\frac{dQ_{l+1}}{dW}<0~~~~{\sf in}~~~C. \eqno(5.64)
$$
Furthermore, from (5.57), $\frac{d^2\eta}{d\xi^2}<0$ in region C.
It follows from Lemma 2 of Section 3, the ratio $\eta/\xi$ is a
decreasing function of $\xi$, which together with (5.64) lead to
$$
\frac{d}{dW}\left(\frac{Q_{m+1}}{Q_{l+1}}\right)>0~~~~{\sf
in}~~~~C. \eqno(5.65)
$$
Thus, we prove case (i) of Lemma 3. Case (ii) of Lemma 3 follows
from case (i) by an exchange of the subscripts $m$ and $l$.
Lemma~3 is then proved.

So far, the above Lemmas 1 and 3 are almost identical copies of
Lemmas 1 and 3 of Section 3, but now applicable to the
$N$-dimensional problem. Difficulty arises when we try to
generalize Lemma 4 of Section 3.

It is convenient to transform the Cartesian coordinates
$q_1,~q_2,~\cdots,~q_N$ to a new set of orthogonal coordinates:
$$
(q_1,~q_2,~\cdots,~q_N)\rightarrow (w({\bf q}),~\beta_1({\bf q}),~
\cdots,~\beta _{N-1}({\bf q})) \eqno(5.66)
$$
with
$$
\nabla w \cdot \nabla\beta_i=0 \eqno(5.67)
$$
and
$$
\nabla\beta_i \cdot \nabla\beta_j=0~~~{\sf for}~~i\neq j,
\eqno(5.68)
$$
where $i$ or $j=1,~2,~\cdots,~N-1$. Introducing
$$
h^2_w=1/(\nabla w)^2~,~~~~~h^2_i=1/(\nabla \beta_i)^2 \eqno(5.69)
$$
$$
\hat{w}=h_w\nabla w~~~{\sf and}~~~~\hat{\beta}_i=h_i\nabla \beta_i
\eqno(5.70)
$$
In terms of the new coordinates, the components of $D_n$ are
$$
(D_n)_w=\hat{w}\cdot D_n~~~~{\sf and}~~~~ (D_n)_i=\hat{\beta}_i
\cdot D_n. \eqno(5.71)
$$
Its divergence is
$$
\nabla \cdot D_n=(h_w
h_{\beta})^{-1}\left\{\frac{\partial}{\partial w}
(h_{\beta}(D_n)_w)+ \sum\limits^{N-1}_{i=1}
\frac{\partial}{\partial \beta_i}(h^{-1}_i h_w h_{\beta}
(D_n)_i)\right\}. \eqno(5.72)
$$
Combining (5.12) with (5.30), we have
$$
Q_n(W)=-\int\limits_0^W dw \int h_w h_{\beta} (\nabla \cdot D_n)
\prod\limits^{N-1}_1 d\beta_i~; \eqno(5.73)
$$
therefore,
$$
Q_n(W)=-\oint (D_n)_w  h_{\beta} \prod\limits^{N-1}_1 d\beta_i,
\eqno(5.74)
$$
in which the integration is along the surface
$$
w({\bf q})=W. \eqno(5.75)
$$
From (5.11) and (5.71), it follows that
$$
(D_n)_w=\frac{1}{2} \kappa \frac{\partial f_n}{\partial w}
=-\frac{1}{2} \kappa h_w \nabla w \cdot \nabla f_n. \eqno(5.76)
$$
In terms of curvilinear coordinates, (5.7) can be written as
$$
F^{av}(W)=K(W)^{-1} \oint F({\bf q}) \kappa({\bf q}) h_w h_{\beta}
\prod\limits^{N-1}_1 d\beta_i~. \eqno(5.77)
$$
Substituting (5.76) into (5.74), we find
$$
Q_n(W)=\frac{1}{2} K(W) (\nabla w \cdot \nabla f_n)^{av}.
\eqno(5.78)
$$
Because $h_w^{-1} (\partial f_n/\partial w) = \hat{w} \cdot \nabla
f_n =h_w (\nabla w \cdot \nabla f_n)$, (5.78) can also be written
as
$$
Q_n(W)=\frac{1}{2} K(W) \left(h_w^{-2} \frac{\partial
f_n}{\partial w}\right)^{av}. \eqno(5.79)
$$
Here comes the difficulty. While the above Lemma 3 transfers
relations between $f^{av}_m/f^{av}_l$ to those between
$Q_{m+1}/Q_{l+1}$, the latter is
$$
\left(h_w^{-2}\frac{\partial f_{m+1}}{\partial w}\right)^{av}\Big/
\left(h_w^{-2}\frac{\partial f_{l+1}}{\partial w}\right)^{av}
\eqno(5.80)
$$
which is quite different from
$\frac{d}{dW}f^{av}_{m+1}/\frac{d}{dW}f^{av}_{l+1}$. This
particular generalization of the lemmas in higher dimensions fails
to establish the Hierarchy Theorem.

For the one-dimensional case discussed in Section~3, we have
$w'<0$ and $x\geq 0$; consequently (5.80) is $f'_{m+1}/f'_{l+1}$.
Therefore, Lemma~4 of Section~3 can also be established by using
(5.80), and the proof of the Hierarchy Theorem can be completed.

\newpage

\section*{\bf References}
\setcounter{section}{9} \setcounter{equation}{0}

\noindent [1] R. Friedberg, T. D. Lee and W. Q. Zhao, IL Nuovo
Cimento A112 (1999), 1195

\noindent [2] R. Friedberg, T. D. Lee and W. Q. Zhao, Ann.Phys.
288 (2001), 52

\noindent
[3] R. Friedberg, T. D. Lee, W. Q. Zhao and A. Cimenser, Ann.Phys. 294 (2001), 67\\
\noindent
[4] R. Friedberg, T. D. Lee, Ann.Phys. 308 (2003), 263

\noindent
[5] A. M. Polyakov, Nucl.Phys. B121 (1977), 429 \\
~[6] G. 't Hooft, in: A. Zichichi, Erice(Eds.), The why's of
subnuclear physics

\noindent ~~~~~~~~~~        Plenum, New York, 1977\\
~[7] E. Brezin, G. Parisi and J. Zinn-Justin, Phys.Rev. D16 (1977), 408\\
~[8] J. Zinn-Justin, J.Math.Phys. 22 (1981), 511 \\
~[9] J. Zinn-Justin, Nucl.Phys. B192 (1981), 125 \\
~[10] J. Zinn-Justin, in: J.-D. Zuber, R. Stora (Eds.), Recent
advances in field theory and

\noindent ~~~~~~~~~statistical mechanics,
Les Houches, session XXXIX, 1982 \\
~[11] J. Zinn-Justin, Private Communication \\
~[12] Sidney Coleman, Aspects of Symmetry, Press Syndicate of

\noindent ~~~~~~~~~the University of Cambridge 1987  \\
~[13] E. Shuryak, Nucl.Phys. B302 (1988), 621

\noindent
[14] S. V. Faleev and P. G. Silvestrov, Phys. Lett. A197
(1995), 372

\newpage

\section*{\bf Appendix }
\setcounter{section}{7} \setcounter{equation}{0}

\noindent
{\bf A.1 A Soluble Example}\\

In this Appendix, we consider a soluble model in which the
potential $V(x)$ of (4.67) is
\begin{eqnarray}\label{eA.1}
~~~~~~~~~~~~~~~~~V(x)= \left\{\begin{array}{lcl}
\infty&&~~\gamma<x\\
\frac{1}{2}\mu^2&&~~\alpha<x<\gamma\\
\frac{1}{2}W^2&~~~~~~~~~~~~~{\sf for}~~~~~~~&-\alpha<x<\alpha\\
0&&-\gamma<x<-\alpha\\
\infty&&~~~~~~~~x<-\gamma,
\end{array}
\right.~~~~~~~~~~~~~~~~~~~~~~~(A.1)\nonumber
\end{eqnarray}
with $W^2>\mu^2$ and
\begin{eqnarray}\label{A.2}
~~~~~~~~~~~~~~~~~~~~~~~~~~~~~~~~~~~~~\gamma=\alpha+\beta.
~~~~~~~~~~~~~~~~~~~~~~~~~~~~~~~~~~~~~~~~~~~~~~~~~~~~~~~~~~(A.2)\nonumber
\end{eqnarray}
Following (4.68), we introduce two symmetric potentials:
\begin{eqnarray}\label{A.3}
~~~~~~~~~~~~~~~~~~~V_a(x) = V_a(-x)~~~~~~~~{\sf and}~~
~~~~~~~~V_b(x)= V_b(-x)~~~~~~~~~~~~~~~~~~~~~~~~~~~(A.3)\nonumber
\end{eqnarray}
with, for $x\geq 0$,
\begin{eqnarray}\label{eA.4}
~~~~~~~~~~~~~~~~V_a(x)= \left\{\begin{array}{lcl}
\infty&&~~\gamma<x\\
\frac{1}{2}\mu^2&~~~~~~~~~~~~~{\sf for}~~~~~~~&~~\alpha<x<\gamma\\
\frac{1}{2}W^2&&~~~0\leq x<\alpha
\end{array}
\right.~~~~~~~~~~~~~~~~~~~~~~~~~~(A.4)\nonumber
\end{eqnarray}
and
\begin{eqnarray}\label{eA.5}
~~~~~~~~~~~~~~~~V_b(x)= \left\{\begin{array}{lcl}
\infty&&~~\gamma<x\\
0&~~~~~~~~~~~~~{\sf for}~~~~~~~&~~\alpha<x<\gamma\\
\frac{1}{2}W^2&&~~~0\leq x<\alpha,
\end{array}
\right.~~~~~~~~~~~~~~~~~~~~~~~~~(A.5)\nonumber
\end{eqnarray}
so that (A.1) can also be written as
\begin{eqnarray}\label{eA.6}
~~~~~~~~~~~~~~~~~V(x)= \left\{\begin{array}{lcl}
V_a(x)&~~~~~~~~~~~~~{\sf for}~~~~~~~&~~x\geq 0\\
V_b(x)&~~~~~~~~~~~~~{\sf for}~~~~~~~&~~x\leq 0.
\end{array}
\right.~~~~~~~~~~~~~~~~~~~~~~~~~~~~~~~(A.6)\nonumber
\end{eqnarray}
Let $\psi(x)$, $\chi_a(x)$ and $\chi_b(x)$ be respectively the
groundstate wave functions of
\begin{eqnarray}\label{A.7}
~~~~~~~~~~~~~~~~~~~(T+V(x))\psi(x)&=&E\psi(x)
~~~~~~~~~~~~~~~~~~~~~~~~~~~~~~~~~~~~~~~~~~~~~~~~~~~~(A.7)\nonumber\\
~~~~~~~~~~~~~~~~~~~(T+V_a(x))\chi_a(x)&=&E_a\chi_a(x)
~~~~~~~~~~~~~~~~~~~~~~~~~~~~~~~~~~~~~~~~~~~~~~~~~~(A.8)\nonumber
\end{eqnarray}
and
\begin{eqnarray}\label{A.9}
~~~~~~~~~~~~~~~~~~~(T+V_b(x))\chi_b(x)&=&E_b\chi_b(x).
~~~~~~~~~~~~~~~~~~~~~~~~~~~~~~~~~~~~~~~~~~~~~~~~~~(A.9)\nonumber
\end{eqnarray}
For $|x|>\gamma$, since $V(x)=\infty$, we have
\begin{eqnarray}
\psi(x)=\chi_a(x)=\chi_b(x)=0.\nonumber
\end{eqnarray}
\newpage

For $|x|<\gamma$, these wave functions are of the form
\begin{eqnarray}\label{eA.10}
~~~~~~~~~~~~~~~~~\psi(x)\propto \left\{\begin{array}{lcl}
\sin k~(-x+\gamma)&&~~\alpha<x<\gamma\\
\cosh q(x-\delta)&~~~~~~~~~~~~~{\sf for}~~~~~~~&-\alpha<x<\alpha\\
\sin p~(x+\gamma)&&-\gamma<x<-\alpha,
\end{array}
\right.~~~~~~~~~~~(A.10)\nonumber
\end{eqnarray}
\begin{eqnarray}\label{eA.11}
~~~~~~~~~~~~~~~~~\chi_a(x)\propto \left\{\begin{array}{lcl}
\sin k_a~(-x+\gamma)&&~~\alpha<x<\gamma\\
\cosh q_a~x&~~~~~~~~~~~~~{\sf for}~~~~~~~&-\alpha<x<\alpha\\
\sin k_a~(x+\gamma)&&-\gamma<x<-\alpha
\end{array}
\right.~~~~~~~~~~(A.11)\nonumber
\end{eqnarray}
and
\begin{eqnarray}\label{eA.12}
~~~~~~~~~~~~~~~~~\chi_b(x)\propto \left\{\begin{array}{lcl}
\sin p_b~(-x+\gamma)&&~~\alpha<x<\gamma\\
\cosh q_b~x&~~~~~~~~~~~~~{\sf for}~~~~~~~&-\alpha<x<\alpha\\
\sin p_b~(x+\gamma)&&-\gamma<x<-\alpha.
\end{array}
\right.~~~~~~~~~~(A.12)\nonumber
\end{eqnarray}
By substituting these solutions to the Schroedinger equations
(A.7)-(A.9), we derive
\begin{eqnarray}\label{A.13}
~~~~~~~~~~~~~~~~~~~~~~~~~~~~~~~~~E&=&\frac{1}{2}p^2=\frac{1}{2}(W^2-q^2)=\frac{1}{2}(\mu^2+k^2),
~~~~~~~~~~~~~~~~~~~~~~~(A.13)\nonumber\\
~~~~~~~~~~~~~~~~~~~~~~~~~~~~~~~~~E_a&=&\frac{1}{2}(W^2-q_a^2)=\frac{1}{2}(\mu^2+k_a^2)
~~~~~~~~~~~~~~~~~~~~~~~~~~~~~~~~~(A.14)\nonumber
\end{eqnarray}
and
\begin{eqnarray}\label{A.15}
~~~~~~~~~~~~~~~~~~~~~~~~~~~~~~~~~E_b=\frac{1}{2}p_b^2=\frac{1}{2}(W^2-q_b^2).
~~~~~~~~~~~~~~~~~~~~~~~~~~~~~~~~~~~~~~~~~~~~~(A.15)\nonumber
\end{eqnarray}
The continuity of $\psi'/\psi$ at $x=\pm\alpha$ relates
\begin{eqnarray}\label{A.16}
~~~~~~~~~~~~~~~~~~~~~~~~~~~~~~~~-k\beta~\cot k\beta=q\beta \tanh
q~(\alpha-\delta)
~~~~~~~~~~~~~~~~~~~~~~~~~~~~~~~~~~~(A.16)\nonumber
\end{eqnarray}
and
\begin{eqnarray}\label{A.17}
~~~~~~~~~~~~~~~~~~~~~~~~~~~~~~~~-p\beta~\cot p\beta=q\beta \tanh
q~(\alpha+\delta)
~~~~~~~~~~~~~~~~~~~~~~~~~~~~~~~~~~~(A.17)\nonumber
\end{eqnarray}
with $\beta$ given by (A.2). In the following, we assume the
barrier heights $W$ and
\begin{eqnarray}\label{A.18}
~~~~~~~~~~~~~~~~~~~~~~~~~~~~~~~~~~~~~~~~~~~\hat{W} \equiv
(W^2-\mu^2)^{1/2}
~~~~~~~~~~~~~~~~~~~~~~~~~~~~~~~~~~~~~~~~~~(A.18)\nonumber
\end{eqnarray}
to be much larger than $k$ and $p$; therefore, the wave function
$\psi$ is mostly contained within the two square-wells; i.e.,
$k\beta$ and $p\beta$ are both near $\pi$. We write
\begin{eqnarray}\label{A.19}
~~~~~~~~~~~~~~~~~~~~~~~~~~~~~~~~~k\beta=\pi-\hat{\theta},~~~~~~~~~~
p\beta=\pi-\theta~~~~~~~~~~~~~~~~~~~~~~~~~~~~~~~~~~~~(A.19)\nonumber
\end{eqnarray}
and expect $\hat{\theta}$ and $\theta$ to be small. Likewise,
introduce
\begin{eqnarray}\label{A.20}
~~~~~~~~~~~~~~~~~~~~~~~~~~~~~k_a\beta=\pi-\hat{\theta}_a,~~~~~~~~~~
p_b\beta=\pi-\theta_b.~~~~~~~~~~~~~~~~~~~~~~~~~~~~~~~~~~~(A.20)\nonumber
\end{eqnarray}

The explicit forms of these angles can be most conveniently
derived by recognizing the separate actions of two related small
parameters: one proportional to the inverse of the barrier height
\begin{eqnarray}\label{A.21}
~~~~~~~~~~~~~~~~~~~~~~~~~~~~~~~~~~~~~~(W\beta)^{-1}<<1
~~~~~~~~~~~~~~~~~~~~~~~~~~~~~~~~~~~~~~~~~~~~~~~~~~(A.21)\nonumber
\end{eqnarray}
and the other
\begin{eqnarray}\label{A.22}
~~~~~~~~~~~~~~~~~~~~~~~~~~~~~~~~~~~~~~~e^{-2W\alpha}<<<1,
~~~~~~~~~~~~~~~~~~~~~~~~~~~~~~~~~~~~~~~~~~~~~~~~(A.22)\nonumber
\end{eqnarray}
denoting the much smaller tunnelling coefficient.

To illustrate how these two effects can be separated, let us
consider first the determination of $\theta_b$ given by (A.20).
The continuity of $\chi'_b/\chi_b$ at $x=\pm\alpha$ gives
\begin{eqnarray}\label{A.23}
~~~~~~~~~~~~~~~~~~~~~~~~~~~~~-p_b\beta \cot p_b\beta=q_b\beta
\tanh q_b\alpha.
~~~~~~~~~~~~~~~~~~~~~~~~~~~~~~~~~~~~~~~~(A.23)\nonumber
\end{eqnarray}
From (A.15), we also have
\begin{eqnarray}\label{A.24}
~~~~~~~~~~~~~~~~~~~~~~~~~~~~~~~~~~~~W^2=p_b^2+q_b^2.
~~~~~~~~~~~~~~~~~~~~~~~~~~~~~~~~~~~~~~~~~~~~~~~~~~~~(A.24)\nonumber
\end{eqnarray}
Although the two small parameters(A.21) and (A.22) are not
independent, their effects can be separated by introducing
$p_\infty$ and $q_\infty$ that satisfy
\begin{eqnarray}\label{A.25}
~~~~~~~~~~~~~~~~~~~~~~~~~~~~~~~~~~-p_\infty\beta \cot
p_\infty\beta=q_\infty\beta
~~~~~~~~~~~~~~~~~~~~~~~~~~~~~~~~~~~~~~~~~~~(A.25)\nonumber
\end{eqnarray}
and
\begin{eqnarray}\label{A.26}
~~~~~~~~~~~~~~~~~~~~~~~~~~~~~~~~~~~~W^2=p_\infty^2+q_\infty^2.
~~~~~~~~~~~~~~~~~~~~~~~~~~~~~~~~~~~~~~~~~~~~~~~~~~~(A.26)\nonumber
\end{eqnarray}
Physically, $p_\infty$ and $q_\infty$ are the limiting values of
$p_b$ and $q_b$ when the distance $2\alpha$ between the two wells
$\rightarrow \infty$, but keeping the shapes of the two wells
unchanged. Hence (A.23) becomes (A.25). Let
\begin{eqnarray}\label{A.27}
~~~~~~~~~~~~~~~~~~~~~~~~~~~~~~~~~~~~~p_\infty\beta=\pi-\theta_\infty~.~~~~~~~~~~
~~~~~~~~~~~~~~~~~~~~~~~~~~~~~~~ ~~~~~~~~~(A.27)\nonumber
\end{eqnarray}
From (A.25), we may expand $\theta_\infty$ in terms of successive
powers of $(W\beta)^{-1}$:
\begin{eqnarray}\label{A.28}
~~~~~~~~~~~~~~\theta_\infty=\frac{\pi}{W\beta}
\left(1-\frac{1}{W\beta}+(1+\frac{\pi^2}{6})(
\frac{1}{W\beta})^2+O( \frac{1}{W\beta})^3\right)~~~~~~~~~
~~~~~~~~~~~~~~~~(A.28)\nonumber
\end{eqnarray}
which determines both $p_\infty$ and $q_\infty$. By substituting
\begin{eqnarray}\label{A.29}
~~~~~~~~~~~~~~~~~~~~~~~~~~~~~\theta_b=\theta_\infty+\nu_1~e^{-2q_\infty
\alpha}+ O(e^{-4q_\infty \alpha})~~~~~~~~~
~~~~~~~~~~~~~~~~~~~~~~~~~~~(A.29)\nonumber
\end{eqnarray}
into (A.23) and using (A.24)-(A.28), we determine
\begin{eqnarray}\label{A.30}
~~~~~~~~~~~~~~~~~~~~~~~~~~~~~~~~~~~~\nu_1= \left( \frac{p_\infty
q_\infty}{W^2}\right)
\frac{2q_\infty\beta}{q_\infty\beta+1}.~~~~~~~~~
~~~~~~~~~~~~~~~~~~~~~~~~~~~~~~~~(A.30)\nonumber
\end{eqnarray}
Likewise, the continuity of $\chi'_a/\chi_a$ at $x=\pm\alpha$
gives
\begin{eqnarray}\label{A.31}
~~~~~~~~~~~~~~~~~~~~~~~~~~~~~-k_a\beta \cot k_a\beta=q_a\beta
\tanh k_a\alpha,
~~~~~~~~~~~~~~~~~~~~~~~~~~~~~~~~~~~~~~~(A.31)\nonumber
\end{eqnarray}
with
\begin{eqnarray}\label{A.32}
~~~~~~~~~~~~~~~~~~~~~~~~~~~~~~~~~~~~\hat{W}^2=W^2-\mu^2=k_a^2+q_a^2.
~~~~~~~~~~~~~~~~~~~~~~~~~~~~~~~~~~~~~~(A.32)\nonumber
\end{eqnarray}
As in (A.25), we introduce $k_\infty$ and $\hat{q}_\infty$ that
satisfy
\begin{eqnarray}\label{A.33}
~~~~~~~~~~~~~~~~~~~~~~~~~~~~~~-k_\infty\beta \cot
k_\infty\beta=\hat{q}_\infty\beta
~~~~~~~~~~~~~~~~~~~~~~~~~~~~~~~~~~~~~~~~~~~~~~~~(A.33)\nonumber
\end{eqnarray}
and
\begin{eqnarray}\label{A.34}
~~~~~~~~~~~~~~~~~~~~~~~~~~~~~~~~~~~~\hat{W}^2=k_\infty^2+\hat{q}_\infty^2.
~~~~~~~~~~~~~~~~~~~~~~~~~~~~~~~~~~~~~~~~~~~~~~~~~~~(A.34)\nonumber
\end{eqnarray}
Similar to (A.27)-(A.28), we define
\begin{eqnarray}\label{A.35}
~~~~~~~~~~~~~~~~~~~~~~~~~~~~~~~~~~~k_\infty\beta=\pi-\hat{\theta}_\infty~.~~~~~~~~~~
~~~~~~~~~~~~~~~~~~~~~~~~~~~~~~~~~~~~~~~~~~(A.35)\nonumber
\end{eqnarray}
and derive
\begin{eqnarray}\label{A.36}
~~~~~~~~~~~~~~\hat{\theta}_\infty=\frac{\pi}{\hat{W}\beta}
\left(1-\frac{1}{\hat{W}\beta}+(1+\frac{\pi^2}{6})(
\frac{1}{\hat{W}\beta})^2 +O(
\frac{1}{\hat{W}\beta})^3\right)~.~~~~~~~~
~~~~~~~~~~~~~~~(A.36)\nonumber
\end{eqnarray}
As in (A.29)-(A.30), we find $\hat{\theta}_a\equiv\pi-k_a\beta$ to
be given by
\begin{eqnarray}\label{A.37}
~~~~~~~~~~~~~~~~~~~~~~~~~~~~~\hat{\theta}_a=\hat{\theta}_\infty
+\hat{\nu}_1~e^{-2\hat{q}_\infty \alpha}+ O(e^{-4\hat{q}_\infty
\alpha})~~~~~~~~~ ~~~~~~~~~~~~~~~~~~~~~~~~~~~(A.37)\nonumber
\end{eqnarray}
with
\begin{eqnarray}\label{A.38}
~~~~~~~~~~~~~~~~~~~~~~~~~~~~~~~~~~~~\hat{\nu}_1= \left(
\frac{k_\infty \hat{q}_\infty}{\hat{W}^2}\right)
\frac{2\hat{q}_\infty\beta}{\hat{q}_\infty\beta+1}.~~~~~~~~~
~~~~~~~~~~~~~~~~~~~~~~~~~~~~~~~~(A.38)\nonumber
\end{eqnarray}
To derive similar expressions for $\theta$ and $\hat{\theta}$ of
(A.19), we first note that the transformation
\begin{eqnarray}\label{A.39}
~~~~~~~~~~~~~~~~~~~~~~~~~~~~~~~~~~~~~~~~~~~~~~~~~~~\alpha
\rightarrow \alpha+\delta ~~~~~~~~~~~~~~~~~~~~~~~~~~~~~~~~~~~~~
~~~~~~~(A.39)\nonumber
\end{eqnarray}
brings (A.23) to (A.17), provided that we also change
\begin{eqnarray}
~~q_b \rightarrow q,~~~~~~~~~p_b \rightarrow p ~~~~~\nonumber
\end{eqnarray}
and therefore
\begin{eqnarray}\label{A.40}
~~~~~~~~~~~~~~~~~~~~~~~~~~~~~~~~~~~~\theta_b \rightarrow \theta.
~~~~~~~~~~~~~~~~~~~~~~~~~~~~~~~~~~~~~~
~~~~~~~~~~~~~~~~~~~~~~~~~~(A.40)\nonumber
\end{eqnarray}
Since according to (A.1), the asymmetry of $V(x)$ is due to the
term $\frac{1}{2} \mu^2>0$ in the positive $x$ region, it is easy
to see that
\begin{eqnarray}\label{A.41}
~~~~~~~~~~~~~~~~~~~~~~~~~~~~~~~~~~~~\delta>0,
~~~~~~~~~~~~~~~~~~~~~~~~~~~~~~~~~~~~~~
~~~~~~~~~~~~~~~~~~~~~~~~~~~(A.41)\nonumber
\end{eqnarray}
as will also be shown explicitly below. Thus, from (A.29) and
through the transformations (A.39)-(A.40), we derive
\begin{eqnarray}\label{A.42}
~~~~~~~~~~~~~~~~~~~~~~~~~~~~~\theta=\theta_\infty+\theta_1
~~~~~~~~~~~~~~~~~~~~~~~~~~~~~~~~~~~~~~
~~~~~~~~~~~~~~~~~~~~~~~~~~(A.42)\nonumber
\end{eqnarray}
where
\begin{eqnarray}\label{A.43}
~~~~~~~~~~~~~~~~~~~~~\theta_1=\nu_1~e^{-2q_{\infty}(\alpha+\delta)}
+O(e^{-4q_{\infty}(\alpha+\delta)})~~~~~~~~~~~~~~~~~
~~~~~~~~~~~~~~~~~~~~~~~~~~(A.43)\nonumber
\end{eqnarray}
with $\nu_1$ given by (A.30). Likewise, we note that the
transformation
\begin{eqnarray}\label{A.44}
~~~~~~~~~~~~~~~~~~~~~~~~~~~~~~~~~~~~~~~~~~~~~~~~~~~\alpha
\rightarrow \alpha-\delta ~~~~~~~~~~~~~~~~~~~~~~~~~~~~~~~~~~~~~
~~~~~~~(A.44)\nonumber
\end{eqnarray}
brings (A.31) to (A.16), provided that we also change
\begin{eqnarray}
~~k_a \rightarrow k,~~~~~~~~~q_a \rightarrow q ~~~~~\nonumber
\end{eqnarray}
and therefore
\begin{eqnarray}\label{A.45}
~~~~~~~~~~~~~~~~~~~~~~~~~~~~~~~~~~~~\hat{\theta}_a \rightarrow
\hat{\theta}. ~~~~~~~~~~~~~~~~~~~~~~~~~~~~~~~~~~~~~~
~~~~~~~~~~~~~~~~~~~~~~~~~(A.45)\nonumber
\end{eqnarray}
Here, we must differentiate three different situations:
\begin{eqnarray}\label{A.46}
~~~~~~~~~~~~~~~~~~~~~~~~~~~~&{\sf (i)}&~~\alpha >\delta,
~~~~~~~~~~~~~~~~~~~~~~~~~~~~~~~~~~~~~ ~~~~~~~\nonumber\\
~~~~~~~~~~~~~~~~~~~~~~~~~~~~&{\sf (ii)}&~~\alpha =\delta
~~~~~~~~~~~~~~~~~~~~~~~~~~~~~~~~~~~~~~~~~~~~~~~~~~~~~~~~~~~~
~~(A.46)\nonumber
\end{eqnarray}
and
\begin{eqnarray}
~~~~~~~~{\sf (iii)}~~\alpha <\delta.
~~~~~~~~~~~~~~~~~~~~~~~~~~~~~~~~~~~~~ ~~~~~~~\nonumber
\end{eqnarray}

In case (i), when
\begin{eqnarray}\label{A.47}
~~~~~~~~~~~~~~~~~~~~~~~~~~~~~~~~e^{-2q_{\infty}(\alpha-\delta)}<<1,
~~~~~~~~~~~~~~~~~~~~~~~~~~~~~~~~~~~~~
~~~~~~~~~~~~~~~~~~(A.47)\nonumber
\end{eqnarray}
from (A.37) and through the transformations given by
(A.44)-(A.45), we find
\begin{eqnarray}\label{A.48}
~~~~~~~~~~~~~~~~~~~~~~~~~~~~~\hat{\theta}=\hat{\theta}_\infty+\hat{\theta}_1
~~~~~~~~~~~~~~~~~~~~~~~~~~~~~~~~~~~~~~
~~~~~~~~~~~~~~~~~~~~~~~~~~~(A.48)\nonumber
\end{eqnarray}
where
\begin{eqnarray}\label{A.49}
~~~~~~~~~~~~~~~~~~~~~\hat{\theta}_1=\hat{\nu}_1~e^{-2\hat{q}_{\infty}(\alpha-\delta)}
+O(e^{-4\hat{q}_{\infty}(\alpha-\delta)})~~~~~~~~~~~~~~~~~
~~~~~~~~~~~~~~~~~~~~~~~~~~~(A.49)\nonumber
\end{eqnarray}
with $\hat{\nu}_1$ given by (A.38). According to (A.13) and
(A.19), we have
\begin{eqnarray}\label{A.50}
~~~~~~~~~~~~~~~~~~~~~~~~\mu^2\beta^2&=&p^2\beta^2-k^2\beta^2=(\pi-\theta)^2-(\pi-\hat{\theta})^2\nonumber\\
                                    &=&(\pi-\theta_\infty-\theta_1)^2-(\pi-\hat{\theta}_\infty-\hat{\theta}_1)^2,
~~~~~~~~~~~~~~~~~~~~~~~~~~~~~(A.50)\nonumber
\end{eqnarray}
which leads to
\begin{eqnarray}\label{A.51}
~~~~~~~\mu^2\beta^2+(2\pi-\theta_\infty-\hat{\theta}_\infty)(\theta_\infty-\hat{\theta}_\infty)
=-2(\pi-\theta_\infty)\theta_1+2(\pi-\hat{\theta}_\infty)\hat{\theta}_1+\theta_1^2-\hat{\theta}_1^2.
~~~~(A.51)\nonumber
\end{eqnarray}
Since in accordance from (A.28) and (A.36), we find
\begin{eqnarray}\label{A.52}
~~~~~~~\hat{\theta}_\infty-\theta_\infty = \frac{\pi}{\beta}\left(
\frac{1}{\hat{W}}-\frac{1}{W}\right)-\frac{\pi}{\beta^2}\left(
\frac{1}{\hat{W}^2}-\frac{1}{W^2}\right)+ \cdots=O\left( \frac{
\mu^2\beta^2}{W^3\beta^3}\right)<<\mu^2\beta^2
~~~~~(A.52)\nonumber
\end{eqnarray}
and
\begin{eqnarray}\label{A.53}
~~~~~~~~~~~~~~~~~~~~~~~~~~~~\theta_\infty+\hat{\theta}_\infty =
\frac{\pi}{\beta}\left( \frac{1}{W}+\frac{1}{\hat{W}}\right)+
\cdots<<2\pi.~~~~~~~~~~~~~~~~~~~~~~~~~~~ ~~~~(A.53)\nonumber
\end{eqnarray}
Thus, the left side of (A.51) is dominated by its first term,
$\mu^2\beta^2$. Since $\theta_1$ and $\hat{\theta}_1$ are
exponentially small, we can neglect $\theta_1^2-\hat{\theta}_1^2$
in (A.51). In addition, because $\theta_\infty$ and
$\hat{\theta}_\infty$ are much smaller than $2\pi$, (A.51) can be
reduced to
\begin{eqnarray}\label{A.54}
~~~~~~~~~~~~\mu^2\beta^2\cong 2\pi(\hat{\theta}_1-\theta_1)\cong
2\pi\left(\hat{\nu}_1~e^{-2\hat{q}_{\infty}(\alpha-\delta)}
-\nu_1~e^{-2q_{\infty}(\alpha+\delta)}\right)
~~~~~~~~~~~~~~~~~~~~~~~~~(A.54)\nonumber
\end{eqnarray}
which gives the dependence of $\delta$ on $\mu^2$. It is important
to note that an exponentially small $\mu^2$ can produce a finite
$\delta$. For $\delta<\alpha$, at $x=\delta$ we have, in
accordance with (A.10)
\begin{eqnarray}\label{A.55}
~~~~~~~~~~~~~~~~~~~~~~~~~~~~~~~~~~~~~~~~~~~~~~\psi'(\delta)=0,
~~~~~~~~~~~~~~~~~~~~~~~~~~~~~~~~~~~~~~~~~~~~~~~
~~~~(A.55)\nonumber
\end{eqnarray}
which gives the minimum of $\psi(x)$. The wave function $\psi(x)$
has two maxima, one for each potential well.

In case (ii), $\alpha=\delta$ and (A.16) gives $\cot k\beta=0$,
and $\hat{\theta}$ takes on the  critical value $\hat{\theta}_c$
with
\begin{eqnarray}\label{A.56}
~~~~~~~~~~~~~~~~~~~~~~~~~~~~~~~~~~~~~~~~~~~~~~~\hat{\theta}_c=\frac{\pi}{2}.
~~~~~~~~~~~~~~~~~~~~~~~~~~~~~~~~~~~~~~~~~~~~~~~~~~
~~~~(A.56)\nonumber
\end{eqnarray}

In case (iii), $\hat{\theta}>\frac{\pi}{2}$,
$k\beta<\frac{\pi}{2}$ and $\psi(x)$ has only one maximum.

As in (4.73) and (4.75) we introduce $\chi(x)$ through
\begin{eqnarray}\label{A.57}
~~~~~~~~~~~~~~~~~~~~~~~~\chi(x) \equiv \left\{\begin{array}{lcc}
\chi_a(x),~~~~~~~~~~~~~~~~~~{\sf for}~~0\leq x\leq \gamma\\
\chi_b(x),~~~~~~~~~~~~~~~~~~{\sf for}~~-\gamma \leq x\leq 0
\end{array}
\right.~~~~~~~~~~~~~~~~~~~~~~~(A.57)\nonumber
\end{eqnarray}
so that
\begin{eqnarray}\label{A.58}
~~~~~~~~~~~~~~~~~~~~~~~~~~~(T+V(x)+\hat{w}(x))\chi(x)=\hat{E}_0\chi(x),.
~~~~~~~~~~~~~~~~~~~~~~~~~~~~~~~~~~~~~~(A.58)\nonumber
\end{eqnarray}
in which, same as (4.76a)-(4.77a),
\begin{eqnarray}\label{A.59}
~~~~~~~~~~~~~~~~~\hat{w}(x)= \left\{\begin{array}{ccc}
0~~~~~~~~~~~~~~~~~~~~~~~~~~~~~~~~{\sf for}~~0<x \leq \gamma\\
E_a-E_b~~~~~~~~~~~~~~~~~~~~~~~~{\sf for}~~-\gamma \leq x<0
\end{array}
\right.~~~~~~~~~~~~~~~~~~~(A.59)\nonumber
\end{eqnarray}
and
\begin{eqnarray}\label{A.60}
~~~~~~~~~~~~~~~~~~~~~~~~~~~~~~~~~~~~~~\hat{E}_0=E_a
~~~~~~~~~~~~~~~~~~~~~~~~~~~~~~~~~~~~~~~~~~~~~~~~~~~~~~~
~~~~(A.60)\nonumber
\end{eqnarray}
with $E_a$ and $E_b$ given by (A.14) and (A.15). Since according
to (A.4)-(A.5), $V_a(x) \geq V_b(x)$, we have
\begin{eqnarray}\label{A.61}
~~~~~~~~~~~~~~~~~~~~~~~~~~~~~~~~~~~~~~E_a>E_b;
~~~~~~~~~~~~~~~~~~~~~~~~~~~~~~~~~~~~~~~~~~~~~~~~~~~~~~
~~~~(A.61)\nonumber
\end{eqnarray}
therefore,
\begin{eqnarray}\label{A.62}
~~~~~~~~~~~~~~~~~~~~~~~~~~~~~~~~~~~~~~\hat{w}'(x) \leq 0.
~~~~~~~~~~~~~~~~~~~~~~~~~~~~~~~~~~~~~~~~~~~~~~~~~~~~~
~~~~(A.62)\nonumber
\end{eqnarray}
Write the Schroedinger equation (A.7) in the form (4.80):
\begin{eqnarray}\label{A.63}
~~~~~~~~~~~~~~~~~~~~~~~~(T+V(x)+\hat{w}(x)-\hat{E}_0)\psi(x)=(\hat{w}(x)-\hat{{\cal
E}})\psi(x)~~~~~~~~~~~~~~~~~~~~~~(A.63)\nonumber
\end{eqnarray}
with
\begin{eqnarray}\label{A.64}
~~~~~~~~~~~~~~~~~~~~~~~~~~~~~~~E=\hat{E}_0-\hat{{\cal
E}}=E_a-\hat{{\cal
E}}.~~~~~~~~~~~~~~~~~~~~~~~~~~~~~~~~~~~~~~~~~~~~~~~~(A.64)\nonumber
\end{eqnarray}
As in (4.82), we have
\begin{eqnarray}\label{A.65}
~~~~~~~~~~~~~~~~~~~~~~~~~~~~\int_{-\gamma}^{\gamma}\chi(x)\psi(x)(\hat{w}(x)-\hat{{\cal
E}})dx=0.~~~~~~~~~~~~~~~~~~~~~~~~~~~~~~~~~~~~~~~(A.65)\nonumber
\end{eqnarray}
In all subsequent equations, we restrict the $x$-axis to
\begin{eqnarray}\label{A.66}
~~~~~~~~~~~~~~~~~~~~~~~~~~~~~~~~~~~~~~|x|\leq \gamma,
~~~~~~~~~~~~~~~~~~~~~~~~~~~~~~~~~~~~~~~~~~~~~~~~~~~~~~~~
~~~~(A.66)\nonumber
\end{eqnarray}
and set $\psi(x)$, $\chi(x)$ positive. Define
\begin{eqnarray}\label{A.67}
~~~~~~~~~~~~~~~~~~~~~~~~~~~~~~~~~f(x)=\psi(x)/\chi(x).
~~~~~~~~~~~~~~~~~~~~~~~~~~~~~~~~~~~~~~~~~~~~~~~~
~~~~(A.67)\nonumber
\end{eqnarray}
We have, as in (4.84)-(4.85),
\begin{eqnarray}\label{A.68}
~~~~~~~~~~~~f(x)=f(\gamma)-2\int_{x}^{\gamma}\chi^{-2}(y)dy
\int_{y}^{\gamma}\chi^2(z)(\hat{w}(z)-\hat{{\cal
E}})f(z)dz,~~~~~~~~~~~~~~~~~~~~~~~~(A.68)\nonumber
\end{eqnarray}
or, on account of (A.65), the equivalent form
\begin{eqnarray}\label{A.69}
~~~~~~~~~~~~f(x)=f(-\gamma)-2\int_{-\gamma}^x\chi^{-2}(y)dy
\int_{-\gamma}^y\chi^2(z)(\hat{w}(z)-\hat{{\cal
E}})f(z)dz.~~~~~~~~~~~~~~~~~~~~~(A.69)\nonumber
\end{eqnarray}

The derivation of $f(x)$ is given by
\begin{eqnarray}\label{A.70}
~~~~~~~~~~~~~~~~~~~~~~~~~~~~~~~~~~~f'(x)=-2\chi^{-2}(x)h(x)~~~~~~~~~~~
~~~~~~~~~~~~~~~~~~~~~~~~~~~~~~~~~(A.70)\nonumber
\end{eqnarray}
where
\begin{eqnarray}\label{A.71}
~~~~~~~~~~~~~~~~~~~~~~~h(x)=\int_{-\gamma}^x\chi^2(z)(\hat{w}(z)-\hat{{\cal
E}})f(z)dz~~~~~~~~~~~~~~~~~~~~~~~~~~~~~~~~~~~~~~~~~(A.71)\nonumber
\end{eqnarray}
or equivalently,
\begin{eqnarray}\label{A.72}
~~~~~~~~~~~~~~~~~~~~~~~h(x)=-\int_{x}^{\gamma}\chi^2(z)(\hat{w}(z)-\hat{{\cal
E}})f(z)dz.~~~~~~~~~~~~~~~~~~~~~~~~~~~~~~~~~~~~~~(A.72)\nonumber
\end{eqnarray}
In order to satisfy (A.65) and by using (A.59), we see that
\begin{eqnarray}\label{A.73}
~~~~~~~~~~~~~~~~~~~~~~~~~~~~~~~~~~~~~~(E_a-E_b)>\hat{{\cal E}}>0.
~~~~~~~~~~~ ~~~~~~~~~~~~~~~~~~~~~~~~~~~~~~~~~~(A.73)\nonumber
\end{eqnarray}
Thus, (A.71)-(A.73) give
\begin{eqnarray}\label{A.74}
~~~~~~~~~~~~~~~~~~~~~~~~~~~~h(-\gamma)&=&h(\gamma)=0,\nonumber\\
~~~~~~~~~~~~~~~~~~~~~~~~~~~~~~~~h(x)&>&0~~~~~~~~~~~{\sf
for}~~~|x|<\gamma,~~~~~~~~~~
~~~~~~~~~~~~~~~~~~~~~~~~(A.74)\nonumber\\
~~~~~~~~~~~~~~~~~~~~~~~~~~~~~~~~h'(x)&>&0~~~~~~~~~~~{\sf
for}~~~-\gamma<x<0~~~~~~~~~~
~~~~~~~~~~~~~~~~~~~~~~~~\nonumber\\
{\sf and}~~~~~~~~~~~~~~~~~~~~~~~~~~~~~~~~~&&\nonumber\\
~~~~~~~~~~~~~~~~~~~~~~~~~~~~~h'(x)&<&0~~~~~~~~~~{\sf
for}~~~0<x<\gamma.~~~~~~~~~~ ~~~~~~~~~~~~~~~~~~~~~~~~\nonumber
\end{eqnarray}
The positivity of $h(x)$ gives
\begin{eqnarray}\label{A.75}
~~~~~~~~~~~~~~~~~~~~~~~~~~~~~~~f'(x)&<&0~~~~~~~~~~{\sf
for}~~~|x|<\gamma.~~~~~~~~~~
~~~~~~~~~~~~~~~~~~~~~~~~~~(A.75)\nonumber
\end{eqnarray}

\noindent
{\bf A.2 A Two-level Model}\\

Before discussing the iterative solutions for $f(x)$ and ${\cal
E}$, it may be useful to first extract some essential features of
the soluble square-well example. Let us first concentrate on Case
(i) of (A.46), with the parameters $\alpha$ and $\delta$
satisfying
$$
e^{-2q_{\infty}\alpha}<<e^{-2q_{\infty}(\alpha-\delta)}<<1.
\eqno(A.76)
$$
We shall also neglect $(W\alpha)^{-1}$ or $(W\beta)^{-1}$, when
compared to $1$. Thus, from (A.27)-(A.28), we have
$$
p_{\infty} \cong \frac{\pi}{\beta}~; \eqno(A.77)
$$
in addition, from(A.30) and (A.38) we find
$$
\hat{\nu}_1 \cong \nu_1 \cong \frac{2p_{\infty}}{W}\cong
\frac{2\pi}{W\beta}~. \eqno(A.78)
$$

From (A.54), we have
$$
\frac{1}{2}\mu^2 \cong \frac{\pi}{\beta^2}\hat{\nu}_1
e^{-2q_{\infty}(\alpha-\delta)} \cong \frac{2\pi^2}{W\beta^3}
e^{-2q_{\infty}(\alpha-\delta)}. \eqno(A.79)
$$
On account of (A.15), (A.20), (A.27) and (A.29),
$$
E_b=\frac{1}{2}p_b^2 \cong
\frac{1}{2}(p_{\infty}-\frac{\nu_1}{\beta}e^{-2q_{\infty}\alpha})^2
\eqno(A.80)
$$
which, for
$$
E_\infty=\frac{1}{2}p_{\infty}^2, \eqno(A.81)
$$
gives
$$
E_b \cong E_\infty-\frac{2
p_{\infty}^2}{W\beta}e^{-2q_{\infty}\alpha}\cong E_\infty-\frac{2
\pi^2}{W\beta^3}e^{-2q_{\infty}\alpha}. \eqno(A.82)
$$
On the other hand, from (A.13), (A.19) and (A.42)-(A.43), we see
that
$$
E=\frac{1}{2}p^2 \cong \frac{1}{2}(p_{\infty} -
\frac{\nu_1}{\beta}e^{-2q_{\infty}(\alpha+\delta)})^2 \cong
E_{\infty}-\frac{2
\pi^2}{W\beta^3}e^{-2q_{\infty}(\alpha+\delta)}. \eqno(A.83)
$$
Thus, under the condition (A.76), we find
$$
\frac{1}{2}\mu^2>>E_\infty-E_b>>E_\infty-E. \eqno(A.84)
$$
As we shall see, these inequalities can be understood in terms of
a simple two-level model.

Introduce
$$
\lambda=E_\infty-E_b. \eqno(A.85)
$$
We note that from (A.82),
$$
\lambda \cong \frac{2 \pi^2}{W\beta^3}e^{-2q_{\infty}\alpha},
\eqno(A.86)
$$
and from (A.79) and (A.83),
$$
\lambda \cong \frac{1}{2}\mu^2 e^{-2q_{\infty}\delta}~~~~~{\sf
and}~~~~~~E_\infty-E \cong \lambda e^{-2q_{\infty}\delta}.
\eqno(A.87)
$$
Consequently, the three small energy parameters in (A.84) are
related by
$$
\frac{1}{2}\mu^2(E_\infty-E) \cong \lambda^2, \eqno(A.88)
$$
From $e^{-2q_{\infty}\delta}<<1$ and (A.76) , we see that
$$
\lambda<<\frac{1}{2}\mu^2<<\frac{2\pi^2}{W\beta^3} \eqno(A.89)
$$
in accordance with (A.79) and (A.84). To understand the role of
the parameter $\lambda$, we may start with the definition of
$V_b(x)$, given by (A.5), keep the parameters
$\beta=\gamma-\alpha$ and $\frac{1}{2}W^2$ fixed, but let the
spacing $2\alpha$ between the two potential wells approach
$\infty$; in the limit $2\alpha \rightarrow \infty$, we have $E_b
\rightarrow E_\infty$. Thus, $\lambda=E_\infty-E_b$ is the energy
shift due to the tunneling between the two potential wells located
at $x<-\alpha$ and $x>\alpha$ in $V_b(x)$.

There is an alternative definition  for $\lambda$, which may
further clarify its physical significance. According to (A.3),
$V_b(x)$ is even in $x$; therefore, its eigenstates are either
even or odd in $x$. In (A.9), $\chi_b(x)$ is the groundstate of
$T+V_b(x)$, and therefore it has to be even in $x$. The
corresponding first excited state $\chi_{od}$ is odd in $x$; it
satisfies
$$
(T+V_b(x))\chi_{od}(x)=E_{od}\chi_{od}(x). \eqno(A.90)
$$
We may define $\lambda$ by
$$
2\lambda \equiv E_{od}-E_b \eqno(A.91)
$$
and regard (A.85) and (A.86) both as approximate expressions, as
we shall see.

Multiplying (A.9) by $\chi_{od}(x)$ and (A.90) by $\chi_b(x)$,
then taking their difference we derive $$
-\frac{1}{2}\left(\chi'_{od}(x)\chi_b(x)-\chi'_b(x)\chi_{od}(x)\right)'
=(E_{od}-E_b)\chi_{od}(x)\chi_b(x). \eqno(A.92) $$ From (A.12), we
may choose the normalization of $\chi_b$ so that
\begin{eqnarray}\label{A.93}
\chi_b(x)=\chi_b(-x)= \left\{
\begin{array}{l}
\left( \frac{ \cosh q_b\alpha}{\sin p_b\beta}\right) \sin
p_b(-x+\gamma) ~~~~~~~~~~~~~~{\sf for}~~~~~~\alpha\leq x \leq
\gamma\\ ~~\\ \cosh q_b x~~~~~~~~~~~~~~~~~~~~~~~~~~~~~~~~~~{\sf
for}~~~~~~~0\leq x \leq \alpha.
\end{array}
\right.~~~~~~~~~~(A.93)\nonumber
\end{eqnarray}
Correspondingly,
\begin{eqnarray}\label{A.94}
\chi_{od}(x)=-\chi_{od}(-x)=\left\{
\begin{array}{l}
\left( \frac { \sinh q_{od}\alpha}{\sin p_{od}\beta}\right) \sin
p_{od}(-x+\gamma)~~~~~~~~~~~{\sf for}~~~~~~\alpha\leq x \leq \gamma\\
~~\\
\sinh q_{od} x~~~~~~~~~~~~~~~~~~~~~~~~~~~~~~~~{\sf
for}~~~~~~~0\leq x \leq \alpha
\end{array}
\right.~~~~~~(A.94)\nonumber
\end{eqnarray}
with

$$ E_b=\frac{1}{2}p_b^2~~~~~~~~{\sf
and}~~~~~~~~~~~~E_{od}=\frac{1}{2}p_{od}^2. \eqno(A.95) $$ As in
(A.25) and (A.26), $q_{od}$ and $p_{od}$ are determined by
$$
-p_{od}\beta \cot p_{od}\beta =q_{od}\beta \coth q_{od}\alpha
\eqno(A.96)
$$
and
$$
W^2=p_{od}^2+q_{od}^2. \eqno(A.97)
$$
\newpage

At $x=0$, we have
$$
\begin{array}{c}
\chi_{od}(0)=0,~~~~~~~~~~~~~~~~\chi_b(0)=1\\
\chi'_{od}(0)=q_{od}~~~~~{\sf and}~~~~~~\chi'_b(0)=0.
\end{array}
\eqno(A.98)
$$
Integrating (A.92) from $x=0$ to $x=\gamma$, we find
$$
\frac{1}{2}q_{od}=(E_{od}-E_b)\int_0^\gamma
\chi_{od}(x)\chi_b(x)dx. \eqno(A.99)
$$
From (A.27)-(A.29), we see that
$$
p_b\beta \equiv\pi-\theta_b \cong \pi~~~~~~~{\sf
and}~~~~~~~\theta_b \cong \theta_\infty \cong \frac{\pi}{W\beta}.
\eqno(A.100)
$$
Likewise, we can also show that
$$
p_{od}\beta \equiv\pi-\theta_{od} \cong \pi~~~~~~~{\sf
and}~~~~~~~\theta_{od} \cong \theta_\infty \cong
\frac{\pi}{W\beta}. \eqno(A.101)
$$
Thus, $q_{od} \cong q_b \cong W$, and the integral in (A.99) is
\begin{eqnarray}\label{A.102}
~~~~~~~~~~~~~~~~~&&\int_0^\gamma \chi_{od}(x)\chi_b(x)dx \cong
\int_\alpha^\gamma
\chi_{od}(x)\chi_{b}(x)dx\nonumber\\
&&~~~~=\frac{\sinh q_{od} \alpha \cosh q_b \alpha}{\sin p_{od}
\beta \sin p_b \beta}
\int_\alpha^\gamma \sin^2 p_\infty(-x+\gamma)dx\nonumber\\
&&~~~~\cong
\frac{e^{(q_{od}+q_b)\alpha}}{4\theta_{od}\theta_b}\cdot\frac{1}{2}\beta
\cong\frac{W^2\beta^3}{8\pi^2}e^{2q_{\infty}\alpha}.~~~~~~~~~~~~~~~~~~
~~~~~~~~~~~~~~~~~~~~(A.102)\nonumber
\end{eqnarray}
Since $q_{od}\cong W$, we derive from (A.91)
$$
\lambda \equiv\frac{1}{2}(E_{od}-E_b)
\cong\frac{2\pi^2}{W\beta^3}e^{-2q_{\infty}\alpha}, \eqno(A.103)
$$
in agreement with (A.86).

We are now ready to introduce the two-level model. We shall
approximate the Hamiltonian $T+V(x)$, $T+V_a(x)$ and $T+V_b(x)$ of
(A.7)-(A.9) by the following three $2\times 2$ matrices:
\begin{eqnarray}\label{A.104}
~~~~~~~~~~~~~~~~~~~~~~~~~~~~~~~h=E_\infty+\left(
\begin{array}{l}
\frac{1}{2}\mu^2~~~~~-\lambda\\
-\lambda~~~~~~~~0
\end{array}
\right),~~~~~~~~~~~~~~~~~~~~~~~~~~~~~~~~~~~~~~(A.104)\nonumber
\end{eqnarray}
\begin{eqnarray}\label{A.105}
~~~~~~~~~~~~~~~~~~~~~~~~~~~~~~~h_a=E_\infty+\frac{1}{2}\mu^2+\left(
\begin{array}{l}
~~0~~~~~~~-\lambda\\
-\lambda~~~~~~~~0
\end{array}
\right)~~~~~~~~~~~~~~~~~~~~~~~~~~~~~(A.105)\nonumber
\end{eqnarray}
and
\begin{eqnarray}\label{A.106}
~~~~~~~~~~~~~~~~~~~~~~~~~~~~~~~h_b=E_\infty+\left(
\begin{array}{l}
~~0~~~~~~~-\lambda\\
-\lambda~~~~~~~~0
\end{array}
\right),~~~~~~~~~~~~~~~~~~~~~~~~~~~~~~~~~~~~(A.106)\nonumber
\end{eqnarray}
with $\psi$, $\chi_a$ and $\chi_b$ as their respective
groundstates which satisfy
$$
h\psi=E\psi,~~~~~~~h_a\chi_a=E_a\chi_a
$$
$$
{\sf and}~~~~~~~~~~h_b\chi_b=E_b\chi_b. \eqno(A.107)
$$
The negative sign in the off-diagonal matrix element $-\lambda$ in
(A.104)-(A.106) is chosen to make
\begin{eqnarray}\label{A.108}
~~~~~~~~~~~~~~~~~~~~~~~~~~~~~~~~~~~~~~~~\chi_a=\chi_b=\frac{1}{\sqrt{2}}\left(
\begin{array}{l}
1\\
1
\end{array}
\right),~~~~~~~~~~~~~~~~~~~~~~~~~~~~~~~~~~~~~(A.108)\nonumber
\end{eqnarray}
simulating the evenness of $\chi_a(x)$ and $\chi_b(x)$. Likewise,
the analog of $\chi_{od}$ is the excited state of $h_b$, with
\begin{eqnarray}\label{A.109}
~~~~~~~~~~~~~~~~~~~~~~~~~~~~~~~~~~~~~~~~\chi_{od}=\frac{1}{\sqrt{2}}\left(
\begin{array}{l}
~1\\
-1
\end{array}
\right)~~~~~~~~~~~~~~~~~~~~~~~~~~~~~~~~~~~~~~~~~~~(A.109)\nonumber
\end{eqnarray}
and
$$
h_b\chi_{od}=E_{od}\chi_{od}. \eqno(A.110)
$$
It is straightforward to verify that
\begin{eqnarray}\label{A.111}
~~~~~~~~~~~~~~~~~~~~~~~~~~~~~~~~~~~~~~~~~~~~\psi=\left(
\begin{array}{l}
\sin \xi \\
\cos \xi
\end{array}
\right)~~~~~~~~~~~~~~~~~~~~~~~~~~~~~~~~~~~~~~~~~~~(A.111)\nonumber
\end{eqnarray}
where
$$
\sin 2\xi=\frac{4\lambda}{\sqrt{(4\lambda)^2+\mu^4}}~~~~{\sf
and}~~~~ \cos 2\xi=\frac{\mu^2}{\sqrt{(4\lambda)^2+\mu^4}},
\eqno(A.112)
$$
\begin{eqnarray}\label{A.113}
~~~~~~~~~~~~~~~~~~~~~~~~~E &=&
E_\infty+\frac{\mu^2}{4}-\sqrt{\lambda^2+\left(\frac{\mu^2}{4}\right)^2},\nonumber\\
E_a &=& E_\infty+\frac{\mu^2}{2}-\lambda~~~~~~~~~~~~~
~~~~~~~~~~~~~~~~~~~~~~~~~~~~~~~~~~~(A.113)\nonumber\\
E_b &=& E_\infty-\lambda\nonumber\\
{\sf and}~~~~~~~~~~~~~~~~~~~~~~~~~~~~~&&\nonumber\\
E_{od} &=& E_\infty+\lambda.\nonumber
\end{eqnarray}
When $\lambda<<\frac{\mu^2}{4}$~, we have
$$
E \cong E_\infty-\frac{2\lambda^2}{\mu^2}~, \eqno(A.114)
$$
in agreement with (A.88).

Next, we wish to examine the relation between the two-level model
and the soluble square-well example when $\lambda$ is
$O(\frac{\mu^2}{4})$. Assume, instead of (A.76),
\begin{eqnarray}\label{A.115}
(W\alpha)^{-1}&<<&1, ~~~~~~~(W\beta)^{-1}<<1,
 ~~~~~~~e^{-2q_\infty\alpha}<<1\nonumber\\
{\sf but}~~~~~~~~~~~~~~~~~~~~~~~~~~~~~&&
~~~~~~~~~~~~~~~~~~~~~~~~~~~~~~~~~~~~~~~~~~~~~~~~~~~~~~~~~~~~~~~(A.115)\nonumber\\
e^{-2q_\infty\delta} &\sim& O(1).\nonumber
\end{eqnarray}
Hence, in the square-well example, (A.83),
$$
E_{\infty}-E \cong \frac{2
\pi^2}{W\beta^3}e^{-2q_{\infty}(\alpha+\delta)},
$$
and (A.86),
$$
\lambda \cong \frac{2 \pi^2}{W\beta^3}e^{-2q_{\infty}\alpha},
$$
remain valid; on the other hand, (A.54) and (A.78) now lead to
$$
\frac{1}{2}\mu^2 \cong \lambda(e^{2q_{\infty}\delta}
-e^{-2q_{\infty}\delta}) = 2\lambda \sinh 2q_\infty \delta.
\eqno(A.116)
$$
Thus, the above expressions for $E_\infty-E$ and $\lambda$    give
\begin{eqnarray}
E_\infty-E &\cong& \lambda~e^{-2q_{\infty}\delta}
=\lambda(\cosh 2q_\infty\delta-\sinh 2q_\infty \delta)\nonumber\\
&=& \lambda(\sqrt{1+\sinh^2 2q_\infty\delta}-\sinh
2q_\infty\delta).\nonumber
\end{eqnarray}
Together with (A.116), this shows that the soluble square-well
example yields
$$
E_\infty-E = \sqrt{\lambda^2+\left(
\frac{\mu^2}{4}\right)^2}-\frac{\mu^2}{4}
$$
in agreement with (A.113) given by the two-level model.

In both the square-well problem and the simple two-level model, we
can also examine the limit, when $\lambda>>\frac{\mu^2}{4}$. In
that case, (A.113) gives
$$
E=E_\infty
-\lambda+\frac{\mu^2}{4}-\frac{1}{32}\frac{\mu^4}{\lambda}+\cdots,
$$
which leads to
$$
E=E_\infty-\lambda=E_b~~~~~~{\sf when}~~~~\mu^2=0,
$$
in agreement with the exact square-well solution. Furthermore, if
we include the first order correction in $O(\mu^2)$, (A.115) gives
$$
E\cong E_\infty+\frac{\mu^2}{4}-\lambda=
E_b+\frac{1}{2}(E_a-E_b)+O(\frac{\mu^4}{\lambda}). \eqno(A.117)
$$
As we shall discuss, for the exact square-well solution, (A.117)
is also valid. Thus, the simple two-level formula (A.113) may
serve as an approximate formula for the exact square-well
solution over the entire range of $\frac{1}{4}\mu^2/\lambda$.\\

\noindent {\bf A.3 Square-well Example (Cont.)}

We return to the soluble square-well example discussed in Section
A.1. As before, $\psi(x)$ is the groundstate of $T+V(x)$ with
energy $E$, which is determined by the Schroedinger equation
(A.7). Likewise, $\chi(x)$ is the trial function given by (A.57);
i.e., the groundstate of $T+V(x)+\hat{w}(x)$ with eigenvalue
$\hat{E}_0=E_a$, in accordance with (A.58)-(A.60). From (A.59) and
(A.65), we see that the energy difference
$$
\hat{{\cal E}}=E_a-E \eqno(A.118)
$$
satisfies
$$
\hat{{\cal E}}=\frac{N}{M+N}(E_a-E_b), \eqno(A.119)
$$
where
$$
M=\int_0^{\gamma} \chi(x)\psi(x) dx \eqno(A.120)
$$
and
$$
N=\int_{-\gamma}^0 \chi(x)\psi(x) dx. \eqno(A.121)
$$
Before we discuss the iterative sequence $\{{\hat{\cal E}}_n\}$
that approaches $\hat{{\cal E}}$, as $n \rightarrow \infty$, it
may be instructive to verify (A.119) by evaluating the integrals
(A.120) and (A.121) directly. Choose the normalization convention
of $\psi$ and $\chi$ so that at $x=\gamma$
$$
\frac{\psi'(\gamma)}{\chi'(\gamma)}=1. \eqno(A.122)
$$
From (A.10)-(A.12) and (A.57) we write
\begin{eqnarray}\label{eA.123}
\psi(x)=\frac{k_a}{k}\left(\frac{\cosh q_a\alpha}{\sin
k_a\beta}\right) \left\{\begin{array}{lcl}
\sin k~(-x+\gamma)&&~~\alpha<x<\gamma\\
\left(\frac{\sin k\beta}{\cosh q(\alpha-\delta)}\right)
\cosh q(x-\delta)&~{\sf for}~~&-\alpha<x<\alpha\\
\left( \frac{ \sin k\beta \cosh q(\alpha+\delta)} {\sin p\beta
\cosh q(\alpha-\delta)}\right) \sin
p~(x+\gamma)&&-\gamma<x<-\alpha
\end{array}
\right.~~~~(A.123)\nonumber
\end{eqnarray}
\begin{eqnarray}\label{A.124}
\chi(x)= \left\{
\begin{array}{l}
\left( \frac{ \cosh q_a\alpha}{\sin k_a\beta}\right) \sin
k_a(-x+\gamma) ~~~~~~~~~~~~~~~~~~~~~~~~~~~\alpha< x <
\gamma\\ ~~\\
\cosh q_a x~~~~~~~~~~~~~~~~~~~~~~~~~~~~~~~~~~~~~~~~~~~~~~~~0< x < \alpha\\
\cosh q_b x~~~~~~~~~~~~~~~~~~~~~~~~~~~~{\sf
for}~~~~~~~~~~~~~~~-\alpha < x < 0\\
~~\\
\left( \frac{ \cosh q_b\alpha}{\sin p_b\beta}\right) \sin
p_b(x+\gamma) ~~~~~~~~~~~~~~~~~~~~~~~~~~~-\gamma < x < -\alpha.
\end{array}
\right.~~~~~~~~~~~~~~~(A.124)\nonumber
\end{eqnarray}
By directly evaluating the integral $~\int\chi(x)\psi(x)dx$, we
can readily verify that for $\gamma \geq x\geq 0$
$$
\frac{1}{2}(k^2-k_a^2)\int_x^{\gamma}\chi(y)\psi(y)dy
=\frac{1}{2}(\chi(x)\psi'(x)-\psi(x)\chi'(x)) \eqno(A.125)
$$
and for $-\gamma \leq x\leq 0$,
$$
\frac{1}{2}(p^2-p_b^2)\int_{-\gamma}^x\chi(y)\psi(y)dy
=-\frac{1}{2}(\chi(x)\psi'(x)-\psi(x)\chi'(x)). \eqno(A.126)
$$
Both relations can also be inferred from the Schroedinger
equations (A.7) and (A.58). Setting $x=0$ and taking the sum
(A.125)+(A.126), we derive
$$
\frac{1}{2}(k^2-k_a^2)M+\frac{1}{2}(p^2-p_b^2)N=0
$$
which, on account of (A.13)-(A.18), lead to the expression for the
energy shift $\hat{{\cal E}}$, in agreement with (A.119).

Next, we proceed to verify directly that $f(x)=\psi(x)/\chi(x)$
satisfies the integral equation (A.68). With the normalization
choice (A.122), we find at $x=\gamma$, since
$\psi(\gamma)=\chi(\gamma)=0$,
$$
f(\gamma)=\frac{\psi'(\gamma)}{\chi'(\gamma)}= 1~, \eqno(A.127)
$$
which gives the constant in the integral equation. The same
equation (A.68) can also be cast in an equivalent form:
$$
f(x)=1+\int_{-\gamma}^{\gamma}\chi^{-1}(x)(x|G|z)\chi(z)(\hat{w}(z)
-\hat{{\cal E}})f(z)dz \eqno(A.128)
$$
where $(x|G|z)$ is the Green's function that satisfies
\begin{eqnarray}\label{A.129}
&(T+V(x)+\hat{w}(x)-\hat{E}_0)(x|G|z)=\delta(x-z)\nonumber\\
{\sf and}~~~~&~~~~ ~~~~~~~~~~~~~~~~~~~~~~~~~~~~~~~~~
~~~~~~~~~~~~~~~~~~~~~~~~~~~~~~~~~~~~~~~~~~~~~~~~~~~~~~~~~~(A.129)\nonumber\\
&(x|G|z)=0~~~~~~~~{\sf for}~~~x>z.\nonumber
\end{eqnarray}
For $x<z$, $(x|G|z)$ is given by
$$
(x|G|z)=-2(\chi(x)\overline{\chi}(z)-\overline{\chi}(x)\chi(z))
\eqno(A.130)
$$
where
$$
\overline{\chi}(x) \equiv \chi(x)\int_0^x\chi^{-2}(y)dy
\eqno(A.131)
$$
is the irregular solution of the same Schroedinger equation
(A.58), satisfied by $\chi(x)$. I.e.,
$$
(T+V(x)+\hat{w}(x))\overline{\chi}(x)=\hat{E}_0\overline{\chi}(x).
\eqno(A.132)
$$
Consequently, over the entire range $-\gamma<x<\gamma$
$$
\overline{\chi}'(x) \chi(x)-\chi'(x)\overline{\chi}(x)=1.
\eqno(A.133)
$$
According to (A.11), (A.12) and (A.57), we have
\begin{eqnarray}\label{A.134}
\overline{\chi}(x)= \left\{
\begin{array}{l}
A \sin k_a(-x+\gamma)+\frac{1}{k_a}\left( \frac{\sin k_a\beta}{
\cosh q_a\alpha}\right)\cos k_a(-x+\gamma) ~~~~~~~~~~~\alpha< x <
\gamma\\ ~~\\
\frac{1}{q_a}\sinh q_a x~~~~~~~~~~~~~~~~~~~~~~~~~~~~~~~~~~~~~~~~~~~~~~~~~~~~~~~~~~0< x < \alpha\\
~~\\
\frac{1}{q_b}\sinh q_b
x~~~~~~~~~~~~~~~~~~~~~~~~~~~~~~~~~~~~~~~~~~~~~~{\sf
for}~~~~~-\alpha < x < 0\\
~~\\
-B \sin p_b(x+\gamma)-\frac{1}{p_b}\left( \frac{\sin p_b\beta}{
\cosh q_b\alpha}\right) \cos p_b(x+\gamma) ~~~~~~~~~~~-\gamma < x
< -\alpha
\end{array}
\right.~~~~(A.134)\nonumber
\end{eqnarray}
where $A$ and $B$ are constants given by
\begin{eqnarray}\label{A.135}
&A=\frac{1}{q_a}\left( \frac{\sinh q_a\alpha}{ \sin
k_a\beta}\right)-
\frac{1}{k_a}\left( \frac{\cos k_a\beta}{ \cosh q_a\alpha}\right)\nonumber\\
{\sf and}~~~~&~~~~ ~~~~~~~~~~~~~~~~~~~~~~~~~~~~~~~~~
~~~~~~~~~~~~~~~~~~~~~~~~~~~~~~~~~~~~~~~~~~~~~~~~~~~~~~~~~~~(A.135)\nonumber\\
&B=\frac{1}{q_b}\left( \frac{\sinh q_b\alpha}{ \sin
p_b\beta}\right)- \frac{1}{p_b}\left( \frac{\cos p_b\beta}{ \cosh
q_b\alpha}\right)~.\nonumber
\end{eqnarray}
Since in (A.128), there are only single integrations of the
products $\chi(z)\psi(z)$ and $\overline{\chi}(z)\psi(z)$, one can
readily verify that $f(x)$ satisfies the integral equation, and
therefore also its equivalent form (A.68).\\

\noindent {\bf A.4 The Iterative Sequence}

The integral equation (A.68), or its equivalent form (A.128), will
now be solved iteratively by introducing
$$
\psi_n(x)=\chi(x)f_n(x). \eqno(A.136)
$$
As in (4.87)-(4.89), $f_n(x)$ and its associated energy
$\hat{{\cal E}}_n$ are determined by
$$
f_n(x)=1-2\int_x^{\gamma}\chi^{-2}(y)dy\int_y^{\gamma}\chi^2(z)(\hat{w}(z)
-\hat{{\cal E}}_n)f_{n-1}(z)dz \eqno(A.137)
$$
and
$$
\int_{-\gamma}^{\gamma}\chi^2(z)(\hat{w}(z) -\hat{{\cal
E}}_n)f_{n-1}(z)dz=0. \eqno(A.138)
$$
When $n=0$, we set
$$
f_0(x)=1. \eqno(A.139)
$$
Introduce
$$
M_n \equiv\int_0^{\gamma}\chi^2(x)f_n(x)dx \eqno(A.140)
$$
and
$$
N_n \equiv\int_{-\gamma}^0\chi^2(x)f_n(x)dx. \eqno(A.141)
$$
From (A.59) and
$$
E_n \equiv E_a-\hat{{\cal E}}_n, \eqno(A.141)
$$
we derive
$$
\hat{{\cal E}}_n=\frac{N_{n-1}}{M_{n-1}+N_{n-1}}(E_a-E_b)
\eqno(A.142)
$$
and
$$
E_n=\frac{M_{n-1}E_a+N_{n-1}E_b}{M_{n-1}+N_{n-1}}~. \eqno(A.143)
$$
For $n=1$, we have from (A.139)-(A.141),
$$
M_0=\frac{1}{2} \left\{\alpha+\frac{1}{2q_a} \sinh
2q_a\alpha+\frac{\cosh^2 q_a\alpha}{\sin^2 k_a\beta}
(\beta-\frac{1}{2k_a}\sin 2k_a\beta)\right\}, \eqno(A.144)
$$
$$
N_0=\frac{1}{2} \left\{\alpha+\frac{1}{2q_b} \sinh
2q_b\alpha+\frac{\cosh^2 q_b\alpha}{\sin^2 p_b\beta}
(\beta-\frac{1}{2p_b}\sin 2p_b\beta)\right\}~~~~~ \eqno(A.145)
$$
and
$$
E_1=E_b+\frac{M_0}{M_0+N_0}(E_a-E_b). \eqno(A.146)
$$
For small $\mu^2$, since $E_a-E_b$ and $M_0-N_0$ are both
$O(\mu^2)$, we find
$$
E_1\cong E_b+\frac{1}{2}(E_a-E_b)+O(\mu^4) \eqno(A.147)
$$
in agreement with (A.117), given by the simple two-level formula.

Next, we examine the integration for $f_n(x)$. Consider first the
region
$$
\alpha<x<\gamma; \eqno(A.148)
$$
(A.137) can be written as
$$
f_n(x)=1+2\hat{{\cal E}}_n\int_x^{\gamma}\chi^{-2}(y)dy
\int_y^{\gamma}\chi^2(z)f_{n-1}(z)dz. \eqno(A.149)
$$
Introduce
$$
\xi=k_a(-x+\gamma) \eqno(A.150)
$$
$$
\epsilon_m=\frac{2\hat{{\cal E}}_m}{k_a^2} \eqno(A.151)
$$
and
$$
\psi_n(x)=\chi(x)f_n(x)= \left\{\frac{\cosh q_a\alpha}{\sin
k_a\beta} \right\}v_n(\xi). \eqno(A.152)
$$
When $n=0$, we set
$$
v_0(\xi)=\sin \xi. \eqno(A.153)
$$
From (A.149), or more conveniently by using $(x|G|z)$ given by
(A.130), one can readily verify that, for $\alpha<x<\gamma$,
\begin{eqnarray}\label{A.154}
~~~~v_1(\xi) &=& (1+\frac{1}{2}~\epsilon_1) \sin \xi
-\frac{1}{2}~\epsilon_1~\xi~\cos \xi,\nonumber\\
~~~~v_2(\xi) &=&
\left\{(1+\frac{1}{2}~\epsilon_2+\frac{3}{8}~\epsilon_1\epsilon_2)
-\frac{1}{8}~\epsilon_1\epsilon_2~\xi^2\right\}\sin \xi-
(\frac{1}{2}~\epsilon_2+\frac{3}{8}~\epsilon_1\epsilon_2)\xi \cos
\xi,
~~~~~(A.154)\nonumber\\
~~~~v_3(\xi) &=&
\left\{(1+\frac{1}{2}~\epsilon_3+\frac{3}{8}~\epsilon_2\epsilon_3
+\frac{5}{16}~\epsilon_1\epsilon_2\epsilon_3)
-\frac{1}{8}(\epsilon_2\epsilon_3
+\epsilon_1\epsilon_2\epsilon_3)~\xi^2\right\}\sin \xi\nonumber\\
&&+\left\{-(\frac{1}{2}~\epsilon_3+\frac{3}{8}~\epsilon_2\epsilon_3
+\frac{5}{16}~\epsilon_1\epsilon_2\epsilon_3)~\xi
+\frac{1}{48}~\epsilon_1\epsilon_2\epsilon_3~\xi^3\right\}\cos
\xi,\nonumber
\end{eqnarray}
etc. These solutions can also be readily derived by directly using
the differential equation satisfied by $\psi_n(x)=\chi(x)f_n(x)$:
$$
(T+V(x)+\hat{w}(x)-E_a)\psi_n(x)=(\hat{w}(x)-\hat{{\cal
E}}_n)\psi_{n-1}(x), \eqno(A.155)
$$
where in accordance with (A.14), $E_a=\frac{1}{2}(\mu^2+k_a^2)$.
For $\alpha<x<\gamma$, we have
$$
V(x)=\frac{1}{2}\mu^2,~~~~~~~\hat{w}(x)=0
$$
and therefore
$$
(T-\frac{1}{2}k_a^2)\psi_n(x)=-\hat{{\cal E}}_n\psi_{n-1}(x).
\eqno(A.156)
$$
Introduce $S_n(\xi)$ and $C_n(\xi)$ to be polynomials in $\xi$,
with
$$
v_n(\xi) \equiv (\prod_1^n\epsilon_m) \left\{S_n(\xi)\sin
\xi+C_n(\xi)\cos \xi\right\}. \eqno(A.157)
$$
From (A.152) and (A.156)-(A.157), we find
\begin{eqnarray}\label{A.158}
~~~~~~~~~~~~~~~~~~~~~~~~~~~~~~~~\ddot{S}_n(\xi)-2\dot{C}_n(\xi)&=& S_{n-1}(\xi)\nonumber\\
{\sf and}~~~~~~~~~~~~~~~~~~~~~~~~~~~~~~~~~~~~~~~~~~~~~~~~~~~~&&
~~~~~~~~~~~~~~~~~~~~~~~~~~~~~~~~~~~~~~~~~~(A.158)\nonumber\\
~~~~~~~~~~~~~~~~~~~~~~~~~~~~~~~~\ddot{C}_n(\xi)+2\dot{S}_n(\xi)&=&
C_{n-1}(\xi),\nonumber
\end{eqnarray}
where the dot denotes $\frac{d}{d\xi}$, so that
$\dot{S}_n=\frac{dS_n}{d\xi}$, etc. At $x=\gamma$, we have
$\xi=0$, $f_n(\gamma)=\psi'_n(\gamma)/\chi'(\gamma)=1$ and
therefore
$$
S_n(0)+\dot{C}_n(0)=\prod_{m=1}^n\epsilon^{-1}_m~. \eqno(A.159)
$$
For $n=0$, $S_0(\xi)=1$ and $C_0(\xi)=0$. Therefore, for $n=1$,
(A.158) becomes
\begin{eqnarray}\label{A.160}
~~~~~~~~~~~~~~~~~~~~~~~~~~~~~~~~\ddot{S}_1-2\dot{C}_1&=& 1\nonumber\\
~~~~~~~~~~~~~~~~~~~~~~~~~~~~~~~~\ddot{C}_1+2\dot{S}_1&=& 0~.
~~~~~~~~~~~~~~~~~~~~~~~~~~~~~~~~~~~~~~~~~~~~~~~~~(A.160)\nonumber
\end{eqnarray}
Assuming $S_1$ and $C_1$ to be both  polynomials of $\xi$, we can
readily verify that $S_1$ is a constant and $C_1$ is proportional
to $\xi$. Using (A.160) and the boundary condition (A.159), we can
establish the first equation in (A.154), and likewise the other
equations for $n>1$.

To understand the structure of $v_1(\xi)$, $v_2(\xi)$,
$v_3(\xi),~\cdots$, we may turn to the exact solution $\psi(x)$
given by (A.123). In analogy to (A.152), we define $v(\xi)$
through
$$
\psi(x) \equiv \frac{\cosh q_a \alpha}{\sin k_a \beta}~v(\xi)~.
\eqno(A.161)
$$
Thus, for $\alpha<x<\gamma$,
$$
v(\xi)=\frac{k_a}{k}~\sin k(-x+\gamma). \eqno(A.162)
$$
From (A.13)-(A.14) and (A.118), we have
$$
\hat{{\cal E}}=E_a-E=\frac{1}{2}~(k_a^2-k^2). \eqno(A.163)
$$
In terms of
$$
\epsilon\equiv ~\frac{2\hat{{\cal E}}}{k_a^2}~ =
1-\frac{k^2}{k_a^2}~, \eqno(A.164)
$$
we write
$$
v(\xi)=\frac{1}{\sqrt{1-\epsilon}}~ \sin(\xi\sqrt{1-\epsilon})
\eqno(A.165)
$$
with $\xi$ given by (A.150), as before. It is straightforward to
expand $v(\xi)$ as a power series in $\epsilon$:
\begin{eqnarray}\label{A.166}
~~~~~~~~~~~~~~v(\xi)&=&\left\{(1+\frac{1}{2}\epsilon+\frac{3}{8}\epsilon^2+\frac{5}{16}\epsilon^3)
-\frac{1}{8}(\epsilon^2+\epsilon^3)\xi^2\right\}\sin \xi\nonumber\\
&&+\left\{-(\frac{1}{2}\epsilon+\frac{3}{8}\epsilon^2+\frac{5}{16}\epsilon^3)\xi
+\frac{1}{48}\epsilon^3\xi^3\right\}\cos \xi +O(\epsilon^4)
~~~~~~~~~~~~~~~~(A.166)\nonumber
\end{eqnarray}
To compare the above series with $v_n(\xi)$ of (A.154), we can
neglect $O(\epsilon^{n+1})$ in (A.166). The replacements of all
linear $\epsilon$-terms by $\epsilon_n$, $\epsilon^2$-terms by
$\epsilon_{n-1}\epsilon_n$, $\epsilon^3$-terms by
$\epsilon_{n-2}\epsilon_{n-1}\epsilon_n$, etc. lead from (A.166)
to $v_n(\xi)$. It is of interest to note that the expansion
(A.166) of $v(\xi)$ in power of $\epsilon$ has a radius of
convergence
$$
|\epsilon|<1. \eqno(A.167)
$$
On the other hand, the iterative sequence $\{v_n(\xi)\}$ is always
convergent, on account of the Hierarchy Theorem. The main
difference between (A.154) and (A.166) is that in (A.154) each
iterative  $\epsilon_n$ is determined by the fraction (A.142).

\newpage

\begin{tabular}{|c|c|c|c|}
\hline
region& $\epsilon_1(x)$ & $\xi(x)$ & $\psi_1(x)$\\
\hline {\rm I} & $\frac{2}{k_a^2}(E_a-E_1)$ & $k_a(-x+\gamma)$ &
      $\frac{\cosh q_a\alpha}{\sin k_a\beta}\left\{(1+\frac{1}{2}\epsilon_1)\sin\xi
      -\frac{1}{2}\epsilon_1\xi\cos\xi\right\}$\\
{\rm II} & $\frac{2}{q_a^2}(E_a-E_1)$ & $q_a x$ &
      $(\kappa_{{\rm II}}+\frac{1}{2}\epsilon_1\xi)\sinh \xi + \rho_{{\rm II}}\cosh \xi$\\
{\rm III} & $\frac{2}{q_b^2}(E_b-E_1)$ & $-q_b x$ &
      $(\kappa_{{\rm III}}+\frac{1}{2}\epsilon_1\xi)\sinh \xi + \rho_{{\rm III}}\cosh \xi$\\
{\rm IV} & $\frac{2}{p_b^2}(E_b-E_1)$ & $p_b(x+\gamma)$ &
      $\frac{\cosh q_b\alpha}{\sin p_b\beta}\left\{(\kappa_{{\rm IV}}+\frac{1}{2}\epsilon_1)\sin\xi
      -\frac{1}{2}\epsilon_1\xi\cos\xi\right\}$\\
\hline
\end{tabular}

\begin{center}
{\small Table 2. The $n=1$ iterative solution $\psi_1(x)$ in the
four
regions:\\
{\rm I} ($\alpha<x<\gamma$), {\rm II} ($0<x<\alpha$),
{\rm III} ($-\alpha<x<0$) and {\rm IV} ($-\gamma<x<-\alpha$).\\
The constants $E_1=E_a-\hat{{\cal E}}_1$, $\kappa_{{\rm
II}},~\kappa_{{\rm III}},~\kappa_{{\rm IV}},~\rho_{{\rm II}}$ and
$\rho_{{\rm III}}$ are given by (A.175)-(A.177).}
\end{center}

In a similar way, we can derive $\psi_n(x)$ in other regions,
$-\alpha<x<\alpha$ and $-\gamma<x<-\alpha$. The results for $n=1$
are given in Table 2. The functions $\epsilon_1(x)$ and $\xi(x)$
are discontinuous from region to region. The constants
$\kappa_{{\rm II}}$ and $\rho_{{\rm II}}$ are determined by
requiring $\psi_1(x)$ and $\psi'_1(x)$ to be continuous at
$x=\alpha$. In region {\rm I}, when $x=\alpha+$, we have
$$
\psi_1(\alpha+)=\frac{\cosh q_a\alpha}{\sin
k_a\beta}\left\{(1+\frac{1}{2}\epsilon_{{\rm I}})\sin
k_a\beta-\frac{1}{2}\epsilon_{{\rm I}}k_a\beta \cos
k_a\beta\right\} \eqno(A.168)
$$
and
$$
\psi'_1(\alpha+)=-k_a\frac{\cosh q_a\alpha}{\sin
k_a\beta}\left\{\cos k_a\beta+\frac{1}{2}\epsilon_{{\rm
I}}k_a\beta \sin k_a\beta\right\} \eqno(A.169)
$$
where the constant
$$
\epsilon_{{\rm I}}\equiv \frac{2}{k_a^2}(E_a-E_1)=\epsilon_1(x)
~~~~{\sf in}~~{\rm I} \eqno(A.170)
$$
with
$$
E_1=E_a-\hat{{\cal E}}_1. \eqno(A.171)
$$
In region {\rm II}, when $x=\alpha-$
$$
\psi_1(\alpha-)=(\kappa_{{\rm II}}+\frac{1}{2}\epsilon_{{\rm
II}}q_a\alpha)\sinh q_a\alpha+\rho_{{\rm II}}\cosh q_a\alpha
\eqno(A.172)
$$
and
$$
\psi'_1(\alpha-)=q_a\left\{(\kappa_{{\rm II}}+\frac{1}{2}
\epsilon_{{\rm II}}q_a\alpha )\cosh q_a\alpha+(\rho_{{\rm
II}}+\frac{1}{2}\epsilon_{{\rm II}}) \sinh q_a\alpha\right\}
\eqno(A.173)
$$
where the constant
$$
\epsilon_{{\rm II}} \equiv \frac{2}{q_a^2}(E_a-E_1)=\epsilon_1(x)
~~~~{\sf in}~~{\rm II}. \eqno(A.174)
$$
The constants $\kappa_{{\rm II}}$ and $\rho_{{\rm II}}$ are
determined by
$$
\psi_1(\alpha-)=\psi_1(\alpha+)~~~~{\sf and}~~~~
\psi'_1(\alpha-)=\psi'_1(\alpha+). \eqno(A.175)
$$
Likewise, the constants $\kappa_{{\rm III}}$ and $\rho_{{\rm
III}}$ are given by
$$
\psi_1(0-)=\psi_1(0+)~~~~{\sf and}~~~~ \psi'_1(0-)=\psi'_1(0+),
\eqno(A.176)
$$
and the constants $\kappa_{{\rm IV}}$ and $E_1$ are determined by
$$
\psi_1(-\alpha-)=\psi_1(-\alpha+)~~~~{\sf and}~~~~
\psi'_1(-\alpha-)=\psi'_1(-\alpha+). \eqno(A.175)
$$

\end{document}